\documentclass[aps,prd,superscriptaddress,
preprint,tightenlines,nofootinbib,floatfix]{revtex4}

\usepackage{graphicx} % Include figure files
\usepackage{dcolumn}  % Align table columns on decimal point
\usepackage{bm}       % bold math

\usepackage{epstopdf}
\usepackage{rotating}
\usepackage{amsmath}

\newcommand{\prt}[1]{\ensuremath{\mathrm{#1}}}
\newcommand{\tabref}[1]{Tab~\ref{#1}}

\begin{document}

%\preprint line(s) will be ignored for PRL/PRD
%\preprint{CBX YYYY-NN}     % For CBX         -- YYYY-NN from the Document Database
%\preprint{CPDRAFT YYYY-NN} % For Paper Draft -- YYYY-NN from the Document Database
%\preprint{Author}          % For CBX & Paper Draft -- one line for each author
%\preprint{CLNS YY/NNNN}       % for CLNS notes -- comes with the author list
%\preprint{CLEO YY-NN}         % for CLNS notes -- comes with the author list

% Use this form if you DO NOT have mathematical symbols in the title
%\title{Your Title Goes Here}

% Add \boldmath if you DO have mathematical symbols in the title
\title{Amplitude analysis of \boldmath $D^0 \to K^+K^-\pi^+\pi^-$}

% for conference papers (ask CLEOAC for appropriate text)
%\thanks{Submitted to the 31$^{\rm st}$ International Conference on High Energy
%Physics, July 2002, Amsterdam}

\preprint{CLNS 11/2082}  % the CLNS number
\preprint{CLEO 11-08}    % the CLEO number

\author{M.~Artuso}
\author{S.~Blusk}
\author{R.~Mountain}
\author{T.~Skwarnicki}
\author{S.~Stone}
\author{L.~M.~Zhang}
\affiliation{Syracuse University, Syracuse, New York 13244, USA}
\author{T.~Gershon}
\affiliation{University of Warwick, Coventry CV4 7AL, United Kingdom}
\author{G.~Bonvicini}
\author{D.~Cinabro}
\author{A.~Lincoln}
\author{M.~J.~Smith}
\author{P.~Zhou}
\author{J.~Zhu}
\affiliation{Wayne State University, Detroit, Michigan 48202, USA}
\author{P.~Naik}
\author{J.~Rademacker}
\affiliation{University of Bristol, Bristol BS8 1TL, United Kingdom}
\author{D.~M.~Asner}
\altaffiliation[Present address: ]{Pacific Northwest National Laboratory, Richland, WA 99352}
\author{K.~W.~Edwards}
\author{K.~Randrianarivony}
\author{G.~Tatishvili}
\altaffiliation[Present address: ]{Pacific Northwest National Laboratory, Richland, WA 99352}
\affiliation{Carleton University, Ottawa, Ontario, Canada K1S 5B6}
\author{R.~A.~Briere}
\author{H.~Vogel}
\affiliation{Carnegie Mellon University, Pittsburgh, Pennsylvania 15213, USA}
\author{P.~U.~E.~Onyisi}
\author{J.~L.~Rosner}
\affiliation{University of Chicago, Chicago, Illinois 60637, USA}
\author{J.~P.~Alexander}
\author{D.~G.~Cassel}
\author{S.~Das}
\author{R.~Ehrlich}
\author{L.~Gibbons}
\author{S.~W.~Gray}
\author{D.~L.~Hartill}
\author{B.~K.~Heltsley}
\author{D.~L.~Kreinick}
\author{V.~E.~Kuznetsov}
\author{J.~R.~Patterson}
\author{D.~Peterson}
\author{D.~Riley}
\author{A.~Ryd}
\author{A.~J.~Sadoff}
\author{X.~Shi}
\author{W.~M.~Sun}
\affiliation{Cornell University, Ithaca, New York 14853, USA}
\author{J.~Yelton}
\affiliation{University of Florida, Gainesville, Florida 32611, USA}
\author{P.~Rubin}
\affiliation{George Mason University, Fairfax, Virginia 22030, USA}
\author{N.~Lowrey}
\author{S.~Mehrabyan}
\author{M.~Selen}
\author{J.~Wiss}
\affiliation{University of Illinois, Urbana-Champaign, Illinois 61801, USA}
\author{J.~Libby}
\affiliation{Indian Institute of Technology Madras, Chennai, Tamil Nadu 600036, India}
\author{M.~Kornicer}
\author{R.~E.~Mitchell}
\affiliation{Indiana University, Bloomington, Indiana 47405, USA }
\author{D.~Besson}
\affiliation{University of Kansas, Lawrence, Kansas 66045, USA}
\author{T.~K.~Pedlar}
\affiliation{Luther College, Decorah, Iowa 52101, USA}
\author{D.~Cronin-Hennessy}
\author{J.~Hietala}
\affiliation{University of Minnesota, Minneapolis, Minnesota 55455, USA}
\author{S.~Dobbs}
\author{Z.~Metreveli}
\author{K.~K.~Seth}
\author{A.~Tomaradze}
\author{T.~Xiao}
\affiliation{Northwestern University, Evanston, Illinois 60208, USA}
\author{L.~Martin}
\author{A.~Powell}
\author{P.~Spradlin}
\author{G.~Wilkinson}
\affiliation{University of Oxford, Oxford OX1 3RH, United Kingdom}
\author{J.~Y.~Ge}
\author{D.~H.~Miller}
\author{I.~P.~J.~Shipsey}
\author{B.~Xin}
\affiliation{Purdue University, West Lafayette, Indiana 47907, USA}
\author{G.~S.~Adams}
\author{J.~Napolitano}
\affiliation{Rensselaer Polytechnic Institute, Troy, New York 12180, USA}
\author{K.~M.~Ecklund}
\affiliation{Rice University, Houston, Texas 77005, USA}
\author{J.~Insler}
\author{H.~Muramatsu}
\author{C.~S.~Park}
\author{L.~J.~Pearson}
\author{E.~H.~Thorndike}
\affiliation{University of Rochester, Rochester, New York 14627, USA}
\author{S.~Ricciardi}
\affiliation{STFC Rutherford Appleton Laboratory, Chilton, Didcot, Oxfordshire, OX11 0QX, United Kingdom}
\author{C.~Thomas}
\affiliation{University of Oxford, Oxford OX1 3RH, United Kingdom}
\affiliation{STFC Rutherford Appleton Laboratory, Chilton, Didcot, Oxfordshire, OX11 0QX, United Kingdom}
\collaboration{CLEO Collaboration}
\noaffiliation

% Please hard code the date when you have a final draft and submit to the arXiv and journal
\date{January 27, 2012}

\begin{abstract} 
% Insert abstract here.
The first flavor-tagged amplitude analysis of the decay $D^{0}$ to the self-conjugate final state %%@
$K^{+}K^{-}\pi^{+}\pi^{-}$ 
is presented. Approximately 3000 signal decays are selected from data acquired by the CLEO II.V, CLEO III, and %%@
CLEO-c detectors. The three most significant amplitudes, which contribute to the model that best fits the data, are %%@
$\phi\rho^{0}$, $K_{1}(1270)^{\pm}K^{\mp}$, and non-resonant $K^{+}K^{-}\pi^{+}\pi^{-}$. Separate amplitude 
analyses of $D^{0}$ and $\overline{D^{0}}$ candidates indicate no $CP$ violation among the amplitudes at the level %%@
of $5\%$ to $30\%$ depending on the mode. In addition, the sensitivity to the $CP$-violating parameter %%@
$\gamma/\phi_3$ from a sample of 2000 $B^{+}\to 
\widetilde{D^{0}}(K^{+}K^{-}\pi^{+}\pi^{-})K^{+}$ decays, where $\widetilde{D}$ is a $D^{0}$ or $\overline{D^{0}}$, %%@
collected at 
LHCb or a future flavor facility, is estimated to be $(11.3\pm 0.3)^{\circ}$ using the favored model. 
\end{abstract}

\maketitle

\section{Introduction}

A rich variety of interesting physics may be explored by investigating the decay  $D^0 \to K^-K^+\pi^+\pi^-$.
Study of the relative contribution of the intermediate resonances participating in the decay can help in the %%@
understanding of the behavior of the strong interaction at low energies. The mode is also of interest for its %%@
application in $CP$ violation studies, both for improving the knowledge of the CKM unitarity triangle, and in %%@
probing for new physics effects through direct $CP$ violation searches in $D$ meson decays.   Although results on %%@
the resonant structure of the decay have been reported by the E791~\cite{E791} and FOCUS~\cite{FOCUS} %%@
collaborations, studies of higher precision are required. In particular, neither of these previous analyses %%@
differentiated between $D^{0}$ and $\overline{D^{0}}$ decays. 

An important goal in flavor physics is the precise determination of the CKM unitarity triangle angle $\gamma$ %%@
(denoted by others as $\phi_3$), the phase of $V_{cb}$ relative to $V_{ub}$.  This parameter can be measured %%@
through the study of interference effects in the decay $B^\pm \to \widetilde{D^0}K^\pm $.  Here $\widetilde{D^0}$ %%@
indicates either a $D^0$ or a $\overline{D^0}$ meson decaying to a hadronic final state.  Experimentally, in order %%@
to obtain the best possible knowledge of $\gamma$ it is important to make use of as many $\widetilde{D^0}$ decay %%@
modes as possible.  The  decay $\widetilde{D^0} \to K^+K^-\pi^+\pi^-$
has been noted as being of  potential interest in this respect~\cite{JONASGUY}, especially since large numbers of %%@
$B^\pm \to  \widetilde{D^0}K^\pm$ decays  involving this mode will be collected by the LHCb collaboration.  The %%@
existing knowledge of the substructure of $D^0 \to K^+K^-\pi^+\pi^-$  is however inadequate to make a reliable %%@
assessment of the potential sensitivity to $\gamma$, and so improved information is required.

A search for direct $CP$ violation in singly-Cabibbo-suppressed (SCS) charm decays is a promising method to test %%@
for the contribution of new physics, which in several plausible scenarios could lead to ${\cal{O}}(1\%)$ effects, %%@
an order of magnitude higher than is expected in the Standard Model~\cite{SCSCHARM}.  Evidence of $CP$ violation %%@
has recently been reported in two-body SCS decays \cite{BIGDOG}, hence it is important to look elsewhere.  The %%@
decay $D^0 \to K^+K^-\pi^+\pi^-$, with a rich structure of intermediate resonances, is a suitable mode in which to %%@
perform such a search.  
$CP$ violation studies involving $D^0 \to K^+K^-\pi^+\pi^-$ have been conducted by the FOCUS~\cite{FOCUSTODD} and %%@
BABAR~\cite{BABARTODD} collaborations, using the method of $T$-odd correlations, and null results have been %%@
reported.
A $CP$ violation search made with an amplitude analysis  remains valuable however, as it probes each intermediate %%@
resonance of the decay separately, and hence can expose effects which may be diluted or concealed by the more %%@
inclusive $T$-odd correlation approach.

This paper describes a flavor-tagged amplitude analysis of the decay $D^0 \to K^+K^-\pi^+\pi^-$ made using data %%@
collected by the CLEO collaboration in $e^+e^-$ collisions at the Cornell Electron Storage Ring (CESR).  An %%@
amplitude model is constructed and a $CP$ violation study performed.  The model is also used to assess the %%@
potential sensitivity of the decay in a  future $B^\pm \to  \widetilde{D^0}K^\pm$  $\gamma$ measurement.  The data %%@
analysed were collected with several different configurations of the CLEO detector, and at different center-of-mass %%@
energies.  Comparison of the results obtained for each data set provides a powerful test of systematic robustness. %%@
One sample consists of $CP$-tagged decays from CLEO-c running at the $\psi(3770)$.  These events provide unique %%@
access to the strong-phase differences between the intermediate resonances in a manner which was not available to %%@
previous studies.

The paper is organized as follows. Section~\ref{sec:dataset} discusses the data sets used in the analysis.  %%@
Section~\ref{sec:fit} describes the amplitude fit procedure and the development of the resonance model.  %%@
Section~\ref{sec:results} presents the final model, summarizes the systematic uncertainties and gives the result of %%@
the $CP$ violation test.  The sensitivity of the decay $D^0 \to K^+K^-\pi^+\pi^-$ in a
measurement of   $\gamma$ with $B^\pm \to  \widetilde{D^0}K^\pm$  decays
 is considered in Sec.~\ref{sec:gamma}, and conclusions are given in Sec.~\ref{sec:conclude}.

\section{Data set and event selection}
\label{sec:dataset}

The data analysed in this paper were produced in symmetric $e^+e^-$ collisions at CESR between 1995 and 2008, and %%@
collected with three different configurations of the CLEO detector:
CLEO II.V, CLEO III, and CLEO-c.

In CLEO II.V~\cite{CLEO2PT5} tracking was provided by a three-layer double-sided silicon vertex detector, and two %%@
drift chambers. Charged particle identification came from $dE/dx$ information in the drift chambers, and %%@
time-of-flight (TOF) counters inserted before the calorimeter.  For CLEO III~\cite{CLEO3} a new silicon  vertex %%@
detector was installed, and a ring imaging Cherenkov (RICH) detector was deployed to enhance the particle %%@
identification abilities~\cite{CLEORICH}.  In CLEO-c, the vertex detector was replaced with a low-mass wire drift %%@
chamber \cite{ZD}.  A superconducting solenoid supplied a 1.5 T magnetic field for CLEO II.V and III, and 1~T for %%@
CLEO-c operation, where the average particle momentum was lower.
In all detector configurations neutral pion and photon identification was provided by a 7800-crystal  CsI %%@
electromagnetic calorimeter.

Four distinct data sets are analysed in the present study:
\begin{enumerate}
\item{
approximately 9~$\rm fb^{-1}$  accumulated at $\sqrt{s} \approx 10$~GeV by the CLEO II.V detector;
}
\item{
a total of 15.3~$\rm fb^{-1}$  accumulated by the CLEO III detector  in an energy range $\sqrt{s} = 7.0-11.2$~GeV, %%@
with over 90\% of this sample taken at  $\sqrt{s} = 9.5-10.6$~GeV;
}
\item{
 818~$\rm pb^{-1}$ collected at the $\psi(3770)$ resonance by the CLEO-c detector;
}
\item{
a further 600~$\rm pb ^{-1}$  taken by CLEO-c at $\sqrt{s} = 4170$~MeV.}
\end{enumerate}
These samples are referred to as the {\bf CLEO II.V}, {\bf  CLEO III}, {\bf CLEO-c 3770} and {\bf CLEO-c 4170} data %%@
sets, respectively.

The analysis considers two classes of signal decays, for both of which information on the quantum numbers of the %%@
meson decaying to the signal mode is provided by an event tag.
\begin{itemize}
\item{
  {\bf  Flavor-tagged decays} are selected from the CLEO II.V and CLEO III data sets,  in which the flavor of the %%@
decaying meson is determined by the charge of the `slow pion', $\pi_s$, in the $D^{*+} \to D^0 \pi^+_s$ decay %%@
chain. Flavor-tagged decays are also selected from the two CLEO-c data sets, where here the tag is obtained through %%@
the charge of a kaon associated with the decay of the other $D$ meson in the event. 
}
\item{
 { $\mathbf {CP}$\bf-tagged  decays} are selected in the CLEO-c 3770 data set alone.  In $\psi(3770)$ decays the %%@
$D-\overline{D}$ pair is produced coherently.  Therefore, the $CP$  of the signal $D$ can be determined if the %%@
other $D$ meson is reconstructed in a decay to a $CP$-eigenstate.  Useful information is also obtained if the %%@
tagging meson is reconstructed decaying into the modes
$K^0_S \pi^+\pi^-$ or $K^0_L \pi^+\pi^-$, for which the relative contribution of $CP$-even and $CP$-odd states is %%@
known~\cite{LIBBY}.
}
\end{itemize}

Detector response is studied with GEANT-based~\cite{GEANT} Monte Carlo simulations of each detector configuration, %%@
in which the Monte Carlo events are processed with the same reconstruction code as used for data.

\subsection{Flavor-tagged CLEO II.V and CLEO III samples}

Selections are run on both the CLEO II.V and CLEO III data sets to identify events containing the fully %%@
reconstructed decay chain  $D^{*+} \to D^0 \pi^+_s$, $D^0 \to K^+K^-\pi^+\pi^-$.  These selections are not %%@
identical for the two data sets on account of the different detector responses. 

Requirements are first placed on the attributes of the individual charged tracks used in the reconstruction.  The %%@
tracks must be well-measured and satisfy criteria based on fit quality. They must also be consistent with coming %%@
from the interaction point in three dimensions. In the CLEO~III analysis, the polar angle of each considered track  %%@
is required to satisfy $|\cos \theta| < 0.9$.  The momentum of the slow pion candidate 
must be above 100 (150)~MeV/$c$ in the CLEO~II.V (CLEO~III) analysis and below 500~MeV/$c$, 
and that of the other final state particles must be between 200~${\rm MeV}/c$ and 5000~MeV/$c$.  

Particle identification information plays an important role in the selection.  In the CLEO~III analysis, candidate %%@
tracks with momentum above 500~MeV/$c$ are classified as kaons if they have at least three associated photons in %%@
the RICH detector, and a ring fit to the photon hits indicates that the kaon hypothesis is more probable than that %%@
of the pion.    Lower momentum tracks, 
and those tracks lying outside the angular acceptance of the RICH,
are identified as kaons if they  have a $dE/dx$ value within 2.1~$\sigma$ 
of that expected for a true kaon.  Pion candidates are required to have a $dE/dx$ value within 
3.2~$\sigma$ 
of that expected for a true pion. 
In the CLEO~II.V study, tracks with $dE/dx$ information are identified as kaon candidates if they have a $dE/dx$ %%@
value lying within 2.1 (2.5)~$\sigma$ of that expected for a true kaon (pion).  When TOF information is available,  %%@
both kaon and pion candidates are required to lie within 2.5~$\sigma$ of their expected value.

A possible background to the signal arises from the decay $D^0 \to K^0_S (\pi^+\pi^-) K^+K^-$, and so a $K^0_S$ %%@
veto procedure is performed. If the two pions have an invariant mass compatible with a $K^0_S$ decay the event is %%@
rejected if the flight distance of the $K^0_S$ candidate from the interaction point, normalized by the assigned %%@
error, is greater than two,  or either pion has an impact parameter in the transverse plane greater than 1.5~mm. 

After all selection criteria 3.0$\%$  of the remaining events in the CLEO~II.V sample are found  to have more than %%@
a unique pair of $D^{*+}$ and $D^0$
candidates.  If there is more than one $D^{*+}$ candidate, the $D^{*+}$ chosen chosen is the one which has a %%@
$D^{*+}-D^0$ invariant mass difference closest to the expected value.
If there is more than one $D^0$ candidate, the one that is chosen is the $D^{0}$ for which the $dE/dx$ information %%@
of  the four daughter tracks best matches the signal hypothesis.    In the CLEO~III sample only 0.3$\%$  of events %%@
contain multiple candidates; these events are discarded.

Two kinematic fits are performed to the decay with the constraint that the four tracks from the $D^0$ meson %%@
candidate originate from a common vertex, and also that the $D^0$ candidate and the slow pion from the $D^{*+}$ %%@
originate from another common vertex.  Loose criteria are placed on the quality of these fits.
The analysis is optimized to 
favor charm mesons produced in the primary interaction, rather than $B$ meson decay, and so it is required that the %%@
$D^{*+}$ momentum is at least half the maximum kinematically allowed value. 
This requirement suppresses combinatoric background.

\begin{figure}[htb]
\begin{center}
\includegraphics[width=1.00\textwidth]{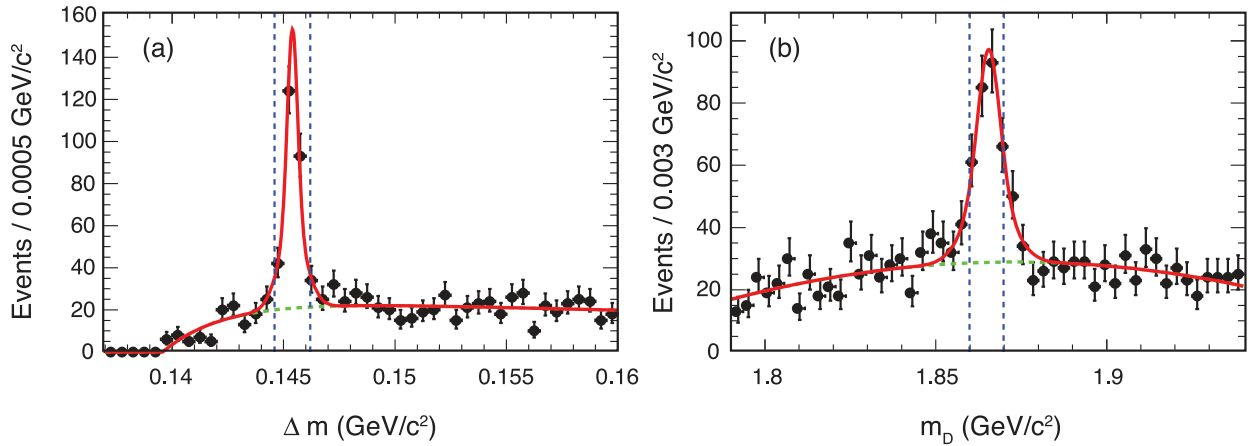}
\caption{ (a) $\Delta m$ (after $m_D$ selection) and (b) $m_D$ (after $\Delta m$ selection) distributions of  CLEO %%@
II.V $D^0$ candidates.}
\label{fig:cleo2pt5spectra}
\end{center}
\end{figure}

\begin{figure}[htb]
\begin{center}
\includegraphics[width=1.00\textwidth]{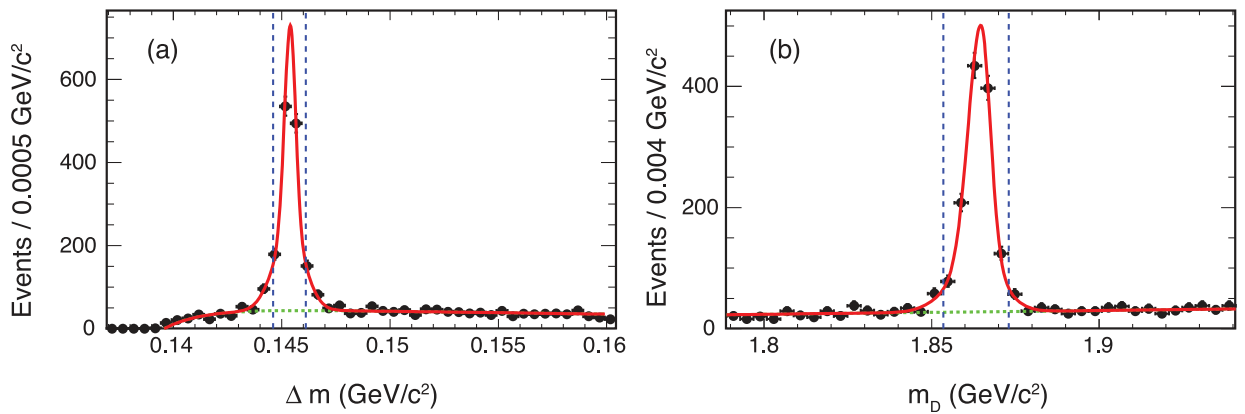}
\caption{ (a) $\Delta m$ (after $m_D$ selection) and (b) $m_D$ (after $\Delta m$ selection) distributions of  CLEO %%@
III $D^0$ candidates.}
\label{fig:cleo3spectra}
\end{center}
\end{figure}

Figure~\ref{fig:cleo2pt5spectra} shows the spectrum of the $D^0$ candidate invariant mass, $m_D$, and that of the %%@
invariant mass difference, $\Delta m$, between the $D^{*+}$ and $D^0$ candidates for the CLEO~II.V selection after %%@
the vertex-constrained fit.  In Fig.~\ref{fig:cleo3spectra} are shown the equivalent plots for the CLEO~III %%@
analysis.    

The $m_D$ and $\Delta m$  distributions are fit with single (sum of bifurcated)  Gaussians for the signal peaks
in the CLEO II.V (CLEO III) sample, and empirical functions to 
describe the background.  A signal region is defined as being within $\pm 5.0$~MeV/$c^2$ %%@
($^{+8.3}_{-11.2}$~MeV/$c^2$) of the world average value of the $D^0$ mass~\cite{PDG} for $m_D$ in the CLEO II.V %%@
(CLEO III) case, and within $\pm 0.80$~MeV/$c^2$ ($^{+0.72}_{-0.80}$~MeV/$c^2$) of the world average value of the %%@
$D^{*+}-D^0$ mass difference for $\Delta m$. 
An analysis of the $\pi^+\pi^-$ invariant mass spectrum, detailed in Sec.~\ref{sec:fit}, indicates that the peaking %%@
background from $D^0 \to K^0_S K^+K^-$ events in the sample is negligible, and so the overall signal yield and %%@
background level can be extracted directly from these fits.  For both data sets the fits to the $m_D$ and $\Delta %%@
m$ spectra give consistent results.  In total 279 events are selected in the signal region from the CLEO~II.V data %%@
set, with an estimated purity of $(74 \pm 3)\%$.  In the CLEO~III analysis  1225 events are selected with a purity %%@
of $(89.2 \pm 0.4)\%$. 

In order to learn about the characteristics of the contamination in the signal region, dedicated background samples %%@
are also selected.  These are taken from three separate regions of $m_D - \Delta m$ space:  one with the same %%@
acceptance on $m_D$ as the signal region, but with 
$148.5 < \Delta m < 155.5 $~MeV/$c^2$,
and two others with the same acceptance on $\Delta m$ as for the signal sample, but with $1.7 < m_D < %%@
1.8$~GeV/$c^2$ and $1.9 < m_D < 1.95$~GeV/$c^2$ respectively.  Simulation studies indicate that these regions %%@
contain negligible contributions from true signal decays, and that the attributes of the selected events are %%@
representative of those of the background events  in the signal sample.

In order to improve the resolution of the four-momenta used in the amplitude analysis, the selected events in the %%@
signal sample are subjected to a kinematical fit in which the $D^0$ candidate daughter particles are constrained to %%@
originate from a common vertex, and to have an invariant mass equal to the nominal $D^0$ mass.  The same procedure %%@
is applied to the selected events in the background sample.

The performance of the flavor tag is calibrated in data using $D^0 \to K^-\pi^+\pi^-\pi^+$ decays.  These decays, %%@
accompanied by a slow pion, are selected with the same procedure as for the signal sample. By comparing the charge %%@
of the tagging slow pion with that of the slow pion from the fully reconstructed $D$ decay it is possible to %%@
determine directly the mistag rate, that is, the fraction of occasions on which the tagging decision is incorrect.  %%@
Small corrections are applied to account for the contribution from doubly-Cabibbo-suppressed decays.  Simulation %%@
studies are used to validate that the mistag rate as determined by this procedure is consistent with that of  $D^0 %%@
\to K^+K^-\pi^+\pi^-$ decays. This study is only performed on the CLEO III sample, yielding a mistag rate of $(0.64 %%@
\pm 0.05)$.  In the subsequent amplitude analysis the same value is taken to apply for the CLEO II.V sample, and a %%@
relative $50\%$ uncertainty assigned to account for this assumption.

\subsection{Flavor-tagged CLEO-c samples}

Flavor tagging is performed in both CLEO-c data sets by searching for another charged kaon in the event, in %%@
addition to those used in reconstructing the signal $D$ decay. Such a tagging kaon originates from the decay of the %%@
other $D$ meson. If this decay is assumed to be Cabibbo-favored, and there are no other additional kaons in the %%@
event, then the charge of the tagging kaon indicates the flavor of the decaying meson, and hence the flavor of the %%@
signal decay can be inferred.

Standard CLEO-c selection criteria, as described in Ref.~\cite{SELECTION}, are imposed on the tracks used in the %%@
$D^0$ reconstruction and for the tagging kaon. Events are only considered with a single tagging kaon candidate, on %%@
which a momentum cut is applied.  Selecting higher momentum kaons is found to be advantageous both in enhancing the %%@
purity of the sample, and in suppressing events where the tagging decision is incorrect. The momentum of the %%@
tagging kaon is required to exceed 400~MeV/$c$ in the CLEO-c 3770 data set, and 600~MeV/$c$ in the CLEO-c 4170 data %%@
set.

It is necessary to apply a more stringent $K^0_S$ veto to suppress $D^0 \to K^0_S K^+K^-$ contamination than in the %%@
CLEO~II.V and CLEO~III selections.  This is because at CLEO-c the $D$ mesons are produced at or close to threshold, %%@
and hence the flight distance of any resulting $K^0_S$ is lower.  Therefore, events are rejected in which a $K^0_S$ %%@
candidate has a flight distance, normalized by the assigned uncertainty, of greater than one,  or in which either %%@
of the daughter pions has an impact parameter in the transverse plane greater than 1~mm. 

\subsubsection{CLEO-c 3770  sample}
\label{sec:cleoc3770}

\begin{figure}[htb]
\begin{center}
\includegraphics[width=1.00\textwidth]{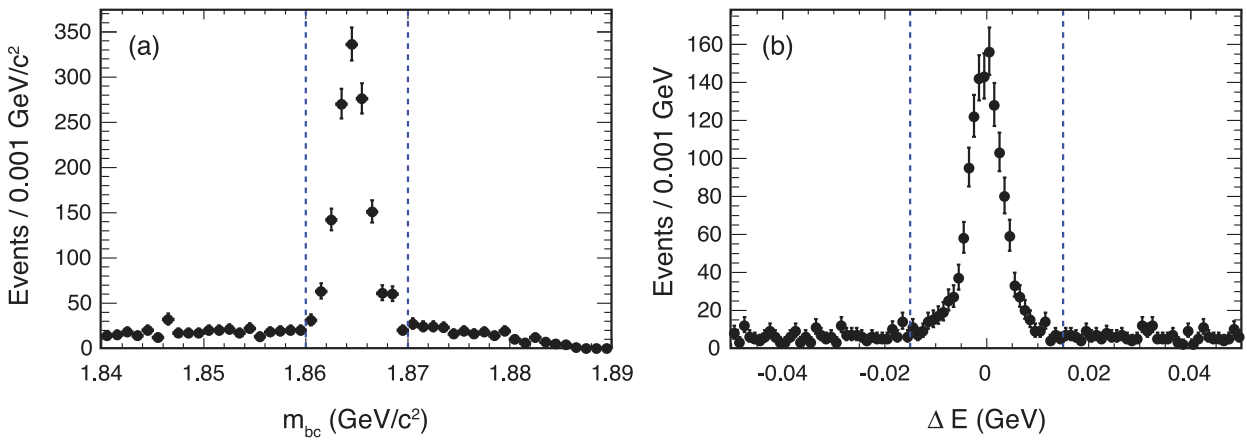}
\caption{ (a) $m_{bc}$ and (b) $\Delta E$ distributions of events passing the kaon-tagged $D^0 \to %%@
K^+K^-\pi^+\pi^-$ selection in the CLEO-c 3770 data set. Each distribution is plotted after applying the selection %%@
cut on the other variable.}
\label{fig:mbced_3770}
\end{center}
\end{figure}

Two kinematical variables are defined: the beam-constrained candidate mass, 
\begin{displaymath}
m_{bc} \equiv \sqrt{s/(4 c^4) -  \mathbf{p}_D^2 / c^2} \; , 
\end{displaymath}
where $\mathbf{p}_D$ is the momentum of the signal $D$ candidate, and $\Delta E \equiv E_D - \sqrt{s}/2$, where %%@
$E_D$ is the sum of the energies of the daughter particles of the signal $D$  candidate.   The distributions of %%@
$m_{bc}$ and $\Delta E$ are shown in Fig.~\ref{fig:mbced_3770} for kaon-tagged candidates in the CLEO-c 3770 data %%@
set.  The signal decays peak at the nominal $D^0$ mass  in $m_{bc}$ and zero in $\Delta E$.  In making the final %%@
selection a window of $\pm 5$~MeV/$c^2$ and $\pm 15$~MeV is placed around these expected values for $m_{bc}$ and %%@
$\Delta E$ respectively.   A sample of 1396 events is selected in the signal region, of which 14 contain two %%@
candidates.

In this latter class of event only one candidate, chosen at random,  is retained for subsequent analysis.
Interpolating the results of fits to the sideband regions into the signal window indicates that the contamination %%@
from non-peaking background is at the level of $(13.5 \pm 0.5)\%$.  The residual contamination from $D^0 \to K^0_S %%@
K^+K^-$ decays is found to constitute $(2.4 \pm 0.4) \%$ of the sample, as determined from the amplitude fit %%@
studies described in Sec.~\ref{sec:fit}.

In addition a sample of 763 events  is selected for non-peaking background studies from the regions defined by $-5 %%@
< (m_{bc} - 1865 \,{\rm MeV}/c^2) < 5$ MeV/$c^2$ and $| \Delta E \pm 45\,{\rm MeV} | < 30$~MeV.
A further sample of 445 events which fail the $K^0_S$ veto, but pass all other signal selection criteria, are %%@
selected in order to characterize the residual $D^0 \to K^0_S K^+K^-$ contamination.

 The performance of the flavor tag is calibrated in data using $D^0 \to K^-\pi^+\pi^-\pi^+$ decays.  These decays, %%@
accompanied by a tagging kaon, are selected with the same procedure as for the signal sample. The method is %%@
validated using simulated data and corrections are made for the doubly-Cabibbo-suppressed decays in the sample as %%@
for the CLEO III calibration. It is concluded that the mistag rate of signal events in data is $(4.5\pm 0.5)\%$.

All selected events are subjected to a kinematical fit with the invariant mass of the candidate constrained to that %%@
of the $D^0$, in order to provide the best possible resolution for the amplitude study.

\subsubsection{CLEO-c 4170 sample}
\label{sec:cleoc4170}

At $\sqrt{s} = 4170$~MeV, pairs of charm mesons can be produced in a variety of configurations, including
 $D \overline{D}$, $D^*\overline{D}$, $D^*\overline{D}{}^*$, $D^* \overline{D}\pi$, $D_s^+D_s^-$ and  %%@
$D_s^{+*}D_s^-$. Several of these configurations may result in events which contain a $D^0$ accompanied by a %%@
$\overline{D^0}$, or a $D^0$ and a $D^-$. Depending on the production process and subsequent strong or %%@
electromagnetic decay, there will be one or more prompt pions or photons also present in the event.  Even without %%@
reconstructing these additional particles it is possible to separate statistically the different production and %%@
decay categories, as they exhibit different distributions in $m_{bc} - \Delta E$ space. This property has been %%@
exploited in Ref.~\cite{PROD4170} to study charm production at these energies.

\begin{figure}[htb]
\begin{center}
\includegraphics[width=1.00\textwidth]{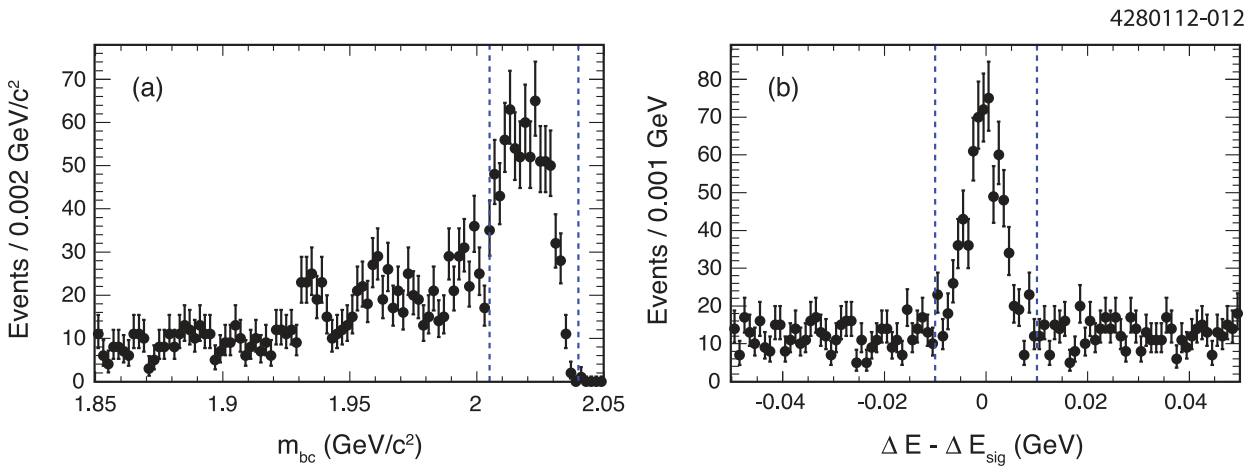}
\caption{ Distributions of (a) $m_{bc}$ (after $\Delta E - \Delta E_{\rm sig}$ selection) and (b) $\Delta E - %%@
\Delta E_{\rm sig}$ (after $m_{bc}$ selection) 
 for events passing the kaon-tagged $D^0 \to K^+K^-\pi^+\pi^-$ selection in the CLEO-c 4170 data set.  }
\label{fig:mbcde_4170}
\end{center}
\end{figure}

$D^*\overline{D}{}^*$  events have the highest rate and intrinsic purity, and so these are isolated for the %%@
amplitude analysis.
A variable $\Delta E_{\rm sig} \equiv a - b~m_{bc}$ is defined, where the coefficients $a = 2.112$~GeV
and $b= 1.12 \, c^2$ are obtained
from a fit to the distribution of simulated signal candidates, and events are selected in the region $2.005 < %%@
m_{bc} < 2.040$~GeV/$c^2$ and $|\Delta E - \Delta E_{\rm sig}|  < 10$~MeV.
In addition, to suppress background further, a restriction is placed on the momentum of the $D^0$ candidate that it %%@
be above 450~MeV/$c$.
Figure~\ref{fig:mbcde_4170}~(a) presents the distribution of $m_{bc}$ with the cut on $\Delta E - \Delta E_{\rm %%@
sig}$ applied,
and Fig.~\ref{fig:mbcde_4170}~(b) the corresponding plot for  $\Delta E - \Delta E_{\rm sig}$,  after selecting on %%@
$m_{bc}$.
A total of 739 events is selected,  of which 5 contain a second candidate. 
In the case of these multiple candidate events,  only one candidate, chosen at random, is retained
for the subsequent analysis.

Two regions are selected to provide non-peaking background samples. These are defined by the criteria $2.005 < %%@
m_{bc} < 2.040$~GeV/$c^2$, $|(\Delta E - \Delta E_{\rm sig}) \pm 45\, {\rm MeV}|  < 20$~MeV and $p_D > %%@
450$~MeV/$c$. This selection yields a total of 978 background candidates. Simulation indicates that the density and %%@
nature of the candidates in this sample is compatible with that of the contamination inside the signal window.  It %%@
is thus estimated that the fraction of non-peaking background in the signal sample is $(33 \pm 1) \%$.
The fraction of $D^0 \to K^0_S K^+K^-$ decays in the signal sample is estimated to be $(1.3 \pm 0.4) \%$.

The performance of the kaon flavor tag is determined using $D^0 \to K^-\pi^+\pi^-\pi^+$ events in the same manner %%@
as for the 3770~MeV sample.  The mistag rate is measured to be $(7.5 \pm 0.7)\%$.

As in the case of the 3770~MeV analysis, the selected candidates are refit with the mass of the $D^0$ imposed as a %%@
constraint.

\subsection{CP-tagged CLEO-c sample}

$CP$-tagged events are selected from the 3770~MeV CLEO-c data set in which both $D$-meson decays are reconstructed, %%@
one through its decay to $K^+K^-\pi^+\pi^-$ and the other to a $CP$-eigenstate.  
The latter decay provides a tag through which the $CP$ eigenvalue of the signal decay can be determined.   The %%@
signal decay is selected in the same manner as for the flavor-tagged sample, and is
required to lie  within $\pm 5$~MeV/$c^2$ of the nominal $D^0$ mass in $m_{bc}$  and within $\pm 20$~MeV of zero in %%@
$\Delta E$.  The selection criteria for the $CP$-tags are identical to those used 
in Ref.~\cite{NORM}.  The $CP$-tags include the modes $D \to K^0_L \pi^0$ and $K_L^0 \omega$, where the presence of %%@
the $K^0_L$ meson is inferred from a missing-mass technique, having reconstructed all 
the other particles in the event.   The number of selected candidates is presented in Table~\ref{tab:cptag}.

Events are also selected in which the tag is provided by either of the decays $K^0_S \pi^+\pi^-$ or $K^0_L %%@
\pi^+\pi^-$, reconstructed with the same requirements as in Ref.~\cite{LIBBY}. These modes can be considered %%@
admixtures of $CP$-odd and $CP$-even eigenstates, and are  exploited in the analysis thanks to available %%@
measurements of the strong-phase variation across the Dalitz space of each decay~\cite{LIBBY}. %%@
Table~\ref{tab:cptag} reports the yields for this category of event.

\begin{table}
\begin{center}
\caption[]{$CP$-tagged $D \to K^+K^-\pi^+\pi^-$ events selected from the CLEO-c 3770~MeV data set.}
\label{tab:cptag}
\begin{tabular}{lcclcclc} \hline \hline
\multicolumn{2}{c}{$CP$-even tags} & \hspace*{0.3cm} & \multicolumn{2}{c}{$CP$-odd tags}  & \hspace*{0.3cm} & %%@
\multicolumn{2}{c}{Admixture} \\
Mode & Yield  &&  Mode & Yield && Mode & Yield \\ \hline
$K^+K^-$                                       & 11    &&    $K^0_S \pi^0$ & 15                               && %%@
$K^0_S \pi^+\pi^-$ & 63  \\
$\pi^+\pi^-$                                   &  8     &&$K^0_S \omega (\pi^+\pi^-\pi^0)$ & 12  &&  $K^0_L %%@
\pi^+\pi^-$ & 87  \\
$K_L^0 \pi^0$                                & 9      &&$K^0_S \phi (K^+K^-)$ & 1                   && \\
$K^0_S \pi^0\pi^0$                       & 7      &&$K^0_S \eta (\gamma \gamma)$ & 2     && \\
$K^0_L\omega(\pi^+\pi^-\pi^0)$ & 14    &&$K^0_S \eta (\pi^+\pi^-\pi^0)$ & 1     && \\
                                                        &         && $K^0_S \eta'(\pi^+\pi^-\eta)$ & 1      && \\ %%@
\hline
Total                                                & 49    &&                                            & 32   %%@
&&           & 150 \\ \hline \hline
\end{tabular}
\end{center}
\end{table}

The level of contamination is estimated from study of  
sidebands in the two-dimensional $m_{bc}$ space of the reconstructed $D$ mesons, in the missing-mass sideband for %%@
the events
containing $K^0_L$ tag
and using Monte Carlo to determine the contribution of peaking backgrounds.
The fraction of contamination is found to be $(35.9 \pm 5.4)\%$  for the $CP$-even tags, $(17.2 \pm 3.4) \%$ for %%@
the $CP$-odd tags,  $(16.5 \pm 2.3)\%$ for the $K^0_S \pi^+\pi^-$ tags and $(19.0\pm 2.2)\%$ for the $K^0_L %%@
\pi^+\pi^-$ tags.
The sidebands do not provide a data set of sufficient size to allow the composition of the background to be %%@
studied.  For this purpose, a simulated background sample
is prepared.

Multiple candidates occur in $1.3$\%  of selected signal events.  In these events only a single candidate is %%@
propagated for further analysis, taking the signal-tag combination
in which the sum of $m_{bc}$ values is closest to twice the nominal $D^0$ mass.  In events where the tag involves a %%@
$K^0_L$, the candidate is selected for which $m_{bc}$ of 
the signal decay  is closest to the $D^0$ mass.  All selected candidates are refit with a $D^0$ mass constraint.

\subsection{Summary of signal samples}

A summary of the tagged samples is presented in Table~\ref{tab:samplesummary}.  A total of 3639 flavor-tagged %%@
events and 231 $CP$-tagged events are selected for the amplitude analysis studies, of which 2959 and 181 events, %%@
respectively, are estimated to be true signal decays.

\begin{table}[htb]
\begin{center}
\caption[]{Summary of signal samples.}
\label{tab:samplesummary}
\begin{tabular}{lccc} \hline \hline
Sample & \hspace*{0.2cm} Yield  \hspace*{0.2cm}& \hspace*{0.2cm} Purity [$\%$]  & Mistag rate %%@
[$\%$]\hspace*{0.2cm}\\  \hline
Flavor tags & & \\
CLEO II.V  & 279  &  $74 \pm 3$ & $0.64\pm 0.32$\\
CLEO III & 1225  &  $89.2 \pm 0.4$ & $0.64\pm 0.05$\\
CLEO-c 3770 & 1396 & $84.1 \pm 0.6$ & $4.5\pm 0.5$\\
CLEO-c 4170 & 739  &  $65.7 \pm 1.1$ & $7.5\pm 0.7$\\ \hline
$CP$-tags & & \\
$CP$-eigenstates & 81   & $71.9 \pm 3.1$ & -- \\
Admixture & 150  &  $82.1 \pm 1.5$ & --\\ \hline\hline
\end{tabular}
\end{center}
\end{table}

\section{Amplitude analysis}
\label{sec:fit}

The relative magnitudes and phases of the intermediate resonances are determined by a maximum-likelihood fit to the %%@
data selected. The formalism and composition of the likelihood is described in Sec.~\ref{subsec:like}. %%@
Sections~\ref{subsec:norm} and \ref{subsec:chi} describe the calculation of the likelihood normalization and the %%@
goodness-of-fit, respectively. 

\subsection{Likelihood fit}
\label{subsec:like}
The likelihood contains probability density functions (PDFs) for the signal amplitude model and background %%@
components as a function of particle four-momenta. These PDFs are modified to account for the variations in %%@
acceptance over phase space. The method to construct the PDFs is similar to that used to describe a decay to three %%@
final-state pseudoscalar mesons (three-body decay) performed previously by the CLEO Collaboration \cite{KOPP}. %%@
However, a decay to four final-state pseudoscalar mesons (four-body decay) is more complicated due to the %%@
non-uniform phase space and the possibility of having two separate intermediate resonances contributing to the %%@
amplitude. 

The PDFs are functions of the four-momentum $p_j$, of the $D^{0}$ and its decay products, where the index $j=$0, 1, %%@
2, 3, and 4, corresponds to $D^{0}$, $K^{+}$, $K^{-}$, $\pi^{+}$, and $\pi^{-}$ mesons, respectively. The four-body %%@
phase space function, $R_4(p_j)$, is not uniform in any set of five independent kinematic variables that are %%@
required to describe the phase space, unlike the uniform three-body phase-space function over the Dalitz plot. %%@
Therefore, an analytic form of $R_{4}(p_j)$ is used that accounts for the kinematic constraints among the %%@
four-momenta that arise from the invariant masses of the $D^{0}$ and its decay products \cite{FOURBODYPS}. 

There are several different amplitude types that can arise in four-body decays.
\begin{itemize}
\item{A single-resonance amplitude such as $D^{0}\to K^{*}(892)^{0}K^{-}\pi^{+},~K^{*}(892)^{0}\to K^{+}\pi^{-}$. %%@
Unless the resonance is a scalar, there are several possible amplitudes with the same intermediate resonance but %%@
with differing  orbital angular momentum among the final state particles.}

\item{A quasi-two-body amplitude such as $D^{0}\to\phi\rho^{0}$, where subsequently $\phi\to K^{+}K^{-}$ and %%@
$\rho^{0}\to\pi^{+}\pi^{-}$. If both intermediate particles are vector mesons than they can be in either an $S$, %%@
$P$, or $D$ wave orbital angular momentum state, which leads to three distinct amplitudes.}
\item{A cascade amplitude such as $D^{0}\to K_{1}(1270)^{+}K^{-}$, where subsequently $K_{1}(1270)^{+}\to %%@
K^{*}(892)^{0}\pi^{+}$ followed by $K^{*}(892)^{0}\to K^{+}\pi^{-}$. If the first and second intermediate %%@
resonances are both spin one and of opposite parity the second intermediate resonance and the pseudoscalar meson %%@
can be in be either in an $S$ or $D$ wave orbital angular momentum state, which leads to two distinct amplitudes.}
\end{itemize}
In contrast, for a three-body decay only single-resonance amplitudes, with unambiguous orbital angular momentum %%@
assignment, are possible. 

The total amplitude for the $D^{0}\to K^{+}K^{-}\pi^{+}\pi^{-}$ decay, $\mathcal{A}_{D^{0}}$, is modeled as a %%@
coherent sum over the $i$ intermediate states considered 
\begin{equation}
 \mathcal{A}_{D^{0}}(a_i,p_j) = \sum_{i} a_{i} \mathcal{A}_{i}(p_j) \;,  
\end{equation}
where $a_i=|a_i|e^{i\phi_i}$ is a complex factor and $\mathcal{A}_i(p_j)$ is a parametrization of the %%@
intermediate-state amplitude. In the likelihood fit the real and imaginary parts of $a_i$ are determined, rather %%@
than $|a_i|$ and $\phi_i$. This is because $|a_i|$ and $\phi_i$ are bounded and cyclic variables, respectively, %%@
which can lead to numerical problems in the fit. The parametrization of a signal-resonance amplitude is given by
\begin{equation}
 \mathcal{A}_{i}(p_j) = \mathcal{G}_{i}(p_j)\mathcal{S}_{i}(p_j)\mathcal{F}_{i}(p_j)\mathcal{F}_{D}(p_j) ,
\end{equation}
where $\mathcal{G}_{i}(p_j)$ and $\mathcal{S}_{i}(p_j)$ are the lineshape and spin factor for the resonance, %%@
respectively. Here $\mathcal{F}_i(p_j)$ and $\mathcal{F}_{D}(p_j)$ are the angular-momentum barrier penetration %%@
factors for the resonance and $D^{0}$, respectively. For most resonances the lineshape is parametrized by a %%@
relativistic Breit-Wigner propagator with a width that depends upon the spin and the daughter momenta of the %%@
resonance \cite{BW}. The only exception is the $f_{0}(980)\to \pi^{+}\pi^{-}$ resonance where a coupled-channel %%@
(Flatt\'{e}) lineshape \cite{FLATTE} is used. The values of the mass and natural width used in the Breit-Wigner %%@
propagators are taken from Ref.~\cite{PDG}. The parameters used to describe the $f_{0}(980)$ resonance are taken %%@
from Ref.~\cite{FOCUS}. Non-resonant states in which there is orbital angular momentum among the daughters are %%@
modelled as a very broad resonance with a mass corresponding to the measured invariant mass and a width that is %%@
very much greater than the mass of the $D^{0}$ meson; this leads to the spin factor alone altering the distribution %%@
of the events over phase space. The spin factors are Lorentz-invariant matrix elements that describe %%@
angular-momentum conservation in the decay and are described in Appendix~\ref{app_form}. The functional form of %%@
$\mathcal{F}$ is that presented by Blatt and Weisskopf in Ref.~\cite{BLATTWEISS}, which depends on both the spin %%@
and daughter momenta of the resonance. 
The parametrization of a quasi-two-body amplitude or a cascade amplitude is given by 
\begin{equation}
\mathcal{A}_{i} = \mathcal{G}_{i}^{1}(p_j) \mathcal{G}_{i}^{2}(p_j) %%@
\mathcal{S}_{i}(p_j)\mathcal{F}_{i}^{1}(p_j)\mathcal{F}_{i}^{2}(p_j)\mathcal{F}_D(p_j),
\end{equation} 
where $\mathcal{G}_{i}^{j}(p_j)$ and $\mathcal{F}_{i}^{j}(p_j)$ $(j=1,2)$ are the lineshape and Blatt-Weisskopf %%@
angular-momentum barrier penetration factors for the two intermediate resonances that participate in the decay, %%@
respectively. 

In the combined fit to events tagged as either $D^{0}$ and $\overline{D^{0}}$ it is assumed that there is no $CP$ %%@
violation in the decay. Therefore, the amplitude for $\overline{D^{0}}\to K^{+}K^{-}\pi^{+}\pi^{-}$, %%@
$\mathcal{A}_{\overline{D^{0}}}(p_j)$, is identical to $\mathcal{A}_{D^{0}}(p_j)$, except that the charges of the %%@
daughters are conjugated; this is equivalent to the following interchanges of the four-momenta: $p_1\leftrightarrow %%@
p_2$ and $p_3\leftrightarrow p_4$.      

The $D^{0}$ signal PDF $\mathcal{S}_{D^{0}}(a_i,p_j)$, is given by 
\begin{equation}
 \mathcal{S}_{D^{0}}(a_i,p_j) = \frac{\epsilon(p_j)|\mathcal{A}_{D^{0}}(a_i,p_j)|^2 R_{4}(p_j)}{\int %%@
\epsilon(p_j)|\mathcal{A}_{D^{0}}(a_i,p_j)|^2 R_{4}(p_j) dp_j} \;,  \label{eq:d0pdf}
\end{equation}
where $\epsilon(p_j)$ is the acceptance parametrized in terms of the four-momenta.
The $\overline{D^{0}}$ signal PDF, $\mathcal{S}_{\overline{{D}^{0}}}$, is identical apart from the substitution of %%@
$\mathcal{A}_{\overline{D^{0}}}$ for $\mathcal{A}_{D^{0}}$. A method that does not require explicit evaluation of %%@
the functional form of $\epsilon$ is used to fit most data sets. This method is described in %%@
Section~\ref{subsec:norm}.

The background PDF, $\mathcal{B}(p_j)$, is determined for each data set either from sideband or simulated data. The %%@
PDF consists of both combinatoric components and those from specific resonances, which are added incoherently. The %%@
results of this parametrization are given in Sec.~\ref{sec:results}. Therefore, the log-likelihood function for a %%@
flavor-tagged data set is
\begin{eqnarray}
 \ln{\mathcal{L}} & = & \sum^{N_{D^{0}}}_k\ln{\left[f_S\left\{(1-\omega)\mathcal{S}_{D^{0}}(a_i,p_j^k) + %%@
\omega\mathcal{S}_{\overline{D^{0}}}(a_i,p_j^k)\right\} + (1-f_S)\mathcal{B}(p_j^k)\right]} \nonumber \\ 
& + & \sum^{N_{\overline{D^{0}}}}_k\ln{\left[f_S\left\{(1-\omega)\mathcal{S}_{\overline{D^{0}}}(a_i,p_j^k) + %%@
\omega\mathcal{S}_{D^{0}}(a_i,p_j^k)\right\} + (1-f_S)\mathcal{B}(p_j^k)\right]} \; ,
\label{eq:flavlog}
\end{eqnarray} 
where $f_S$ is the fractional amount of signal in the data sample, $\omega$ is the mistag rate, $p_j^{k}$ are the %%@
four-momenta of the $D^{0}$ and its daughters for the $k^{\mathrm{th}}$ event, and $N_{D^{0}}$ %%@
$(N_{\overline{D^{0}}})$ is the number of events tagged as $D^{0}$ $(\overline{D^{0}})$.

The signal PDF for $CP$-even $(\mathcal{S}_+)$ and $CP$-odd $(\mathcal{S}_-)$ tagged data is given by:
\begin{equation}
 \mathcal{S}_{\pm}(a_i,p_j) = %%@
\frac{\epsilon(p_j)|\mathcal{A}_{D^{0}}(a_i,p_j)\pm\mathcal{A}_{\overline{D^{0}}}(a_i,p_j)|^2 R_{4}(p_j)}{\int %%@
\epsilon(p_j)|\mathcal{A}_{D^{0}}(a_i,p_j)\pm\mathcal{A}_{\overline{D^{0}}}(a_i,p_j)|^2 R_{4}(p_j) dp_j} \;.  
\label{eq:sigpdf}
\end{equation}
Hence, the log-likelihood function fit to $CP$-tagged data is
\begin{eqnarray}
 \ln{\mathcal{L}_{CP}} & = & \sum^{N_{+}}_k\ln{\left[f_S^{+}\mathcal{S}_{+}(a_i,p_j^k) + %%@
(1-f_S^{+})\mathcal{B}_{+}(p_j^k)\right]} \nonumber \\ 
& + & \sum^{N_{-}}_{k}\ln{\left[f_S^{-}\mathcal{S}_{-}(a_i,p_j^k) + (1-f_S^{-})\mathcal{B}_{-}(p_j^k)\right]} \; ,
\label{eq:cplog}
\end{eqnarray}
where $N_{+}$ $(N_{-})$, $f_{S}^{+}$ $(f_{S}^{-})$ and $\mathcal{B}_{+}$ $(\mathcal{B}_{-})$ are the number of %%@
tagged decays, fraction of signal, and background PDF for the $CP$-even ($CP$-odd) eigenstate, respectively. 

The final type of data used in the fit is the CLEO-c 3770 sample of mixed $CP$ tagged by %%@
$K^{0}_{S,L}\pi^{+}\pi^{-}$. The quantum-correlated amplitude for $K^{+}K^{-}\pi^{+}\pi^{-}$ $vs.$ %%@
$K^{0}_{S}\pi^{+}\pi^{-}$ decays is given by: 
\begin{equation}
 \frac{1}{\sqrt{2}}\left[\mathcal{A}_{D^{0}}(p_j^k)\mathcal{A}(m_{-}^{2},m_{+}^{2}) - %%@
\mathcal{A}_{\overline{D^{0}}}(p_j^k)\mathcal{A}(m_{+}^{2},m_{-}^{2}) \right] \; ,
\end{equation} 
where $\mathcal{A}(m_{+}^{2},m_{-}^2)$ is the amplitude for $D^{0}\to K^{0}_{S}\pi^{+}\pi^{-}$ as a function of %%@
Dalitz plot variables $m_{+}^2$ and  $m_{-}^2$, which are the invariant-mass squared of the $K^{0}_{S}\pi^{+}$ and %%@
$K^{0}_{S}\pi^{-}$ pairs. (The amplitude for $\overline{D^{0}}\to K^{0}_{S}\pi^{+}\pi^{-}$ is equal to %%@
$\mathcal{A}(m_{-}^{2},m_{+}^2)$ assuming there is no $CP$ violation in the decay.) Following Refs.~\cite{GGSZ,BP},
dividing the $D^{0}\to K^{0}_{S}\pi^{+}\pi^{-}$ Dalitz plot into bins symmetrically about the line $m_+^2 = m_-^2$ %%@
allows the mixed-$CP$ amplitude squared for $D^{0}\to K^{+}K^{-}\pi^{+}\pi^{-}$ in the $m^{\mathrm{th}}$ bin to be %%@
written as   
\begin{equation}
 |\mathcal{A}_{m}(p_j)|^2 \propto |\mathcal{A}_{D^{0}}|^2 K_{-m} +  |\mathcal{A}_{\overline{D^{0}}}|^2 K_{m} - 
2\sqrt{K_{m}K_{-m}}\left[c_m\mathrm{Re}(A_{D^{0}}\mathcal{A}^{*}_{\overline{D^{0}}})+s_m\mathrm{Im}(\mathcal{A}_{D^%%@
{0}}\mathcal{A}^{*}_{\overline
{D^{0}}})\right] \; ,
\end{equation}
where $K_{m}$ is the fraction of $D^{0}\to K^{0}_{S}\pi^{+}\pi^{-}$ decays in the $m^{\mathrm{th}}$ bin and $c_m$ %%@
$(s_m)$ is the amplitude-weighted average of the cosine (sine) of the strong-phase difference between $D^{0}$ and %%@
$\overline{D^{0}}$ decays to $K^{0}_{S}\pi^{+}\pi^{-}$ within the bin. The pairs of symmetric bins have index $m$ %%@
if they lie in the region $m_{-}^{2}<m_{+}^{2}$ and $-m$ in the region $m_{-}^{2}>m_{+}^{2}$. The values of $K_m$, %%@
$c_m$, and $s_m$ used in this analysis are those measured by the CLEO Collaboration~\cite{LIBBY}. There are several %%@
binnings presented in Ref.~\cite{LIBBY}; this analysis uses the binning in equal intervals of the strong-phase
    difference, which is referred to as the {\bf equal $\Delta\delta_D$ binning} in Ref.~\cite{LIBBY} and herein. %%@
There is an equivalent expression for the amplitude of events tagged by $D^{0}\to K^{0}_{L}\pi^{+}\pi^{-}$:  
\begin{equation}
 |\mathcal{A}_m^{\prime}(p_j)|^2 \propto |\mathcal{A}_{D^{0}}|^2 K_{-m}^{\prime} +  %%@
|\mathcal{A}_{\overline{D^{0}}}|^2 K_{m}^{\prime} + 
2\sqrt{K_{m}^{\prime}K_{-m}^{\prime}}\left[c_m^{\prime}\mathrm{Re}(A_{D^{0}}\mathcal{A}^{*}_{\overline{D^{0}}})+s_m
^{\prime}\mathrm{Im}(\mathcal{A}_{D^{0}}\mathcal{A}^{*}_{\overline
{D^{0}}})\right] \; ,
\end{equation}
where $K_{m}^{\prime}$, $c_{m}^{\prime}$, and $s_{m}^{\prime}$ are analogous parameters to those defined for %%@
$D^{0}\to K^{0}_{S}\pi^{+}\pi^{-}$ decays.  The value of these parameters used are those for the equal %%@
$\Delta\delta_D$ binning reported in Ref.~\cite{LIBBY}. The signal PDF for $K^{0}_{S}\pi^{+}\pi^{-}$ %%@
$(K^{0}_{L}\pi^{+}\pi^{-})$, $\mathcal{S}_{m}^{(\prime)}$, is given by:
\begin{equation}
 \mathcal{S}_{m}^{(\prime)}(a_i,p_j) = \frac{\epsilon(p_j)|\mathcal{A}_{m}^{(\prime)}(a_i,p_j)|^2 R_{4}(p_j)}{\int  %%@
\epsilon(p_j)|\mathcal{A}_{m}^{(\prime)}(a_i,p_j)|^2 R_{4}(p_j) dp_j} \;.  
\end{equation} 
The log-likelihood function that is maximized for the $K^{0}_{L,S}\pi^{+}\pi^{-}$ tagged event sample is 
\begin{eqnarray}
 \ln{\mathcal{L}_{CP-\mathrm{mixed}}} & = & \sum_{m=-8,m\neq 0}^{8}
 \left\{ 
 \sum^{N_m}_{k}\ln{\left[f_{S}^{CP-\mathrm{mix}}\mathcal{S}_{m}(a_i,p_j^k) + %%@
(1-f_{S}^{CP-\mathrm{mix}})\mathcal{B}_{CP-\mathrm{mix}}(p_j^k)\right]} \right. \nonumber \\ 
& + & \left. \sum^{N_{m}^{\prime}}_k\ln{\left[f_{S}^{CP-\mathrm{mix}\prime}\mathcal{S}_{m}^{\prime}(a_i,p_j^k) + %%@
(1-f_{S}^{CP-\mathrm{mix}\prime})\mathcal{B}_{CP-\mathrm{mix}}^{\prime}(p_j^k)\right]}\right\}\; ,
\label{eq:cpmixlog}
\end{eqnarray}
where $N_{m}^{(\prime)}$, $f_{S}^{CP-\mathrm{mix}(\prime)}$ and $\mathcal{B}_{CP-\mathrm{mix}}^{(\prime)}$  are the %%@
number of tagged decays in each bin, fraction of signal, and background PDF for the $K^{0}_{S}\pi^{+}\pi^{-}$ %%@
($K^{0}_{L}\pi^{+}\pi^{-}$) tagged events, respectively. 

The combined log-likelihood function to be maximized is the sum of the expression in Eq.~(\ref{eq:flavlog}) for %%@
each flavor-tagged data set, plus the log-likelihood functions given in Eqs.~(\ref{eq:cplog}) and %%@
(\ref{eq:cpmixlog}). While performing the fit, one intermediate resonance component has $a_i$ fixed to unity such %%@
that the amplitude and phase of the other components are determined relative to it.

\subsection{Efficiency parametrization and normalization}      
\label{subsec:norm}
The efficiency parametrization and normalization methods are presented together in this section as they are closely %%@
linked. Consider first the logarithm of the flavor-tagged $D^{0}$ signal PDF given by Eq.~(\ref{eq:d0pdf}):
\begin{equation}
 \ln{\mathcal{S}_{D^{0}}(a_i,p_j)} = \ln{\epsilon(p_j)}+\ln{|\mathcal{A}_{D^{0}}(a_i,p_j)|^2
}+\ln{R_4(p_j)}-\ln{\left[\int \epsilon(p_j)|\mathcal{A}_{D^{0}}(a_i,p_j)|^2R_4(p_j)dp_j\right]}\; .  
\end{equation}
Only the second and last terms depend on the amplitude model parameters, which means that only these need to be %%@
computed while performing the fit. Therefore, the acceptance function, $\epsilon(p_j)$, is only required to compute %%@
the normalization integral. The normalization integral is evaluated by a Monte Carlo integration method, as %%@
described below, which does not require an analytic form of $\epsilon(p_j)$. Not requiring an analytic form is %%@
desirable, given such acceptance functions can be difficult to parametrize even over a three-body phase space, due %%@
to the rapid change in efficiency at the edge of phase space. These problems are compounded for four-body decays %%@
due to the higher dimensionality of the phase space. 

However, the treatment is less straightforward in the presence of background. The combined PDF, ignoring the mistag %%@
rate, is
\begin{equation}
 f_S\mathcal{S}+(1-f_S)\mathcal{B}  =  f_S \frac{\epsilon(p_j)|\mathcal{A}_{D^{0}}(a_i,p_j)|^2 R_{4}(p_j)}{\int %%@
\epsilon(p_j)|\mathcal{A}_{D^{0}}(a_i,p_j)|^2 R_{4}(p_j) dp_j}
+  (1-f_S) \frac{B(p_j) R_{4}(p_j)}{\int B(p_j) R_{4}(p_j) dp_j} \; ,
\end{equation}
where $B(\mathrm{s})$ is a function that describes the background distribution relative to phase space. Given that %%@
$\epsilon$ no longer factorizes from the second term it has to be computed for each event to minimize the %%@
log-likelihood function. The combined PDF becomes
\begin{eqnarray}
 f_S\mathcal{S}+(1-f_S)\mathcal{B}  =  \epsilon(p_j) R_4(p_j) & & \left[  f_S %%@
\frac{|\mathcal{A}_{D^{0}}(a_i,p_j)|^2 }{\int \epsilon(p_j)|\mathcal{A}_{D^{0}}(a_i,p_j)|^2 R_{4}(p_j) dp_j} %%@
\right.\nonumber \\
  && \left.  + (1-f_S) \frac{B_{\epsilon}(p_j) }{\int\epsilon(p_j)B_\epsilon(p_j) R_{4}(p_j) dp_j}\right] \; ,
\end{eqnarray}
where
$B_{\epsilon}(p_j)$ is defined to be the background distribution relative to the acceptance corrected phase space %%@
[$\epsilon(p_j)R_4(p_j)$]. 
Now that $\epsilon$ factorizes from the complete PDF, it only has to be computed as part of the normalization. 

The normalization integrals are determined by a Monte Carlo technique. The simulated events are generated according %%@
to the distribution
\begin{equation}
|\mathcal{A}_{D^{0}}(a_i^{\mathrm{gen}},p_j)|^2R_4(p_j) \nonumber \; ,
\end{equation}
where $a_i^{\mathrm{gen}}$ are a fixed set of parameters. Using this set of simulated events the normalization %%@
integral can be approximated by
\begin{equation}
 \int\epsilon(p_j)|\mathcal{A}_{D^{0}}(a_i,p_j)|^2 R_{4}(p_j) \approx
 \frac{1}{N_{\mathrm{gen}}}\sum^{N_{\mathrm{gen}}}_k %%@
\epsilon(p_j^k)\frac{|\mathcal{A}_{D^{0}}(a_i,p_j^k)|^2}{|\mathcal{A}_{D^{0}}(a_i^{\mathrm{gen}},p_j^k)|^2} \; ,
\end{equation}
where $N_{\mathrm{gen}}$ is the number of simulated events. 

Furthermore, the effect of the acceptance function can be incorporated in determining this integral by summing over %%@
events, that once passed through the CLEO detector simulation, satisfy the selection criteria described in %%@
Sec.~\ref{sec:dataset} for a given data set. This is equivalent to generating events according to the PDF: 
\begin{equation}
\epsilon(p_j)|\mathcal{A}_{D^{0}}(a_i^{\mathrm{gen}},p_j)|^2R_4(p_j) \nonumber \; .
\end{equation} 
Therefore, the normalization integral is given by
\begin{equation}
 \int\epsilon(p_j)|\mathcal{A}_{D^{0}}(a_i,p_j)|^2 R_{4}(p_j) \approx
 \frac{1}{N_{\mathrm{sel}}}\sum^{N_{\mathrm{sel}}}_k %%@
\frac{|\mathcal{A}_{D^{0}}(a_i,p_j^k)|^2}{|\mathcal{A}_{D^{0}}(a_i^{\mathrm{gen}},p_j^k)|^2} \; ,
\end{equation}
where $N_{\mathrm{sel}}$ is the number of simulated events selected. The background samples are also fit using %%@
these normalization events so that the function $B_{\epsilon}$ is determined. 

The number of integration events is chosen to ensure the uncertainty on the integral is less than 0.3\%. Two %%@
separate samples of one million selected events each are used to perform the integration of the fits to CLEO III %%@
and CLEO-c data. It is no longer possible to generate an additional signal 
     simulation sample of this size for the CLEO II.V data. Given the similarity between the CLEO II.V and CLEO III %%@
detectors the CLEO III integration events are used. To account for the small differences in particle identification %%@
and tracking performance between CLEO II.V and CLEO III, the integration events are reweighted by
\begin{equation}
 \prod^{4}_{j=1}{\frac{\epsilon_{\mathrm{II.V}}(|\mathbf{p}_{j}|)}{\epsilon_{\mathrm{III}}(|\mathbf{p}_{j}|)}} \; 
\end{equation}  
where $\epsilon_{\mathrm{II.V}}(|\mathbf{p}_{j}|)$ [$\epsilon_{\mathrm{III}}(|\mathbf{p}_{j}|)$] is the efficiency %%@
as a function of the momentum of the daughters in the laboratory frame, $|\mathbf{p}_{j}|$, for CLEO II.V [CLEO %%@
III]. The ratios $\epsilon_{\mathrm{II.V}}/\epsilon_{\mathrm{III}}$ are computed from simulation. 
The mean value of the weight applied is  0.87 with an R.M.S. of 0.04. Such a reweighting does not account for %%@
correlations among the daughters and the dependence of the acceptance on other variables; therefore, a conservative %%@
systematic uncertainty is assigned to account for this approximation, which is discussed further in %%@
Sec.~\ref{subsec:syst}.

\subsection{Goodness-of-fit}
\label{subsec:chi}
In order to quantify the quality of a given fit a $\chi^2$ value is computed. The four-body phase space can be %%@
described completely by any five invariant-mass-squared variables $s_{ij}=(p_i+p_j)^{2}$ and %%@
$s_{ijk}=(p_i+p_j+p_k)^2$. Therefore, the events are binned in terms of $s_{12}$, $s_{123}$, $s_{23}$, $s_{234}$, %%@
and $s_{34}$ to compute the $\chi^2$. Initially, the phase space is divided into equal bins. At least fifty events %%@
are required in each bin so that the $\chi^2$ calculation is robust. Therefore, after the initial equal division of %%@
the phase space, bins are merged until they satisfy the minimum number of events criterion. 

 The $\chi^2$ is given by
\begin{equation}
 \sum^{n}_{p=1} \frac{\left[N_p-N_p^{\mathrm{exp}}(a_i)\right]^2}{N_p^{\mathrm{exp}}(a_i)} \; ,
\end{equation}
where $N_{p}$ and $N_p^{\mathrm{exp}}(a_i)$ are the observed and expected number of events per bin, respectively, %%@
and $n$ is the number of bins. In the general case the value of $N_{p}^{\mathrm{exp}}(a_i)$ is given by 
\begin{equation}
 N_p^{\mathrm{exp}}(a_i) = N\int_{\mathrm{bin}~p}\left[{f_S}\mathcal{S}_{D^{0}}(a_i)+(1-f_S)\mathcal{B}\right]dp_j %%@
\;, 
\end{equation}
where $N$ is the total number of events in a particular sample and the integral is over the $p^{\mathrm{th}}$ bin. %%@
The Monte Carlo integration events are used to compute this integral such that
\begin{equation}
  N_j^{exp}(a_i) = %%@
\frac{N}{N_{\mathrm{sel}}}\sum^{N_\mathrm{sel}^p}_{p=1}\left[\frac{{f_S}|\mathcal{A}_{D^{0}}(a_i)|^2+(1-f_S)
B_{\epsilon}}{|\mathcal{A}_{D^{0}}(a_i^{\mathrm{gen}})|^2}\right] \;  ,
\end{equation}
where $N_\mathrm{sel}^p$ is the number of Monte Carlo integration events in the $p^{\mathrm{th}}$ bin. 

The number of degrees of freedom, $\nu$, is given by
\begin{equation}
 \nu = (n-1) - n_{\mathrm{par}} \;
\end{equation} 
where $n_{\mathrm{par}}$ is the number of free parameters in the amplitude or background model being fit. The %%@
number of bins is reduced by one because the number of expected events in the final bin considered is determined by %%@
the overall normalization and as such does not represent a degree of freedom. The compatibility of the combined fit %%@
hypothesis to each individual data set is estimated by determining the $\chi^2$ per bin for that data set. The %%@
$CP$-tagged samples are not used in computing the $\chi^2$ because of the limited statistics in the four %%@
sub-samples.  

\section{Results}
\label{sec:results}
The results of the amplitude analysis of $D^{0}\to K^{+}K^{-}\pi^{+}\pi^{-}$ are presented in this section. The %%@
fits to the sideband samples used to parameterize the background in each data set are presented in %%@
Sec.~\ref{subsec:back}. The method used to arrive at the components in the baseline model and the results are %%@
presented in Sec.~\ref{subsec:modselres}. Robustness tests of the fitting method and cross checks of the final %%@
result are given in Sec.~\ref{subsec:robust}. The procedures to evaluate systematic uncertainties are outlined in %%@
Sec.~\ref{subsec:syst}. A search for $CP$ violation in the decay is presented in Sec.~\ref{subsec:cp}.  

\subsection{Background parameterization}
\label{subsec:back}
The fraction of signal in each data set considered in the amplitude fit has been estimated for the various %%@
selections and is given in Table~\ref{tab:samplesummary}. As well as the different relative amount of background in %%@
each data set, the composition of the background is also different due to the variation in the value of $\sqrt{s}$, %%@
the detector configuration and tagging method. In addition, the type of backgrounds must be classified into those %%@
that form candidates that are peaking or non-peaking at the nominal $D^{0}$ meson mass. 

Non-peaking backgrounds are random combinations of four particles that do not originate from the same $D^{0}$ %%@
decay, but some may form a resonance such as a $\phi$ or $\rho$. Separate sideband samples have been selected as %%@
described in Sec.~\ref{sec:dataset}, which are fit to determine the non-peaking background PDFs for each data set. %%@
The model used to describe the background is an incoherent sum of resonances along with a non-resonant component, %%@
which is a constant, such that
\begin{equation}
 B_{\epsilon}(\mathbf{s},b_i,b_{nr}) = b_{nr}+\sum_i b_{i}|A_i(\mathbf{s})|^2 \;, 
\end{equation} 
where $b_{nr}$ and $b_i$ are real parameters determined by the fit. Various combinations of amplitudes are tested. %%@
The combination with the lowest $\chi^2/\nu$, which does not contain any components that contribute less than 0.5\% %%@
to the total, is selected. Table~\ref{tab:back} gives the fractional contributions of the different components for %%@
each flavor-tagged data set. All data sets have a significant non-resonant component in the background but the %%@
resonances that contribute vary significantly among the different data sets.

\begin{table}[htbp]
\begin{center}
\caption{Fractional contribution of each component to the background for each flavor-tagged data set. The %%@
$\chi^2/\nu$ of the background fit is also given.}\label{tab:back}
\begin{tabular}{lcccc}
\hline\hline
  &  \hspace{0.0cm} CLEO II.V  \hspace{0.0cm} & \hspace{0.0cm} CLEO III  \hspace{0.0cm} & \hspace{0.0cm} CLEO-c %%@
$3770$ \hspace{0.0cm} & \hspace{0.0cm} CLEO-c $4170$ \hspace{0.0cm}   \\
\hline
$K_{1}(1270)^{-}(\overline{K^{*}_{0}}(1430)^{0}\pi^{-})K^{+}$ &---&---&0.019$\pm$0.064&---\\
$K_{1}(1270)^{+}(K^{*0}\pi^{+})K^{-}$ &0.053$\pm$0.015&---&---&---\\
$K_{1}(1270)^{-}(\overline{K^{*0}}\pi^{-})K^{+}$ &0.006$\pm$0.015&0.032$\pm$0.015&0.033$\pm$
0.021&---\\
$K_{1}(1270)^{+}(\rho K^{+})K^{-}$&---&---&0.005$\pm$0.013&---\\
$K_{1}(1400)^{+}(K^{*0}\pi^{+})K^{-}$ &---&0.055$\pm$0.014&0.015$\pm$0.023&---\\
$\phi\pi^{+}\pi^{-}$ &---&0.079$\pm$0.011&0.143$\pm$0.022&0.102$\pm$0.013\\
$K^{*0} \overline{K^{*0}}$ &0.007$\pm$0.012&0.045$\pm$0.014&0.101$\pm$0.019&0.010$\pm$0.017\\
$\overline{K^{*0}}K^{+}\pi^{-}$ &---&---&---&0.032$\pm$0.023\\
$f_0(980)K^{+}K^{-}$&0.033$\pm$0.047&0.128$\pm$0.049&0.293$\pm$0.118&---\\
$\rho K^{+}K^{-}$ &---&---&---&0.243$\pm$0.034\\
$K^{*0}K^{-}\pi^{+}$&---&---&0.017$\pm$0.035&0.098$\pm$0.025\\
Non-resonant &0.899$\pm$0.048&0.661$\pm$0.046&0.373$\pm$0.103&0.516$\pm$0.078\\
\hline
 $\chi^2/\nu$ & 1.32& 1.20& 1.17& 2.13 \\
\hline\hline
\end{tabular}
\end{center}
\end{table}

The only significant peaking background comes from $D^{0}\to K^{0}_{S}(\pi^{+}\pi^{-})K^{+}K^{-}$ decays in the %%@
CLEO-c datasets. The larger average momentum of the $K^{0}_{S}$ mesons in the CLEO II.V and CLEO III data sets %%@
leads to a significant displacement of most $K^{0}_{S}$ decay vertices from the interaction point, which allows the %%@
efficient rejection of this background. The lower average momenta at CLEO-c means that the such a separation is %%@
less effective. 
The fraction of this component is left as a free parameter in the fit to data. A four-body model of the %%@
distribution of the $K^{0}_{S} (\pi^+\pi^-)K^{+}K^{-}$ events over the kinematic variables is obtained by fitting %%@
events that fail the  $K^{0}_{S}$ veto. The resonant components of the model considered are a subset of those %%@
reported in Ref.~\cite{BABARKSKK} and these are fit coherently to this sample. Only those with a significant %%@
non-zero contribution are retained. The model parameters found are then used in the fits to the signal sample data %%@
to yield the  background fractions reported in Sec.~\ref{sec:cleoc3770} and~\ref{sec:cleoc4170}.
 
The non-peaking background distribution for the CLEO-c $CP$-tagged data is estimated from a fit to a combination of %%@
generic simulation events and data sidebands, because of the limited statistics. There is also a peaking $D^{0}\to %%@
K^{0}_{S}(\pi^{+}\pi^{-})K^{+}K^{-}$ contribution, which is estimated from the simulation alone. Due to the %%@
reliance of these background estimates on simulated events the strategy to determine the systematic uncertainty %%@
related to the background in the $CP$-tagged sample is more conservative than the flavor-tagged samples; this is %%@
discussed further in Sec.~\ref{subsec:syst}.  

\subsection{Model selection and results}
\label{subsec:modselres}
\begin{figure}
 \begin{center}
\includegraphics[width=1.00\textwidth]{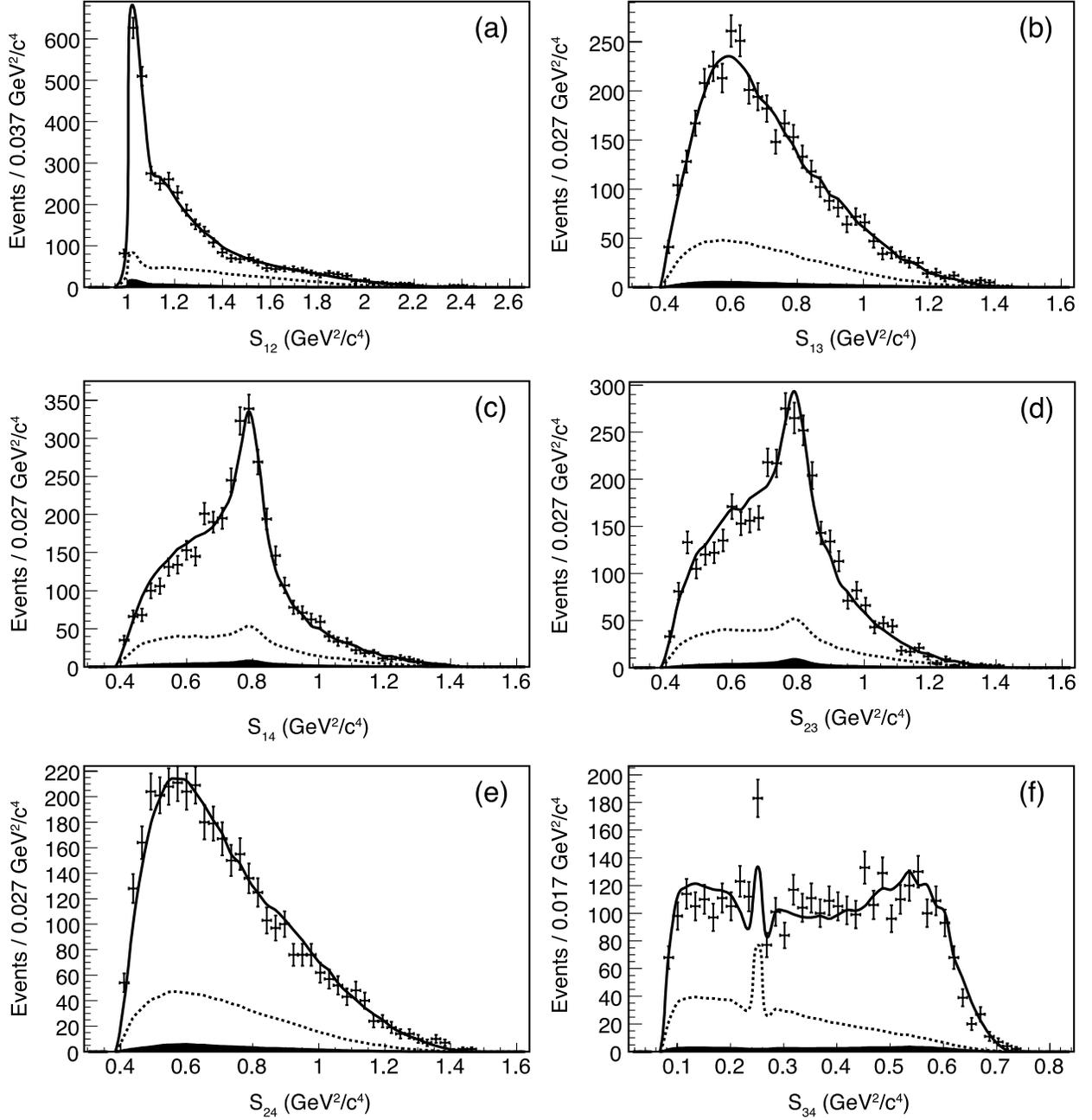}
 \end{center}
 \caption{The (a) $s_{12}$, (b) $s_{13}$, (c) $s_{14}$, (d) $s_{23}$, (e) $s_{24}$, and (f) $s_{34}$ projections %%@
for all flavor-tagged data (points with error bars) with the best fit (solid line) superimposed. The indices %%@
correspond to $K^{+}=1$, $K^{-}=2$, $\pi^{+}=3$, and $\pi^{-}=4$. The contributions from mistag (filled region) and %%@
background plus mistag (dashed line) are also shown.}\label{fig:flavfit2}
\end{figure}

\begin{figure}
 \begin{center}
\includegraphics[width=1.00\textwidth]{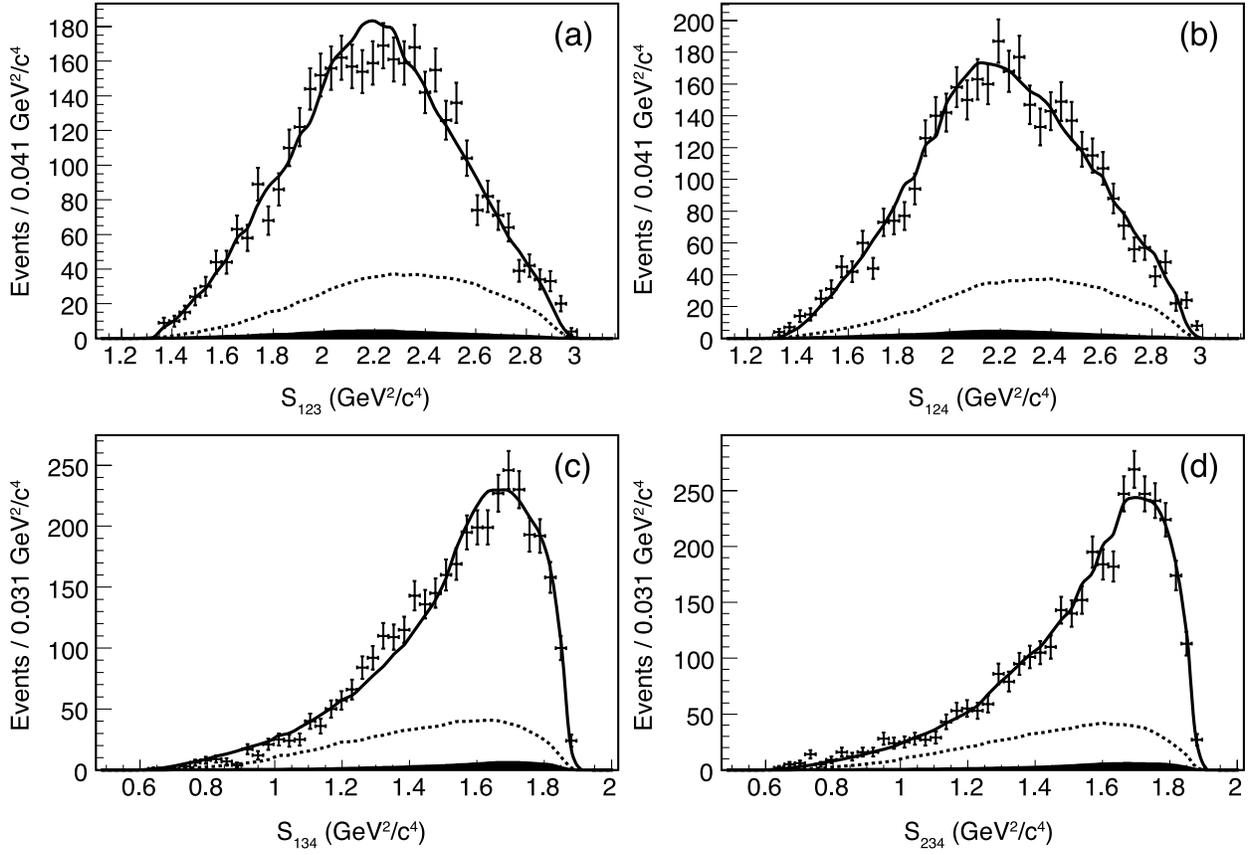}
 \end{center}
 \caption{Distribution of  for the three-body invariant-mass observables: The (a) $s_{123}$, (b) $s_{124}$, (c) %%@
$s_{134}$, and (d) $s_{234}$ projections for all flavor-tagged data (points with error bars) with the best fit %%@
(solid line) superimposed. The contributions from mistag (filled region) and background plus mistag (dashed line) %%@
are also shown.}\label{fig:flavfit3}
\end{figure}

There are many possible amplitudes which can contribute to the $D^{0}\to K^{+}K^{-}\pi^{+}\pi^{+}$ decay, therefore %%@
a strategy to determine the best combination is defined. From inspection of the invariant-mass squared projections %%@
(see Figs.~\ref{fig:flavfit2} and \ref{fig:flavfit3}) there is clear evidence for intermediate $\phi\to %%@
K^{+}K^{-}$, $\overline{K^{*0}}\to K^{-}\pi^{+}$, $\rho^{0}\to\pi^{+}\pi^{-}$ and a broad high-mass kaon decaying %%@
to $K^{-}\pi^{+}\pi^{-}$ ($K_1(1270)^{-}$, $K_1(1400)^{-}$, $K^{*}(1410)^{-}$, $K_2^{*}(1430)^{-}$, or %%@
$K^{*}(1680)^{-}$). Therefore, all models considered contain at least one amplitude with one of these intermediate %%@
resonances. In addition, for states that are not self conjugate, the conjugate amplitude is always included as %%@
well; for example, if $K_1(1270)^{-}K^{+}$ is a component in the model so is $K_1(1270)^{+}K^{-}$.

Models containing seven components are tested and fit to the combined data set. The five models with the smallest %%@
$\chi^2/\nu$ are considered further. An additional component is then added, which either contains one of the %%@
principal resonances discussed above or is non-resonant. All models, including those from the previous iteration, %%@
are compared and the best five are retained. This process continues until the five models with the lowest %%@
$\chi^{2}/\nu$ are the same as those in the previous iteration.

At this point amplitudes containing intermediate resonances not already considered, such as $\omega$, $f_{0}(980)$, %%@
and $f_2(1270)$, are added. The best fifty models are then retained. Any models that contain components %%@
contributing less than 5\% are then simplified by removing these components and the revised model is fit to the %%@
data. Testing these simplified models ensures that the improvement in the $\chi^2/\nu$ by including these small %%@
components is significant. Finally, all models that have been tested are ranked according to their $\chi^2/\nu$. %%@
The different amplitudes that have been included in the models tested are listed in Appendix \ref{app:amptested}.

The components of the model with the lowest $\chi^2/\nu$ are shown in Table~\ref{tab:fitres}. For some amplitudes, %%@
pairs of particles do not decay via a resonance but are in a state of relative orbital angular momentum $(L)$; each %%@
such pair is surrounded by curly brackets with an $S$ or $P$ subscript, indicating an $L=0$ or $L=1$ state, %%@
respectively. The relative orbital angular momentum state - $S$, $P$, or $D$ wave - of intermediate resonances and %%@
pairs of particles is also given for an amplitude if more than one is possible. There are other models with similar %%@
$\chi^2/\nu$; the principal variations are different or additional angular momentum states for the %%@
$\phi\pi^{+}\pi^{-}$ decay. More information on these alternative models can be found in %%@
Appendix~\ref{app:bestfits}. The $\chi^2/\nu$ for the combined fit and the $\chi^2/\mathrm{bin}$ for each %%@
flavor-tagged data set are presented in Table~\ref{tab:bestfit}.

\begin{table}[t]
\begin{center}
\caption{Real and imaginary parts of $a_i$ for combined fit to all data. Only the statistical uncertainties are %%@
given. The daughters of the $K_1(1270)^\pm$ are assumed to be in an $S$-wave state.}
 \label{tab:fitres}
\begin{tabular}{lcc}
\hline\hline
 Amplitude & $\mathrm{Re}(a_i)$ & $\mathrm{Im}(a_i)$  \\ \hline
$K_{1}(1270)^{+}(K^{*0}\pi^{+}) K^{-}$ & 1.0  & 0.0  \\
$K_{1}(1270)^{-}(\overline{K^{*0}}\pi^{-})K^{+}$ & 
\hspace{0.5cm} $\phantom{-}0.16\pm 0.08$ \hspace{0.5cm} & \hspace{0.5cm} $-0.31\pm 0.06$ \hspace{0.5cm} \\
$K_{1}(1270)^{+}(\rho^{0} K^{+})K^{-}$ & 
$\phantom{-}4.07\pm 0.64$ & $\phantom{-}4.22\pm 0.89$\\
$K_{1}(1270)^{-}(\rho^{0} K^{-})K^{+}$ &
$\phantom{-}6.90\pm 0.59$ & $\phantom{-}0.20\pm 1.10$\\
$ K^{*}(1410)^{+} (K^{*0} \pi^{+})K^{-}$ & 
$\phantom{-} 4.62\pm 0.56$ & $-4.10 \pm 0.72$ \\
$ K^{*}(1410)^{-} (\overline{K^{*0}}\pi^{-}) K^{+}$ &
$\phantom{-}2.61\pm 0.93$&$-6.25\pm 0.59$ \\
$K^{*0} \overline{K^{*0}}$ $S$ wave &
$\phantom{-}0.32\pm 0.04$ &$\phantom{-}0.13\pm 0.04$ \\
$\phi\rho^{0}$ $S$ wave & 
$-0.32\pm 0.15$ & $\phantom{-}0.99\pm 0.09$\\
$\phi\rho^{0}$ $D$ wave & 
$\phantom{-}0.20\pm 0.33$ & $-1.43\pm 0.18$\\
$\phi \left\{\pi^{+}\pi^{-}\right\}_S$ &
$-1.70 \pm 0.83$ & $-5.93\pm 0.48$ \\
$\left\{K^{-}\pi^{+}\right\}_P\left\{K^{+}\pi^{-}\right\}_S$ & 
$\phantom{-}82.6\pm 6.8\phantom{0}$ & $\phantom{-}11.4\pm 9.7\phantom{0}$ \\
\hline\hline
\end{tabular}
\end{center}
\end{table}

\begin{table}
 \begin{center}
  \caption{$\chi^2/\nu$ for the combined fit or $\chi^2/n$ for each flavor-tagged data set. For the combined fit %%@
there are 22 free parameters.}\label{tab:bestfit}
	\begin{tabular}{lcc} \hline\hline
	Data set & $\chi^2/\nu$  or $\chi^2/n$ & $\nu$ or $n$\\ \hline
	Combined & 1.63 & 113\\
	CLEO-c 3770 & 1.29 & 55 \\
	CLEO-c 4170 & 1.20 & 26\\
	CLEO III & 1.54 &  49 \\
	CLEO II.V & 2.00 & 6\\ \hline\hline
  \end{tabular}
 \end{center}
\end{table}
     
The best fit projected on to the distributions of $s_{ij}$ and $s_{ijk}$ for the combined flavor-tagged data set %%@
are shown in Figs.~\ref{fig:flavfit2} and \ref{fig:flavfit3}, repsectively. Reasonable agreement is seen between %%@
the data and the fit for most distributions. An exception is the narrow peak in the background $s_{34}$ %%@
distribution, which is due to the peaking $K^{0}_{S}K^{+}K^{-}$ background found in the CLEO-c data. The best fit %%@
underestimates the data in this region; therefore, an additional systematic uncertainty, described in %%@
Sec.~\ref{subsec:syst}, is assigned to account for this discrepancy.

The best fit real and imaginary components of $a_i$ are given in Table~\ref{tab:fitres}. The statistical %%@
correlations among 
the real and imaginary components of $a_i$ are given in Ref.~\cite{EPAPS}.
The values of the magnitude and phase of the amplitude derived from the fitted parameters are given in %%@
Table~\ref{tab:fitres2} along with the associated statistical and systematic uncertainties. The fit fraction, which %%@
is defined as
\begin{equation}
 \frac{\int |\mathcal{A}_i|^2 dp_j}{\int |\mathcal{A}_{D^{0}}|^2 dp_j} \; ,
\end{equation}
is also given for each component. The fit fraction indicates the relative contribution of a component to the total %%@
branching fraction. The individual component fractions do not have to sum to 100\% due to interference. %%@
(Information about 
the interference among the amplitudes is given in Ref.~\cite{EPAPS}.) For the best fit the sum of the fit fractions %%@
is $(96.7\pm 2.6\pm 9.8)\%$ where the first uncertainty is statistical and the second is systematic. The dominant %%@
intermediate state is the $\phi\rho$ quasi-two-body decay. There are three other decay modes which contribute over %%@
10\% to the total branching fraction: $K_1(1270)^{+}K^{-}$, $\phi\pi^{+}\pi^{-}$, and the non-resonant %%@
$(K^{-}\pi^{+})_P(K^{+}\pi^{-})_S$. 
\begin{table}[t]
\begin{center}
\caption{Modulus, phase, and fit fraction for each component of the baseline model. The first uncertainty is %%@
statistical and the second is systematic.}
 \label{tab:fitres2}
\begin{tabular}{lccc}
\hline\hline
 Component & $|a_i|$ & $\phi_i$ (rad) & Fit Fraction (\%) \\
\hline
$K_{1}(1270)^{+}(K^{*0}\pi^{+}) K^{-}$ 
& 1.0  & 0.0 & $\phantom{0}7.3\pm 0.8 \pm 1.9$ \\
$K_{1}(1270)^{-}(\overline{K^{*0}}\pi^{-})K^{+}$ 
& \hspace{0.3cm} $0.35\pm 0.06\pm 0.03$ \hspace{0.3cm} & \hspace{0.3cm} $1.10\pm 0.22\pm 0.23$ \hspace{0.3cm} & %%@
\hspace{0.3cm} $\phantom{0}0.9\pm 0.3 \pm 0.4 $ \hspace{0.3cm} \\
$K_{1}(1270)^{+}(\rho^{0} K^{+})K^{-}$ 
&  $5.86\pm 0.77 \pm 2.03$ & $0.80\pm 0.13\pm 0.08$ & $\phantom{0}4.7\pm 0.7 \pm 0.8$\\
$K_{1}(1270)^{-}(\rho^{0} K^{-})K^{+}$ 
&  $6.90\pm 0.59\pm 3.07$& $0.03\pm 0.16 \pm 0.23$ & $\phantom{0}6.0\pm 0.8 \pm 0.6 $\\
$ K^{*} (1410)^{+} (K^{*0} \pi^{+})K^{-}$ 
&  $6.18\pm 0.64\pm 0.75$& $0.73 \pm 0.11 \pm 0.33$ & $\phantom{0}4.2\pm 0.7 \pm 0.8$\\
$ K^{*}(1410)^{-} (\overline{K^{*0}}\pi^{-}) K^{+}$ 
&  $6.78\pm 0.65 \pm 1.25$& $1.18\pm 0.13 \pm 0.48$ & $\phantom{0}4.7\pm 0.7 \pm 0.7$\\
$K^{*0} \overline{K^{*0}}$ $S$ wave 
&  $0.34\pm 0.04\pm 0.14$& $0.39\pm 0.12\pm 0.18$ & $\phantom{0}6.1\pm 0.8 \pm 0.9$ \\
$\phi\rho^{0}$ $S$ wave 
& $1.04\pm 0.10\pm 0.31$& $1.89\pm 0.14\pm 0.35$ & $38.3\pm 2.5\pm 3.8$\\
$\phi\rho^{0}$ $D$ wave 
& $1.44\pm 0.19 \pm 0.38$ & $1.43\pm 0.22 \pm 0.48$ & $\phantom{0}3.4\pm 0.7\pm 0.6$\\
$\phi \left\{\pi^{+}\pi^{-}\right\}_S$ 
& $6.17\pm 0.52 \pm 1.58$ & $1.85\pm 0.13 \pm 0.37$& $10.3\pm 1.0\pm 0.8$ \\
$\left\{K^{-}\pi^{+}\right\}_P\left\{K^{+}\pi^{-}\right\}_S$
& $83.4\pm \phantom{0}6.8\pm 29.3$ &$0.14\pm 0.12\pm 0.28$ & $10.9\pm 1.2 \pm 1.7$ \\
\hline\hline
\end{tabular}
\end{center}

\end{table}

Detailed comparison of these results to those presented previously \cite{E791,FOCUS} is not straightforward given %%@
the lack of flavor tagging in both analyses and the absence of spin factors in the E791 study. However, there is %%@
agreement with the previous findings in the presence of a significant $\phi\rho^{0}$ contribution. Further, a %%@
significant $\phi\pi^{+}\pi^{-}$ is observed by the E791 collaboration \cite{E791}. Significant contributions from %%@
both the $\phi\rho$ and $\phi\pi^{+}\pi^{-}$ modes are anticipated because there is only one %%@
singly-Cabibbo-suppressed diagram that contributes to the rate \cite{E791}. However, for other modes such as %%@
$K^{*0}K^{+}\pi^{-}$ there are two leading-order amplitudes of opposite sign, which result in a suppression of the %%@
decay rate. In addition, there is evidence of a large $K_1(1270)^{+}K^{-}$ contribution, which was also observed by %%@
the FOCUS collaboration \cite{FOCUS}. However, the expectation that $K_1(1270)$ mesons with the same charge as the %%@
$W$ boson in the decay will be the dominant amplitude \cite{FOCUS} is not observed. This may be a consequence of %%@
final-state interactions playing a significant role. Further evidence of final-state interactions is given by the %%@
presence of a statisticallly significant $K^{*0}\overline{K^{*0}}$ contribution, because in the SU(3)-flavor limit, %%@
the two $W$-exchange amplitudes contributing to this final state cancel \cite{E791}.  

 There are two significant differences from the FOCUS model. Firstly, no significant $f_{0}(980)\pi^{+}\pi^{-}$ %%@
contribution is found. Secondly, a non-resonant contribution with angular momentum structure is required to fit the %%@
data. (If there is no angular momentum among the particles in the non-resonant component the $\chi^2/\nu$ increases %%@
to 2.1.)    

 It is of interest to assess the impact of the $CP$-tagged data given that the quantum-correlated states should %%@
provide additional information about $\phi_i$ compared to the flavor-tagged data. Given the limited number of %%@
events in the sample it is not possible to fit the $CP$-tagged data alone. However, the CLEO II.V flavor-tagged %%@
data sample has a similar size and purity to the CLEO-c $CP$-tagged data sample. Therefore, the statistical impact %%@
of these two samples is compared by determining how much the mean relative statistical
uncertainty on $\mathrm{Re}(a_i)$ and $\mathrm{Im}(a_i)$ changes when either the $CP$-tagged CLEO-c data
set or the flavor-tagged CLEO II.V data set is excluded from the combined fit. The
mean relative statistical uncertainty of the $a_i$ increases by 12.3\% when the CP-tagged
data set is removed compared to 7.5\% when the CLEO II.V data set is removed, indicating that a small sample %%@
$CP$-tagged data is more powerful than an additional sample of flavor-tagged events of similar size.  

\clearpage

\subsection{Robustness tests}
\label{subsec:robust}
Tests of the result are made by dividing the data into different subsamples. Firstly, separate fits to the data %%@
collected at center-of-mass energies around the $\Upsilon$ resonances (CLEO II.V and CLEO III) and data collected %%@
at CLEO-c are made. This tests the compatibility of the results using data produced at different center-of-mass %%@
energies and selected with different tagging techniques. The resulting fit fractions are compared to one another in %%@
Table~\ref{tab:flavtag_comp}. The number of standard deviation difference between the two results takes into %%@
account the uncorrelated systematic uncertainties between the two data samples. There is good agreement between the %%@
two samples.

\begin{table}
 \begin{center}
 \caption{Fit fractions for the CLEO II.V/CLEO III and CLEO-c 3770/4170 data sets fit separately and the difference %%@
between the two results, normalized by the uncorrelated uncertainty.}\label{tab:flavtag_comp}
  \begin{tabular}{lccc} \hline\hline
   Component & \multicolumn{2}{c}{Fit fraction (\%)} & Difference $(\sigma)$ \\
             & CLEO-II.V/III & CLEO-c \\ \hline
	$K_{1}(1270)^{+}(K^{*0}\pi^{-})K^{-}$ 
& \hspace{0.3cm} $\phantom{0}4.4\pm 0.9\pm 0.8$ \hspace{0.3cm} &  \hspace{0.3cm} $\phantom{0}9.8\pm 1.2\pm 4.6$  %%@
\hspace{0.3cm} &  \hspace{0.3cm} $1.1$  \hspace{0.3cm} \\
	$K_{1}(1270)^{-}(\overline{K^{*0}}\pi^{-})K^{+}$ 
& $\phantom{0}3.6\pm 0.9\pm 1.1$ & $\phantom{0}0.2\pm 0.2\pm 2.3$ & $1.2$ \\
$K_{1}(1270)^{+}(\rho^{0} K^{+})K^{-}$ 
& $\phantom{0}2.6\pm 0.8 \pm 0.5$ & $ \phantom{0}8.5\pm 1.4\pm 4.2$ & $1.3$ \\
$K_{1}(1270)^{-}(\rho^{0} K^{-})K^{+}$ 
& $\phantom{0}7.9\pm 1.2\pm 0.8$ & $\phantom{0}3.5\pm 1.1 \pm 3.4 $& $1.2$\\
$ K^{*}(1410)^{+}(K^{*0} \pi^{+})K^{-}$ 
& $\phantom{0}4.5\pm 0.9\pm 0.6$ & $\phantom{0}4.5\pm 1.1 \pm 1.2 $ & $0.0$ \\
$ K^{*}(1410)^{-} (\overline{K^{*0}}\pi^{-}) K^{+}$ 
& $\phantom{0}5.5\pm 1.0\pm 0.8$ & $\phantom{0}4.6\pm 0.9 \pm 1.1$ & $0.4$ \\
$K^{*0} \overline{K^{*0}}$ $S$ wave 
& $\phantom{0}7.5\pm 1.8 \pm 1.8$ & $\phantom{0}5.2\pm 1.0\pm 2.0$ & $0.7$ \\
$\phi\rho^{0}$ $S$ wave 
& $39.8\pm 2.7\pm 1.4$& $36.9\pm 3.2\pm 4.0$ & $0.5$ \\
$\phi\rho^{0}$ $D$ wave 
& $\phantom{0}4.7\pm 1.0\pm 1.1$ & $\phantom{0}1.7\pm 0.8\pm 2.3$ & $1.0$ \\
$\phi \left\{\pi^{+}\pi^{-}\right\}_S$ 
& $\phantom{0}8.3\pm 1.2 \pm 0.5$ & $10.9\pm 1.7\pm 3.0$ & $0.7$\\
$\left\{K^{-}\pi^{+}\right\}_P\left\{K^{+}\pi^{-}\right\}_S$
& $14.7\pm 1.8\pm 4.3$ & $\phantom{0}7.8\pm 1.4\pm 5.3$ & $1.1$ \\
\hline\hline		 
  \end{tabular}
 \end{center}
\end{table}

\begin{figure}
 \begin{center}
 \includegraphics[width=1.00\textwidth]{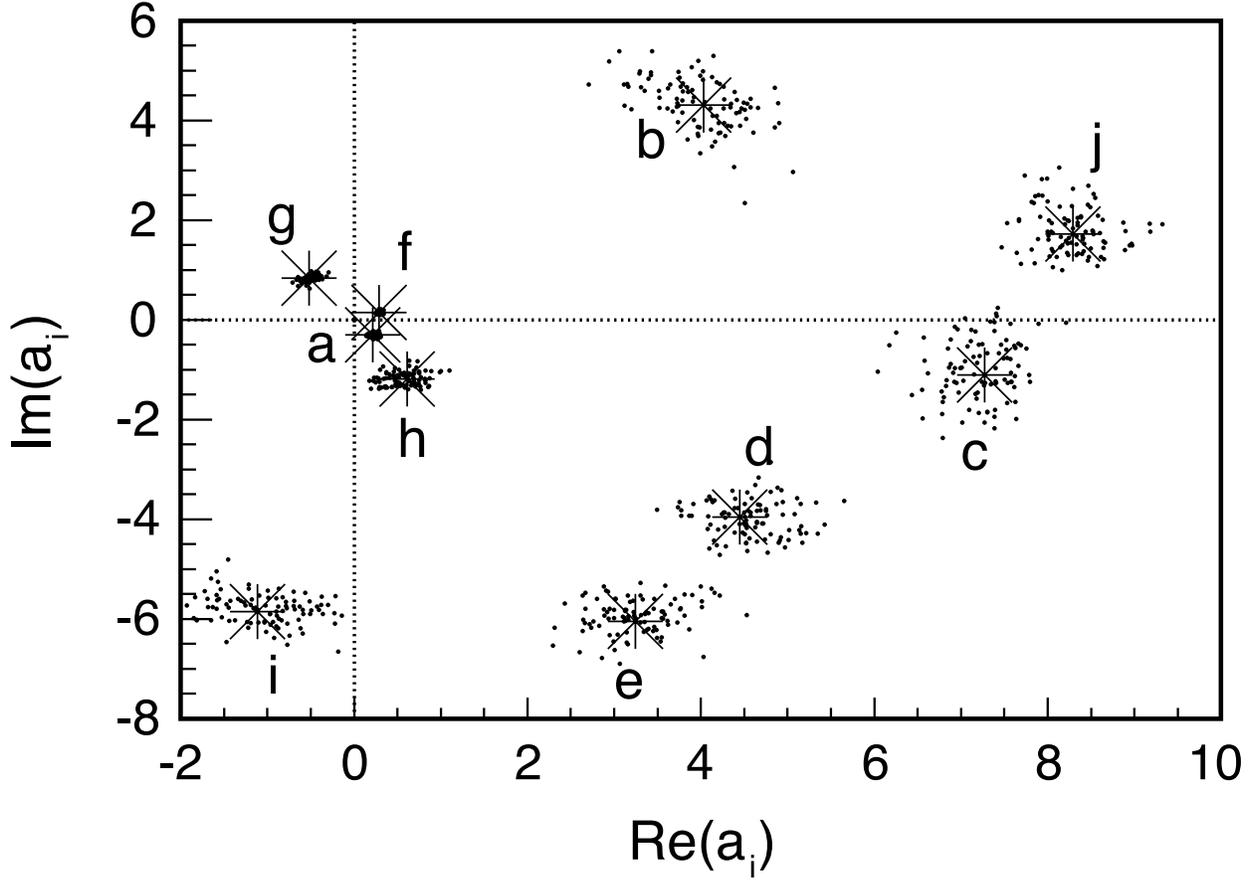}
  \caption{Fitted $a_i$ (points) distribution for the ensemble test of 200 simulated data sets. Also shown is the %%@
generated value of $a_i$ (asterisk). The labels correspond to: 
(a) $K_1(1270)^{-}(\overline{K^{*0}}\pi^{-})K^{+}$, 
(b) $K_1(1270)^{+}(\rho^{0}K^{+})K^{-}$, 
(c) $K_1(1270)^{-}(\rho^{0}K^{-})K^{+}$, 
(d) $K^{*}(1410)^{+}(K^{*0}\pi^{+})K^{-}$, 
(e) $K^{*}(1410)^{-}(\overline{K^{*0}}\pi^{-})K^{+}$, 
(f) $K^{*0}\overline{K^{*0}}$ $S$ wave, 
(g) $\phi\rho^{0}$ $S$ wave, 
(h) $\phi\rho^{0}$ $D$ wave, 
(i) $\phi\left\{ \pi^{+}\pi^{-}\right\}_{S}$, and 
(j) $\left\{K^{-}\pi^{+}\right\}_{P}\left\{K^{+}\pi^{-}\right\}_{S}$. 
The values of $\mathrm{Re}(a_i)$ and $\mathrm{Im}(a_i)$ are scaled by 0.1 for the %%@
$\left\{K^{-}\pi^{+}\right\}_{P}\left\{K^{+}\pi^{-}\right\}_{S}$ amplitude.
}\label{fig:argand}
 \end{center}
\end{figure}

The fitter is also tested on an ensemble of simulated data sets to identify any bias in the fit and determine the %%@
reliability of the statistical uncertainties returned by the fit. The ensemble is 200 sets of ~4000 events %%@
generated with the CLEO-c simulation, which are fit individually. The distribution of $a_i$ for the ensemble of %%@
experiments is shown in Fig.~\ref{fig:argand}; the values are seen to be scattered about the generated values with %%@
no significant biases. The distribution of the pull, defined as the difference between the fitted and generated %%@
parameters divided by the uncertainty on the parameter returned by the fit, is formed for the real and imaginary %%@
part of $a_i$ for the ensemble of simulation experiments. If the fit is unbiased and the uncertainties are %%@
correctly determined the pull distribution will be normal. This is tested by fitting each pull distribution with a %%@
Gaussian function. The width and mean found by the fit to the pull distributions are given in Table~\ref{tab:pull}. %%@
The ensemble study is performed with the fit using the generated and reconstructed four-momenta of the particles. %%@
As the resolution is neglected in the fit the difference between the results obtained with the generated and %%@
reconstructed four-momenta allows the systematic uncertainty related to resolution to be determined.
Small biases are observed but only one is greater than three standard deviations from zero. The widths of the pull %%@
distributions are all compatible with unity indicating the uncertainties are correctly evaluated.  The small bias %%@
is accounted for in evaluating the systematic uncertainties, which is described in Sec.~\ref{subsec:syst}. The %%@
average $\chi^2/\nu$ for the ensemble of fits is $0.96\pm 0.01$, which is a further indication that the fitting %%@
algorithm is well behaved.

\begin{table}[htbp]
\begin{center}
\caption{Mean $(\mu)$ and width $(\sigma)$ of the pull distributions from simulated data for all of the fitted %%@
parameters using either the generated or reconstructed four-momenta.}
\label{tab:pull}
\begin{tabular}{lcccc} \hline\hline
Parameter & \multicolumn{2}{c}{Generated $p_i$} & \multicolumn{2}{c}{Reconstructed $p_i$} \\ 
          & $\mu$ & $\sigma$ & $\mu$ & $\sigma$ \\ 
\hline
$K_{1}(1270)^{-}(\overline{K}^{*0}\pi^{-})K^{+}$  $\mathrm{Re}(a_i)$ 
& \hspace{0.1cm} $-0.43\pm 0.13$ \hspace{0.1cm} & \hspace{0.1cm} $1.02\pm 0.14$ \hspace{0.1cm} & \hspace{0.1cm} %%@
$-0.36\pm 0.12$ \hspace{0.1cm} & \hspace{0.1cm} $0.99\pm 0.13$ \hspace{0.1cm} \\
$K_{1}(1270)^{-}(\overline{K}^{*0}\pi^{-})K^{+}$  $\mathrm{Im}(a_i)$
& $\phantom{-}0.03\pm 0.13$ & $1.14 \pm 0.12$ & $-0.10\pm 0.14$ & $1.13\pm 0.13$\\
$K_{1}(1270)^{+}(\rho^{0}K^{+})K^{-}$ $\mathrm{Re}(a_i)$ 
& $\phantom{-}0.08\pm 0.12$ & $1.06 \pm 0.12$ & $\phantom{-}0.16\pm 0.19$ & $1.35 \pm 0.19$ \\
$K_{1}(1270)^{+}(\rho^{0}K^{+})K^{-}$ $\mathrm{Im}(a_i)$
& $-0.15\pm 0.09$ & $0.85\pm 0.07$ & $-0.29 \pm 0.10$ & $0.83 \pm 0.10$ \\
$K_{1}(1270)^{-}(\rho^{0}K^{-})K^{+}$ $\mathrm{Re}(a_i)$
& $-0.01\pm 0.11$ & $0.88\pm 0.10$ & $-0.03\pm 0.12$ & $1.01\pm 0.11$ \\
$K_{1}(1270)^{-}(\rho^{0}K^{-})K^{+}$  $\mathrm{Im}(a_i)$ 
& $-0.24\pm 0.14$ & $1.17\pm 0.14$ & $-0.22\pm 0.12$ & $1.08\pm 0.12$ \\
$K^{*}(1410)^{+} (K^{*0}\pi^{+}), K^{-}$ $\mathrm{Re}(a_i)$ 
& $-0.32 \pm 0.13$ & $1.11 \pm 0.12$ & $-0.41\pm 0.11$ & $0.98 \pm 0.09$ \\
$K^{*}(1410)^{+} (K^{*0}\pi^{+}) K^{-}$ $\mathrm{Im}(a_i)$ 
& $\phantom{-}0.08\pm 0.14$ & $1.08\pm 0.12$ & $\phantom{-}0.11\pm 0.11$ & $0.89\pm 0.10$ \\
$K^{*}(1410)^{-} (\overline{K}^{*0}\pi^{-})K^{+}$ $\mathrm{Re}(a_i)$ 
& $-0.05\pm 0.11$ & $0.91\pm 0.10$ & $-0.08\pm 0.13$ & $0.96\pm 0.11$ \\
$K^{*}(1410)^{-} (\overline{K}^{*0} \pi^{-}), K^{+}$ $\mathrm{Im}(a_i)$
& $-0.35 \pm 0.11$ & $0.95\pm 0.12$ & $\phantom{-}0.23\pm 0.13$ & $1.06 \pm 0.12$ \\
$K^{*0}\overline{K}^{*0}$ S wave  $\mathrm{Re}(a_i)$ 
& $\phantom{-}0.28\pm 0.11$ & $0.94\pm 0.09$ & $\phantom{-}0.21\pm 0.14$ & $1.13\pm 0.12$ \\
$K^{*0}\overline{K}^{*0}$ S wave  $\mathrm{Im}(a_i)$ 
& $-0.15\pm 0.11$ & $0.99\pm 0.10$ & $-0.02\pm 0.10$ & $0.83\pm 0.09$ \\
$\phi\rho^{0}$ S wave  $\mathrm{Re}(a_i)$ 
& $-0.29\pm 0.11$ & $0.87\pm 0.15$ & $-0.26\pm 0.10$ & $0.90\pm 0.12$ \\
$\phi\rho^{0}$ S wave  $\mathrm{Im}(a_i)$ 
& $-0.02\pm 0.10$ & $0.85\pm 0.09$ & $\phantom{-}0.17\pm 0.10$ & $0.86\pm 0.10$ \\
$\phi\rho^{0}$ D wave $\mathrm{Re}(a_i)$
& $\phantom{-}0.20\pm 0.13$ & $1.05\pm 0.11$ & $\phantom{-}0.31\pm 0.13$ & $1.11\pm 0.12$ \\
$\phi\rho^{0}$ D wave $\mathrm{Im}(a_i)$ 
& $-0.06\pm 0.11$ & $0.96\pm 0.09$ & $\phantom{-}0.01\pm 0.12$ & $0.91\pm 0.10$ \\
$\phi\left\{\pi^{+}\pi^{-}\right\}_{S}$  $\mathrm{Re}(a_i)$ 
& $-0.20\pm 0.14$ & $1.14 \pm 0.12$ & $-0.13\pm 0.14$ & $1.16 \pm 0.12$ \\
$\phi\left\{\pi^{+}\pi^{-}\right\}_{S}$  $\mathrm{Im}(a_i)$ & 
$-0.16\pm 0.09$ & $0.85\pm 0.10$ & $-0.24 \pm 0.13$ & $1.08 \pm 0.12$ \\
$\left\{K^{-}\pi^{+}\right\}_{P}\left\{K^{+}\pi^{-}\right\}_{S}$ $\mathrm{Re}(a_i)$ 
& $\phantom{-}0.17\pm 0.10$ & $0.87\pm 0.10$ & $\phantom{-}0.01\pm 0.12$ & $0.98\pm 0.11$ \\
$\left\{K^{-}\pi^{+}\right\}_{P}\left\{K^{+}\pi^{-}\right\}_{S}$ $\mathrm{Im}(a_i)$  
& $\phantom{-}0.03\pm 0.16$ & $1.15\pm 0.21$ & $-0.07\pm 0.12$ & $0.86\pm 0.11$ \\
\hline\hline
\end{tabular}
\end{center}
\end{table}

\subsection{Systematic uncertainties}
\label{subsec:syst}

The systematic uncertainties on the results fall into several categories: amplitude model assumptions, %%@
parameterization of the background, modeling of experimental effects, and the fitter performance. 
Each of these categories is discussed below. The resulting systematic uncertainties for the amplitudes, phases, and %%@
fit fractions due to each contribution are given in Tables~\ref{tab:syst_ampphase} and \ref{tab:syst_ff}.

\begin{table}[t]
\begin{center}
\caption{Systematic uncertainties on $|a_i|$ and $\phi_i$ in units of statistical standard deviations ($\sigma$). %%@
The different contributions are: (I) mass and width of resonances; (II) Blatt-Weisskopf penetration factors; (III) %%@
quantum correlations; (IV) background fractions; (V) flavor-tagged background parameterization; (VI) $CP$-tagged %%@
background parameterization ; (VII) $K^{0}_{S}K^{+}K^{-}$ background; (VIII) acceptance; (IX) resolution; (X) %%@
mistag rate; and (XI) fitter bias.} \label{tab:syst_ampphase}
\footnotesize
\begin{tabular}{lcccccccccccc}
\hline\hline
 Parameter & \multicolumn{11}{c}{Source $(\sigma)$} & Total $(\sigma)$ \\
      & I & II & III & IV & V & VI & VII & VIII & IX & X & XI &  \\ \hline
$K_{1}(1270)^{-}(\overline{K^{*0}}\pi^{-})K^{+}$ $|a_i|$ 
& $\phantom{}$0.03$\phantom{}$ & $\phantom{}$0.17$\phantom{}$ & $\phantom{}$0.01$\phantom{}$ & %%@
$\phantom{}$0.31$\phantom{}$ & $\phantom{}$0.15$\phantom{}$ & $\phantom{}$0.08$\phantom{}$& %%@
$\phantom{}$0.01$\phantom{}$ & $\phantom{}$0.23$\phantom{}$ & $\phantom{}$0.00$\phantom{}$ & %%@
$\phantom{}$0.03$\phantom{}$ & $\phantom{}$0.24$\phantom{}$ & $\phantom{}$0.52$\phantom{}$ \\
$K_{1}(1270)^{-}(\overline{K^{*0}}\pi^{-})K^{+}$ $\phi_i$ 
& 0.83 & 0.36 & 0.04 & 0.06 & 0.18 & 0.02 & 0.02 & 0.30 & 0.00 & 0.04 & 0.40 & 1.05 \\ 
$K_{1}(1270)^{+}(\rho^{0}K^{+})K^{-}$ $|a_i|$ 
& 2.52 & 0.23 & 0.17 & 0.06 & 0.30 & 0.03 & 0.04 & 0.43 & 0.31 & 0.01 & 0.07 & 2.61 \\
$K_{1}(1270)^{+}(\rho^{0}K^{+})K^{-}$ $\phi_i$ 
& 0.46 & 0.14 & 0.11 & 0.13 & 0.17 & 0.06 & 0.06 & 0.07 & 0.14 & 0.02 & 0.16 & 0.58 \\
$K_{1}(1270)^{-}(\rho^{0}K^{-})K^{+}$ $|a_i|$ 
& 5.10 & 0.30 & 0.04 & 0.22 & 0.21 & 0.01 & 0.02 & 0.76 & 0.05 & 0.02 & 0.01 & 5.17 \\
$K_{1}(1270)^{-}(\rho^{0}K^{-})K^{+}$ $\phi_i$ 
& 1.27 & 0.42 & 0.07 & 0.11 & 0.07 & 0.04 & 0.06 & 0.51 & 0.22 & 0.04 & 0.13 & 1.47 \\
$K^{*}(1410)^{+} (K^{*0}\pi^{+})K^{-}$ $|a_i|$ 
& 1.04 & 0.31 & 0.18 & 0.06 & 0.11 & 0.06 & 0.06 & 0.27 & 0.18 & 0.02 & 0.27 & 1.18 \\
$K^{*}(1410)^{+} (K^{*0}\pi^{+})K^{-}$ $\phi_i$ 
& 3.04 & 0.47 & 0.09 & 0.11 & 0.22 & 0.09 & 0.04 & 0.29 & 0.15 & 0.04 & 0.11 & 3.10 \\
$K^{*}(1410)^{-} (\overline{K^{*0}}\pi^{-}) K^{+}$ $|a_i|$ 
& 1.66 & 0.76 & 0.07 & 0.07 & 0.22 & 0.01 & 0.02 & 0.45 & 0.15 & 0.02 & 0.26 & 1.92 \\
$K^{*}(1410)^{-} (\overline{K^{*0}}\pi^{-}) K^{+}$ $\phi_i$ 
& 3.46 & 1.04 & 0.13 & 0.23 & 0.22 & 0.11 & 0.01 & 0.23 & 0.12 & 0.01 & 0.13 & 3.64 \\
$K^{*0} \overline{K^{*0}}$ $S$ wave $|a_i|$ 
& 3.86 & 0.27 & 0.23 & 0.20 & 0.17 & 0.07 & 0.07 & 0.55 & 0.00 & 0.01 & 0.19 & 3.93 \\
$K^{*0} \overline{K^{*0}}$ $S$ wave $\phi_i$ 
& 1.20 & 0.64 & 0.01 & 0.22 & 0.26 & 0.11 & 0.02 & 0.37 & 0.00 & 0.02 & 0.21 & 1.47 \\
$\phi\rho^{0}$ $S$ wave $|a_i|$ 
& 3.04 & 0.16 & 0.22 & 0.23 & 0.16 & 0.07 & 0.06 & 0.43 & 0.00 & 0.01 & 0.16 & 3.10 \\
$\phi\rho^{0}$ $S$ wave $\phi_i$ 
& 2.42 & 0.47 & 0.24 & 0.14 & 0.13 & 0.15 & 0.12 & 0.35 & 0.00 & 0.02 & 0.33 & 2.54 \\
$\phi\rho^{0}$ $D$ wave $|a_i|$ 
& 1.89 & 0.13 & 0.27 & 0.19 & 0.14 & 0.03 & 0.08 & 0.59 & 0.07 & 0.01 & 0.11 & 2.02 \\
$\phi\rho^{0}$ $D$ wave $\phi_i$ 
& 2.05 & 0.35 & 0.15 & 0.10 & 0.11 & 0.12 & 0.06 & 0.27 & 0.24 & 0.02 & 0.19 & 2.14 \\
$\phi\left\{\pi^{+}\pi^{-}\right\}_S$ $|a_i|$ 
& 3.01 & 0.02 & 0.07 & 0.18 & 0.20 & 0.02 & 0.09 & 0.37 & 0.11 & 0.01 & 0.28 & 3.06 \\
$\phi\left\{\pi^{+}\pi^{-}\right\}_S$ $\phi_i$ 
& 2.64 & 0.68 & 0.19 & 0.09 & 0.12 & 0.16 & 0.12 & 0.38 & 0.02 & 0.03 & 0.16 & 2.77 \\
$\left\{K^{-}\pi^{+}\right\}_P\left\{K^{+}\pi^{-}\right\}_S$ $|a_i|$ 
& 4.25 & 0.13 & 0.02 & 0.10 & 0.30 & 0.06 & 0.05 & 0.43 & 0.00 & 0.02 & 0.20 & 4.29 \\ 
$\left\{K^{-}\pi^{+}\right\}_P\left\{K^{+}\pi^{-}\right\}_S$ $\phi_i$ 
& 2.14 & 1.04 & 0.08 & 0.06 & 0.16 & 0.11 & 0.06 & 0.17 & 0.00 & 0.02 & 0.05 & 2.39 \\ 
\hline\hline
\end{tabular}
\normalsize
\end{center}
\end{table}

\begin{table}[t]
\begin{center}
\caption{Systematic uncertainties on the fit fraction in units of statistical standard deviations ($\sigma$). The %%@
different contributions are: (I) mass and width of resonances; (II) Blatt-Weisskopf penetration factors; (III) %%@
quantum correlations; (IV) background fractions; (V) flavor-tagged background parameterization; (VI) $CP$-tagged %%@
background parameterization; (VI) $K^{0}_{S}K^{+}K^{-}$ background; (VIII) acceptance; (IX) resolution; (X) mistag %%@
rate; and (XI) fitter bias.} \label{tab:syst_ff}
\footnotesize
\begin{tabular}{lcccccccccccc}
\hline\hline
 Fit fraction & \multicolumn{11}{c}{Source $(\sigma)$} & Total $(\sigma)$ \\
      & I & II & III & IV & V & VI & VII & VIII & IX & X & XI & \\ \hline
${K_{1}}(1270)^{+}(K^{*0}\pi^{+})K^{-}$  
& $\phantom{}$2.23$\phantom{}$ & $\phantom{}$0.57$\phantom{}$ & $\phantom{}$0.16$\phantom{}$ & %%@
$\phantom{}$0.25$\phantom{}$ & $\phantom{}$0.27$\phantom{}$ & $\phantom{}$0.06$\phantom{}$ & %%@
$\phantom{}$0.05$\phantom{}$ & \phantom{}0.57$\phantom{}$ & $\phantom{}$0.00$\phantom{}$ & %%@
$\phantom{}$0.01$\phantom{}$ & $\phantom{}$0.05$\phantom{}$ & $\phantom{}$2.40$\phantom{}$ \\
$K_{1}(1270)^{-}(K^{*0}\pi^{-})K^{+}$  
& 1.13 & 0.41 & 0.05 & 0.28 & 0.11 & 0.09 & 0.01 & 0.15 & 0.23 & 0.04 & 0.28 & 1.30 \\
$K_{1}(1270)^{+}(\rho^{0}K^{+})K^{-}$  
& 0.81 & 0.30 & 0.13 & 0.09 & 0.32 & 0.01 & 0.03 & 0.34 & 0.08 & 0.02 & 0.03 & 1.00 \\
$K_{1}(1270)^{-}(\rho^{0}K^{-})K^{+}$  
& 0.41 & 0.32 & 0.08 & 0.13 & 0.09 & 0.06 & 0.04 & 0.51 & 0.09 & 0.03 & 0.05 & 0.75 \\
$ K^{*}(1410)^{+} (K^{*0}\pi^{+})K^{-}$  
& 0.84 & 0.16 & 0.14 & 0.14 & 0.28 & 0.05 & 0.05 & 0.33 & 0.26 & 0.02 & 0.35 & 1.08 \\
$ K^{*}(1410)^{-} (\overline{K^{*0}}\pi^{-}) K^{+}$  
& 0.81 & 0.34 & 0.05 & 0.25 & 0.13 & 0.04 & 0.03 & 0.47 & 0.24 & 0.01 & 0.05 & 1.06 \\
$K^{*0} \overline{K^{*0}}$ $S$ wave  
& 0.83 & 0.22 & 0.20 & 0.12 & 0.27 & 0.06 & 0.06 & 0.64 & 0.01 & 0.01 & 0.05 & 1.13 \\
$\phi\rho^{0}$ $S$ wave  
& 1.34 & 0.35 & 0.41 & 0.29 & 0.05 & 0.10 & 0.10 & 0.22 & 0.23 & 0.02 & 0.23 & 1.53 \\
$\phi\rho^{0}$ $D$ wave  
& 0.43 & 0.28 & 0.28 & 0.13 & 0.04 & 0.03 & 0.09 & 0.66 & 0.10 & 0.02 & 0.09 & 0.90 \\
$\phi \left\{\pi^{+}\pi^{-}\right\}_S$  
& 0.43 & 0.57 & 0.19 & 0.16 & 0.09 & 0.10 & 0.16 & 0.19 & 0.00 & 0.05 & 0.19 & 0.83 \\
$\left\{K^{-}\pi^{+}\right\}_P\left\{K^{+}\pi^{-}\right\}_S$  
& 1.00 & 0.23 & 0.17 & 0.22 & 0.20 & 0.06 & 0.03 & 0.22 & 0.00 & 0.04 & 0.22 & 1.43 \\  
\hline\hline
\end{tabular}
\normalsize
\end{center}
\end{table}
Three assumptions of the amplitude model are tested: the mass and width of resonances, the barrier-penetration %%@
factors, and the absence of quantum correlations in the modeling of CLEO-c flavor tagged data. 
The mass and width assumed for the resonances in the model, $K_1(1270)^{+}$, $K^{*0}$, $K(1410)^{+}$, $\phi$, and %%@
$\rho^{0}$, are taken from Ref.~\cite{PDG}; these are varied by the quoted uncertainties to determine the related %%@
shift of the fit parameters. These shifts are added in quadrature to obtain the total uncertainty related to the %%@
mass and width parameters. The uncertainty from this source is of the same size or larger than the statistical %%@
error for most parameters. The uncertainty on the $K_1(1270)^{+}$ mass and width dominates. Determining the %%@
$K_1(1270)^{+}$ mass and width from data does not improve the overall uncertainty.

The amplitude model includes spin- and momentum-dependent Blatt-Weisskopf orbital-angular-momentum barrier %%@
penetration factors. These factors are set to unity and the fit repeated. The resulting shift in the fitted %%@
parameter values and fit fractions is taken as the systematic uncertainty. 

Quantum correlations in the CLEO-c flavor-tagged data are ignored in the fit. Since many different final states %%@
containing the tagging kaon are summed over in the analysis the effect of correlations is diluted. However, an %%@
alternative signal PDF is tested, which accounts for the correlations at the cost of two additional parameters (see %%@
Appendix~\ref{app:results}). The resulting changes
in the central values of the fitted parameters and fit fractions, taken as
the systematic uncertainties from this source, are found to be less than
half a statistical standard deviation in all cases. 

Both the level and shape of the background are considered when evaluating the systematic uncertainty. The fraction %%@
of non-peaking background in each flavor-tagged sample is estimated from data. The value of $f_S$ is varied by its %%@
statistical uncertainty for each data sample in turn and the shifts in the results are added in quadrature to %%@
estimate the systematic error from this source. The components in the background model for the flavor-tagged data %%@
are changed such that those that are not statistically significant, defined as those contributing a fraction less %%@
than three standard deviations from zero, are removed from the model and the fit repeated. Each such component is %%@
removed in turn and the individual shifts are summed in quadrature to obtain the total uncertainty. In the case of %%@
the $CP$-tagged data where the background model is derived from data and simulation the fit is repeated ignoring %%@
the background. The shifts in the fit parameters with respect to the nominal fit are conservatively taken as the %%@
systematic uncertainty. In Sec. \ref{subsec:modselres} it is noted that the fit underestimates the level of the %%@
$K^{0}_{S}K^{+}K^{-}$ background. To assess the systematic uncertainty related to this discrepancy the fit is %%@
repeated with the fraction of the $K^{0}_{S}K^{+}K^{-}$ background fixed to double that found in the data, which %%@
leads to reasonable agreement with data in the region of the $K^{0}_{S}$. The difference in parameter values %%@
between the nominal fit and that with the $K^{0}_{S}K^{+}K^{-}$ background fraction doubled is taken as the %%@
systematic uncertainty from this source.   
The largest background-related uncertainties are due to the statistical precision on the signal fraction. 

The uncertainty related to the modeling of experimental effects has three separate components: the acceptance, the %%@
resolution, and the mistag rate. The acceptance is incorporated in the fit using simulated data, as described in %%@
Sec.~\ref{subsec:norm}, for CLEO III and CLEO-c data. To evaluate a systematic uncertainty related to the %%@
acceptance an alternative technique is used that is based on the product of individual particle efficiencies as a %%@
function of momentum in the laboratory frame. This is an almost identical procedure to the weighting used for the %%@
CLEO II.V normalization with simulated events. The only difference is an additional factor for CLEO-c data which is %%@
a function of $\pi^{+}\pi^{-}$ invariant mass to account for the $K^{0}_{S}$ veto. Such an approach is known to be %%@
simplistic compared to that using the fully simulated events given the integration over other variables on which %%@
the acceptance depends, such as polar angle, and the fact that it ignores correlations among the momenta. %%@
Therefore, it is considered a conservative approach to evaluating the systematic uncertainty related to the %%@
acceptance. The full difference in fit results for the two different techniques for incorporating the acceptance is %%@
taken as the systematic uncertainty. For the CLEO II.V data a uniform acceptance is assumed as an alternative model %%@
because the product of efficiencies is the technique used in the nominal fit. Most uncertainties due to the %%@
acceptance are around half a statistical standard deviation.

The effects of resolution are ignored in the fit. The ensemble tests of simulated data reported in %%@
Sec.~\ref{subsec:robust} are used to estimate the effect of resolution. The systematic variance due to the %%@
resolution is taken as the difference in the pull means squared for the fits performed with the generated and %%@
reconstructed four-momenta. The effect of resolution is found to be very small or negligible on the fit parameters %%@
and fractions.

The mistag rate for the different flavor-tag samples is varied within the uncertainties reported in %%@
Sec.~\ref{sec:dataset}. The resulting shift in the parameters is taken as the systematic uncertainty.

The final source of systematic uncertainty considered is related to the overall performance of the fitter as %%@
demonstrated in Sec.~\ref{subsec:robust}. Small biases in the pull mean for some parameters are observed when %%@
fitting with the generated four-momenta. The largest bias is 40\% of a statistical uncertainty with a significance %%@
of $3.5\sigma$ from zero. The pull mean is assigned as a systematic uncertainty related to this fitter bias for %%@
each parameter. The source of such a bias could be the finite Monte Carlo statistics used to compute the %%@
normalization or the overall numerical precision of the fit. 

The systematic uncertainty is dominated by the mass and width assumed for the resonances; this leads to the %%@
measurements of the amplitudes and phases being systematically dominated. For the fit fractions the systematic and %%@
statistical uncertainties are approximately of equal magnitude. 

\subsection{$CP$ violation search}
\label{subsec:cp}
The baseline model is fit to the data allowing different values of $\mathrm{Re}(a_i)$ and $\mathrm{Im}(a_i)$ for %%@
$D^{0}$ and $\overline{D^{0}}$ amplitudes. The fit is to the combined $D^{0}$ and $\overline{D^{0}}$ flavor-tagged %%@
data to account correctly for the mistag rate. A comparison of the values of $|a_i|$ and $\phi_i$ for $D^{0}$ and %%@
$\overline{D^{0}}$ amplitudes is given in Table~\ref{tab:cpamp}. The systematic uncertainties are considered to be %%@
fully correlated between the $D^{0}$ and $\overline{D^{0}}$ samples. The fit fractions are compared in %%@
Table~\ref{tab:cpff}. The fit fractions are used to determine the direct-$CP$ asymmetry $A_{CP}$, for each %%@
amplitude, which is defined as
\begin{displaymath}
A_{CP} = \frac{F_{D^{0}}-F_{\overline{D^{0}}}}{F_{D^{0}}+F_{\overline{D^{0}}}} \; ,
\end{displaymath}
where $F_{D^{0}}$ and $F_{\overline{D^{0}}}$ are the fit fractions for $D^{0}$ and $\overline{D^{0}}$ decays, %%@
respectively.
There is no evidence for a significant $CP$ asymmetry between any of the amplitudes. The sensitivity to $A_{CP}$ %%@
varies between $5\%$ to $30\%$ among the amplitudes; therefore, the level of precision is not at the sub-percent %%@
level at which evidence of $CP$ violation has been found in two-body $D^{0}$ decay \cite{BIGDOG}.

\begin{table}[ht]
\begin{center}
\caption{$|a_i|$ and $\phi_i$ for the $D^{0}$ and $\overline{D^{0}}$ amplitudes. The difference between the $D^{0}$ %%@
and $\overline{D^{0}}$ parameter is also given in units of standard deviations $(\sigma)$.}\label{tab:cpamp}
\begin{tabular}{lccc}
\hline\hline
 Parameter & $D^{0}$ decays & $\overline{D^{0}}$ decays & Difference $(\sigma)$  \\ 
\hline
$K_{1}(1270)^{-}(\overline{K^{*0}}\pi^{-})K^{+}$  $|a_i|$ 
& \hspace{1cm} $0.35 \pm 0.08$ \hspace{1cm} & \hspace{1cm} $0.39 \pm 0.09$ \hspace{1cm} & 0.3\\
$K_{1}(1270)^{-}(\overline{K^{*0}}\pi^{-})K^{+}$  $\phi_i$ 
&$1.52 \pm 0.33$ & $0.98 \pm 0.24$ & 1.3 \\

$K_{1}(1270)^{+}(\rho^{0}K^{+})K^{-}$ $|a_i|$
&$5.58 \pm 0.98$ & $5.96 \pm 0.84$ & 0.3\\
$K_{1}(1270)^{+}(\rho^{0}K^{+})K^{-}$ $\phi_i$
&$0.86 \pm 0.17$ & $0.71 \pm 0.15$ & 0.7\\

$K_{1}(1270)^{-}(\rho^{0}K^{-})K^{+}$  $|a_i|$
&$7.03 \pm 1.03$ & $6.39 \pm 0.82$ & 0.5\\
$K_{1}(1270)^{-}(\rho^{0}K^{-})K^{+}$  $\phi_i$
&$0.41 \pm 0.19$ & $0.30 \pm 0.18$ & 0.4\\

$ K^{*}(1410)^{+}(K^{*0}\pi^{+})K^{-}$ $|a_i|$
&$5.39 \pm 0.91$ & $6.51 \pm 0.86$ & 0.9\\
$ K^{*}(1410)^{+}(K^{*0}\pi^{+})K^{-}$ $\phi_i$
&$0.99 \pm 0.18$ & $0.68 \pm 0.13$ & 1.4\\

$ K^{*}(1410)^{-}(\overline{K^{*0}}\pi^{-})K^{+}$ $|a_i|$
&$6.69 \pm 0.82$ & $6.75 \pm 0.92$ & 0.0\\
$ K^{*}(1410)^{-}(\overline{K^{*0}}\pi^{-})K^{+}$ $\phi_i$
&$1.38 \pm 0.20$ & $0.95 \pm 0.15$ & 1.7\\

$K^{*0}\overline{K^{*0}}$ $S$ wave  $|a_i|$
&$0.36 \pm 0.05$ & $0.33 \pm 0.04$ & 0.5\\
$K^{*0}\overline{K^{*0}}$ $S$ wave  $\phi_i$
&$0.47 \pm 0.17$ & $0.24 \pm 0.14$ & 1.1\\

$\phi\rho^{0}$ $S$ wave  $|a_i|$
&$1.02 \pm 0.16$ & $1.05 \pm 0.08$ & 0.2\\
$\phi\rho^{0}$ $S$ wave  $\phi_i$
&$2.08 \pm 0.19$ & $1.77 \pm 0.07$ & 1.5\\

$\phi\rho^{0}$ $D$ wave $|a_i|$
&$1.14 \pm 0.30$ & $1.69 \pm 0.23$ & 1.5\\
$\phi\rho^{0}$ $D$ wave $\phi_i$
&$1.18 \pm 0.36$ & $1.48 \pm 0.14$ & 0.8\\

$\phi\left\{\pi^{+}\pi^{-}\right\}_S$  $|a_i|$
&$5.76 \pm 0.65$ & $6.22 \pm 0.65$ & 0.5\\
$\phi\left\{\pi^{+}\pi^{-}\right\}_S$ $\phi_i$
&$1.75 \pm 0.19$ & $1.98 \pm 0.09$ & 1.1\\

$\left\{K^{-}\pi^{+}\right\}_{P}\left\{K^{+}\pi^{-}\right\}_S$ $|a_i|$
&$84.7 \pm \phantom{0}9.2$ & $81.7 \pm \phantom{0}8.3$ & 0.2\\
$\left\{K^{-}\pi^{+}\right\}_{P}\left\{K^{+}\pi^{-}\right\}_S$ $\phi_i$
&$0.16 \pm 0.16$ & $0.13 \pm 0.10$ & 0.2\\
\hline\hline
\end{tabular}
\end{center}
\end{table}

\begin{table}[ht]
\begin{center}
\caption{$D^{0}$ and $\overline{D^{0}}$ fit fractions. The value of $A_{CP}$ is also given.}\label{tab:cpff}
\begin{tabular}{lccr}
\hline\hline
 & \multicolumn{2}{c}{Fit fraction $(\%)$} & \multicolumn{1}{c}{$A_{CP}$} \\ 
 & $D^{0}$ Decays & $\overline{{D}^{0}}$ Decays & \multicolumn{1}{c}{$(\%)$}  \\ \hline
$K_{1}(1270)^{+}(K^{*0}\pi^{+})K^{-}$ 
&$\phantom{0}7.4 \pm 1.1$ & $\phantom{0}7.5 \pm 1.1$ & $-0.7\pm 10.4$\\
$K_{1}(1270)^{-}(\overline{K^{*0}}\pi^{-})K^{+}$ 
&$\phantom{0}0.9 \pm 0.4$ & $\phantom{0}1.1 \pm 0.5$ & $-10.0\pm 31.5$\\
$K_{1}(1270)^{+}(\rho^{0} K^{+})K^{-}$
&$\phantom{0}4.3 \pm 1.1$ & $\phantom{0}4.9 \pm 1.1$ & $-6.5\pm 16.9$\\
$K_{1}(1270)^{-}(\rho^{0}K^{-})K^{+}$
&$\phantom{0}6.3 \pm 1.1$ & $\phantom{0}5.2 \pm 1.0$ & $\phantom{-}9.6\pm 12.9$\\
$ K^{*}(1410)^{+}(K^{*0}\pi^{+})K^{-}$
&$\phantom{0}3.2 \pm 0.9$ & $\phantom{0}4.8 \pm 1.0$ & $-20.0\pm 16.8$\\
$ K^{*}(1410)^{-} (\overline{K^{*0}}\pi^{-})K^{+}$
&$\phantom{0}4.6 \pm 0.9$ & $\phantom{0}4.7 \pm 0.9$ & $-1.1\pm 13.7$\\
$K^{*0}\overline{K^{*0}}$ $S$ wave 
&$\phantom{0}6.9 \pm 1.2$ & $\phantom{0}5.7 \pm 1.2$ & $\phantom{-}9.5\pm 13.5$\\
$\phi\rho^{0}$ $S$ wave 
&$37.9 \pm 2.9$ & $40.0 \pm 2.9$ & $-2.7\pm \phantom{0}5.3$\\
$\phi\rho^{0}$ $D$ wave
&$\phantom{0}2.2 \pm 0.8$ & $\phantom{0}4.8 \pm 1.2$ & $-37.1\pm 19.0$\\
$\phi\left\{\pi^{+}\pi^{-}\right\}_{S}$ 
&$\phantom{0}9.0 \pm 1.4$ & $10.7 \pm 1.5$ & $-8.6\pm 10.4$\\
$\left\{K^{-}\pi^{+}\right\}_{P}\left\{K^{+}\pi^{-}\right\}_{S}$
&$11.3 \pm 1.7$ & $10.7 \pm 1.6$ & $\phantom{-}2.7\pm 10.6$\\
\hline\hline
\end{tabular}
\end{center}
\end{table}

\clearpage

\section{$\mathbf \gamma$ sensitivity studies}
\label{sec:gamma}
The first estimates of the sensitivity of $B^{\pm}\to \widetilde{D^{0}}(K^{+}K^{-}\pi^{+}\pi^{-})K^{\pm}$ decays to %%@
the $CP$-violating parameter $\gamma$ were very promising \cite{JONASGUY}. These studies found that an uncertainty %%@
of approximately $10^{\circ}$ is expected for a data set corresponding to an integrated luminosity of %%@
$2~\mathrm{fb}^{-1}$ collected by LHCb. However, limited conclusions could be drawn as the model does not %%@
distinguish between $D^{0}$ and $\overline{D^{0}}$ decays \cite{E791,FOCUS}. Therefore, these $\gamma$ sensitivity %%@
studies are repeated for the amplitude model presented in this paper, which is determined from flavor-tagged %%@
$D^{0}$ decays. 

An amplitude fit to simulated $B^{\pm}\to \widetilde{D^{0}}(K^{+}K^{-}\pi^{+}\pi^{-})K^{\pm}$ data is used to %%@
determine $\gamma$. The fit is identical to that described in Ref.~\cite{JONASGUY} and similar to the %%@
amplitude-model-dependent analyses of $B^{+}\to \widetilde{D}(K^{0}_{S}\pi^{+}\pi^{-})K^{+}$ \cite{GGSZ}, which %%@
yield the most precise measurements of $\gamma$ to date \cite{KHHBABAR,KHHBELLE}. The amplitude fit determines %%@
$\gamma$ from the distributions of $\widetilde{D}\to K^{+}K^{-}\pi^{+}\pi^{-}$ events over four-body phase space, %%@
which are different for $\widetilde{D}$ mesons arising from $B^{+}$ or $B^{-}$ decay. The distribution depends on %%@
the ratio between the CKM-suppressed and color-suppressed $B^{-}\to \overline{D^{0}}K^{+}$ amplitude and the %%@
Cabibbo-favored and color-allowed $B^{-}\to D^{0}K^{-}$ amplitude and is parametrized as %%@
$r_{B}e^{i(\delta_B-\gamma)}$, 
where $r_B$ is the magnitude of the amplitude ratio and $\delta_B$ is the $CP$-invariant strong-phase difference %%@
between the amplitudes. The values of $r_B$ and $\delta_B$ are also determined by the amplitude fit.

An ensemble study of 200 simulated data sets containing 2000 $B^{\pm}\to %%@
\widetilde{D}(K^{+}K^{-}\pi^{+}\pi^{-})K^{\pm}$ events each, split evenly between the $B$ meson charges, is used to %%@
estimate the sensitivity to $\gamma$. The number of events in each data set corresponds approximately to that %%@
expected in a few years running of LHCb \cite{LHCBROADMAP}. The values of $\gamma$, $r_B$, and $\delta_B$ are %%@
assumed to be $70^\circ$, 0.1, and $130^{\circ}$, respectively, which are approximately the world average values %%@
\cite{HFAG}. These data samples are then fit assuming the amplitude model used in the generation, with $\gamma$, %%@
$r_B$, and $\delta_B$ as free parameters. The average uncertainty on $\gamma$ from the ensemble of experiments is %%@
$(11.3\pm 0.3)^{\circ}$ and the pull distribution formed from the fitted and generated values follows a normal %%@
distribution, indicating the results are unbiased. The pull distributions for $r_B$ and $\delta_B$ are also normal. %%@
Ensemble studies for the alternative models given in Appendix~\ref{app:bestfits} are also performed. Most %%@
alternative models yield a sensitivity to $\gamma$ similar to that of the baseline model. However, model 6 leads to %%@
an average uncertainty approximately 50\% worse than the other models. Similar model-dependent variations in %%@
uncertainty are reported in Ref.~\cite{JONASGUY}. However, overall these results confirm that there is significant %%@
sensitivity to $\gamma$ in $B^{\pm}\to \widetilde{D}(K^{+}K^{-}\pi^{+}\pi^{-})K^{\pm}$ decays. 

\section{Conclusions}
\label{sec:conclude}
The first amplitude model for $D^{0}\to K^{+}K^{-}\pi^{+}\pi^{-}$ decay derived from flavor-tagged data has been %%@
presented. The data used are from $e^{+}e^{-}$ collisions at center-of-mass energies close to $c\overline{c}$ %%@
threshold and in the region of the $\Upsilon$ resonances. $CP$-tagged quantum-correlated data recorded at the %%@
$\psi(3770)$ resonance are also used in the fit. The model indicates that the quasi-two-body decay %%@
$D^{0}\to\phi\rho^{0}$ is dominant, with significant contributions from the following intermediate states: %%@
$D^{0}\to K_1(1270)^{\pm}K^{\mp}$, $D^{0}\to K^{*}(1410)^\pm K^{\mp}$, and $D^{0}\to\phi\pi^{+}\pi^{-}$. In %%@
addition, there is a significant $D^{0}\to K^{*0}\overline{K^{*0}}$ contribution indicating that final-state %%@
interactions play a significant role in the decay. There is also a non-resonant contribution of around 10\%; the %%@
best fit to data is achieved when there is relative angular momentum among the particles in this contribution.
The accuracy of the model parameters is limited by the uncertainties on the $K_1(1270)^{-}$ resonance parameters.

The amplitude model presented has been used to search for $CP$ violation in the decay by determining the fit %%@
fractions separately for $D^{0}$ or $\overline{D^{0}}$ decays. The fit fractions are found to agree within the %%@
uncertainties, indicating no $CP$ violation in the decay at the level of a few percent. The amplitude model has %%@
also been used in a sensitivity study of $B^{\pm}\to \widetilde{D^{0}}(K^{+}K^{-}\pi^{+}\pi^{-})K^{\pm}$ decays to %%@
the $CP$-violating parameter $\gamma$. The study indicates that $\gamma$ can be determined with a precision of %%@
$(11.3\pm 0.3)^{\circ}$, assuming the baseline model, using this decay at LHCb. A similar precision would be %%@
expected at future flavor facilities.  

\begin{acknowledgments}
We gratefully acknowledge the effort of the CESR staff
in providing us with excellent luminosity and running conditions.
This work was supported by
the National Science Foundation,
the U.S. Department of Energy,
the Natural Sciences and Engineering Research Council of Canada, and
the U.K. Science and Technology Facilities Council.
\end{acknowledgments}

\appendix

\label{app:efficiency}
\section{Spin factor definitions}
\label{app_form}
\newcounter{spinFactorCounter}
\newcommand{\spc}{\refstepcounter{spinFactorCounter} \arabic{spinFactorCounter}}
\begin{table}
\caption{Spin Factors $(\mathcal{S})$ for various amplitudes. In the decay chains, S, P, V and A stand for
  scalar, pseudoscalar, vector, and axial vector, respectively. Letters in square
  brackets refer to whether the decay products are in a relative $S$, $P$, or $D$ wave state of relative orbital %%@
angular momentum. 
  Spin factor number~\ref{SF_DtoV1V2_V1toP0P1_V1toP2P3_D} with \prt{D[\it{D}] \to  V_1 V_2} actually corresponds to %%@
a
  superposition of $D$~and $S$~wave, following the choice of basis
  in Ref.~\cite{MARKIII}.
  If no angular momentum is specified, the lowest angular  momentum state
  compatible with angular momentum conservation and, where
  appropriate, parity conservation, is used.
  \label{tab:spinFactors}}
\begin{tabular}{cll}
\hline\hline
Number     & Decay chain & $\mathcal{S}$  \\ \hline
\spc\label{SF_DtoAP0_AtoVP1_VtoP2P3} & \prt{D \to AP_1, A[\it{S}]\to \rm VP_2, V\to P_3 P_4} &
\( p_1^{\mu}\; P(A)_{\mu\nu}\,  Z_1(V)^{\nu} \) \\
\spc\label{SF_DtoV1V2_V1toP0P1_V1toP2P3_S} & \prt{D[\it{S}] \to  \rm V_1 V_2, V_1\to P_1 P_2, V_2 \to P_3 P_4} &
\( Z_1(V_1)_{\nu}  Z_1(V_2)^{\nu} \)            
\\
\spc\label{SF_DtoV1V2_V1toP0P1_V1toP2P3_P} & \prt{D[\it{P}] \to  \rm V_1 V_2, V_1\to P_1 P_2, V_2 \to P_3 %%@
P_4}\hspace{1cm} &
\(
\epsilon_{\alpha\beta\gamma\delta}\, p_D^{\alpha} q_D^{\beta} q_{V_1}^{\gamma} q_{V_2}^{\delta}
\)                                                            
\\
\spc\label{SF_DtoV1V2_V1toP0P1_V1toP2P3_D} & \prt{D[\it{D}] \to  \rm V_1 V_2, V_1\to P_1 P_2, V_2 \to P_3 P_4} &
\( Z_1(V_1)_{\alpha}\; p_{V_2}^{\alpha}\;\;
   Z_1(V_2)_{\beta} \; p_{V_1}^{\beta} \)                      
\\
\spc\label{SF_DtoVS_VtoP0P1_StoP2P3} & \prt{D \to VS, V\to P_1 P_2, S \to P_3 P_4}
&
\( p_S^{\mu}\; Z_1(V)_{\mu}\)                                  
\\
\spc\label{SF_DtoV1P0_V1toV2P1_V2toP2P3} & \prt{D \to  V_1 P_1, V_1\to V_1 P_2, V_2 \to P_3 P_4} &
\(
\epsilon_{\alpha\beta\gamma\delta}\, p_{V_1}^{\alpha} q_{V_1}^{\beta} p_{P_1}^{\gamma} q_{V_2}^{\delta}
\)                                                           
\\\hline\hline
\end{tabular}
\end{table}

The spin factors used are those calculated
in~\cite{MARKIII}. 
These are expressed in terms of the four-momenta of
the particles involved. For the decay $R \to A,B$ the following
notation is used:
\begin{itemize}
\item $p_R, p_A$, and $p_B$ are the four-momenta of the resonance $R$, 
  its decay products $A$ and, $B$, respectively. This are related such that
  \begin{equation}
    p_R = p_A + p_B \; .
  \end{equation}
\item $q_R$ represents the difference of the four-momenta between the decay
  products of $R$:
\begin{equation}
  q_R \equiv p_A - p_B \; .
\end{equation}
Here the ordering of the particles is important; 
the momentum of the second daughter particle listed in the decay chain is always subtracted 
from the momentum of the first.
\end{itemize}

For the purpose of representing the spin factors in a concise way,
the following functions are defined:

\begin{equation}
\begin{split}
P(p, m^2)^{\mu\alpha} &=  g^{\mu\alpha} - \frac{p^{\mu} p^{\alpha}}{m^2} \;\mathrm{and} \\
Z_1(q, p, m^2)^{\alpha} &= q_{\mu} P(p, m^2)^{\mu\alpha}  \; .\\
\end{split}
\end{equation}
\\
For operators related to the resonance $R$ the expressions are simplified by using the definitions  
\begin{equation}
\begin{split}
   P(R)^{\mu\alpha} &\equiv P(p_R, m_R^2)^{\mu\alpha} \;  \mathrm{and} 
\\ Z_1(R)^{\alpha} &\equiv Z_1(q_R, p_R, m_R^2)^{\alpha} \; .
\end{split}
\end{equation}
For non-resonant contributions $m_R^2$ is replaced with $p^2$, as in $P(p, p^2)^{\mu\alpha}$ for example. The spin %%@
factors used are listed in Table~\ref{tab:spinFactors}.

\begin{table}
\caption[Spin factors used for different decay chains]{Spin factors
  used for different decay chains, including the particle numbering
  scheme. The $2^{\mathrm{nd}}$ column refers to the spin factors as numbered in
  \tabref{tab:spinFactors}, and the particles $P_1$, $P_2$, $P_3$, and $P_4$
  refer to the numbers as defined in \tabref{tab:spinFactors}.
  \label{tab:decayChainsForSpinFactors}}
\begin{tabular}{lcllll} \hline\hline
Decay chain & Spin factor number & $P_1$ & $P_2$ & $P_3$ & $P_4$ \\ \hline
$K_1(1270)^+(K^{*0}\pi^+)K^-$ & \ref{SF_DtoAP0_AtoVP1_VtoP2P3} & $K^-$ & $\pi^+$ & $K^+$ & $\pi^-$ \\ 
$K_1(1270)^-(\overline{K^{*0}}\pi^-)K^+$ & \ref{SF_DtoAP0_AtoVP1_VtoP2P3} & $K^+$ & $\pi^-$ & $K^-$ & $\pi^+$ \\
$K_1(1270)^+(\rho K^+)K^-$ & \ref{SF_DtoAP0_AtoVP1_VtoP2P3} & $K^-$ & $K^+$ & $\pi^+$ & $\pi^-$ \\
$K_1(1270)^-(\rho K^{-})K^{+}$ & \ref{SF_DtoAP0_AtoVP1_VtoP2P3} & $K^+$ & $K^-$ & $\pi^-$ & $\pi^+$ \\
$K^*(1410)^+(K^{*0}\pi^+)K^-$ & \ref{SF_DtoV1P0_V1toV2P1_V2toP2P3} & $K^-$ & $\pi^+$ & $K^+$ & $\pi^-$ \\ 
$K^*(1410)^-(\overline{K^{*0}}\pi^-)K^+$ & \ref{SF_DtoV1P0_V1toV2P1_V2toP2P3} & $K^+$ & $\pi^-$ & $K^-$ & $\pi^+$ %%@
\\
$K^{*0}\overline{K^{*0}}$ $S$ wave & \ref{SF_DtoV1V2_V1toP0P1_V1toP2P3_S} & $K^+$ & $\pi^-$ & $K^-$ & $\pi^+$ \\
$K^{*0}\overline{K^{*0}}$ $P$ wave & \ref{SF_DtoV1V2_V1toP0P1_V1toP2P3_P} & $K^+$ & $\pi^-$ & $K^-$ & $\pi^+$ \\
$K^{*0}\left\{K^{-}\pi^{+}\right\}_P$ $P$ wave & \ref{SF_DtoV1V2_V1toP0P1_V1toP2P3_P} & $K^+$ & $\pi^-$ & $K^-$ & %%@
$\pi^+$ \\
$\overline{K^{*0}}\left\{K^{+}\pi^{-}\right\}$ $P$ wave & \ref{SF_DtoV1V2_V1toP0P1_V1toP2P3_P} & $K^+$ & $\pi^-$ & %%@
$K^-$ & $\pi^+$ \\
$K^{*0}\overline{K^{*0}}$ $D$ wave & \ref{SF_DtoV1V2_V1toP0P1_V1toP2P3_D} & $K^+$ & $\pi^-$ & $K^-$ & $\pi^+$ \\
$\left\{K^{-}\pi^{+}\right\}_{P}\left\{K^{+}\pi^{-}\right\}_{P}$ $D$ wave & \ref{SF_DtoV1V2_V1toP0P1_V1toP2P3_D} & %%@
$K^+$ & $\pi^-$ & $K^-$ & $\pi^+$ \\
$\phi\rho^{0}$ $S$ wave & \ref{SF_DtoV1V2_V1toP0P1_V1toP2P3_S} & $K^+$ & $K^-$ & $\pi^+$ & $\pi^-$ \\ 
$\phi\left\{\pi^{+}\pi^{-}\right\}_{P}$ $S$ wave & \ref{SF_DtoV1V2_V1toP0P1_V1toP2P3_S} & $K^+$ & $K^- $ & $\pi^+ %%@
$& $\pi^-$ \\
$\rho^{0}\left\{K^{+}K^{-}\right\}_{P}$ $S$ wave & \ref{SF_DtoV1V2_V1toP0P1_V1toP2P3_S} & $K^+$ & $K^- $ & $\pi^+ %%@
$& $\pi^-$ \\
$\phi\left\{\pi^{+}\pi^{-}\right\}_{P}$ $P$ wave & \ref{SF_DtoV1V2_V1toP0P1_V1toP2P3_P} & $K^+$ & $K^- $ & $\pi^+ %%@
$& $\pi^-$ \\
$\phi\rho^{0}$ $D$ wave & \ref{SF_DtoV1V2_V1toP0P1_V1toP2P3_D} & $K^+$ & $K^-$ & $\pi^+$ & $\pi^-$ \\
$\phi\left\{\pi^{+}\pi^{-}\right\}_{P}$ $D$ wave & \ref{SF_DtoV1V2_V1toP0P1_V1toP2P3_D} & $K^+$ & $K^- $ & $\pi^+ %%@
$& $\pi^-$ \\
$\left\{\pi^{+}\pi^{-}\right\}_{P}\left\{K^{+}K^{-}\right\}_P$ $D$ wave & \ref{SF_DtoV1V2_V1toP0P1_V1toP2P3_D} & %%@
$K^+$ & $K^- $ & $\pi^+ $& $\pi^-$ \\
%\prt{D0\to \phi(1020)(\to K^+,K^-),\omega(782)(\to \pi^+,\pi^-)} & \ref{SF_DtoV1V2_V1toP0P1_V1toP2P3_S} & %%@
%\prt{K^+, K^-, \pi^+, \pi^-} \\ \hline
$\phi\left\{\pi^+,\pi^-\right\}_S$ & \ref{SF_DtoVS_VtoP0P1_StoP2P3} & $K^+$ & $K^- $ & $\pi^+ $& $\pi^-$ \\
$\left\{K^-\pi^+\right\}_{P}\left\{K^+\pi^-\right\}_{S}$ & \ref{SF_DtoVS_VtoP0P1_StoP2P3} & $K^-$ & $\pi^+$ & $K^+$ %%@
& $ \pi^-$ \\
$\left\{K^+\pi^-\right\}_{P}\left\{K^-\pi^+\right\}_{S}$ & \ref{SF_DtoVS_VtoP0P1_StoP2P3} & $K^+$ & $\pi^-$ & $K^-$ %%@
& $\pi^+$ \\
 \hline\hline
\end{tabular}
\end{table}

It is clear that the exact matching of the particles $P_1$, $P_2$, $P_3$, and 
$P_4$ in the spin factor definition to the final state particles in the
decay is important, as many spin factors change sign under swapping a
pair of particles due to terms such as $q_R = p_1 - p_2$. For the
amplitudes used in this paper, the particle ordering is given in
Table \ref{tab:decayChainsForSpinFactors}.

\section{Amplitudes tested}
\label{app:amptested}
Below is a list of all the different amplitudes that are tested when determining the best model. For final states %%@
that are  flavor specific the charge-conjugate amplitude is not listed but is one of the amplitudes that is tested.
\begin{itemize}
\item{\bf Cascade amplitudes containing a higher $K^*$ resonance}
\begin{itemize}
 \item{$K_1(1270)^{+}(K^{*0}\pi^{+})K^{-}$, $K_1(1270)^{+}(K_{0}^{*}(1430)\pi^{+})K^{-}$, %%@
$K_1(1270)^{+}(\rho^{0}K^{+})K^{-}$, and $K_1(1270)^{+}(\omega K^{+})K^{-}$}
 \item{$K_1(1400)^{+}(K^{*0}\pi^{+})K^{-}$}
 \item{$K_2(1430)^{+}(K^{*0}\pi^{+})K^{-}$ and $K_2(1430)^{+}(\rho^{0}K^{+})K^{-}$}
 \item{$K^*(1680)^{+}(K^{*0}\pi^{+})K^{-}$ and $K^*(1680)^{+}(\rho^{0}K^{+})K^{-}$}
\end{itemize}
\item{\bf Quasi-two-body amplitudes}
\begin{itemize}
\item{$K^{*0}\overline{K^{*0}}$ $S$, $P$, and $D$ wave}
\item{$\phi\rho^{0}$ $S$, $P$, and $D$ wave}
\item{$\phi\omega$ $S$ wave}
\item{$\phi f_2(1270)^{0}$ $P$ and $D$ wave}
\end{itemize}
\item{\bf Single resonance amplitudes}
\begin{itemize}
\item{$\rho \{K^+K^{-}\}_{S}$; $\rho\{K^+K^{-}\}_{P}$ $S$, $P$, and $D$ wave; and $\rho\{K^+K^{-}\}_{D}$ $P$ and %%@
$D$ wave}
\item{$K^{*0}\{K^-\pi^{+}\}_{S}$; $K^{*0}\{K^-\pi^{+}\}_{P}$ $S$, $P$, and $D$ wave; and $K^{*0}\{K^-\pi^{+}\}_{D}$ %%@
$P$ and $D$ wave}
\item{$\phi\{\pi^{+}\pi^{-}\}_{S}$; $\phi\{\pi^{+}\pi^{-}\}_{P}$ $S$, $P$, and $D$ wave; and %%@
$\phi\{\pi^{+}\pi^{-}\}_{D}$ $P$ and $D$ wave}
\item{$f_0(980)^{0}\{\pi^{+}\pi^{-}\}_{S}$ and $f_0(980)^{0}\{K^{+}K^{-}\}_{S}$}
\item{$f_2(1270)^{0}\{K^{+}K^{-}\}_{S}$}
\item{$\omega \{K^{+}K^{-}\}_{S}$}
\end{itemize}
\item{\bf Non-resonant amplitudes}
\begin{itemize}
 \item{$\{K^{+}K^{-}\}_{S}\{\pi^{+}\pi^{-}\}_{S}$; $\{K^{+}K^{-}\}_{S}\{\pi^{+}\pi^{-}\}_P$; %%@
$\{K^{+}K^{-}\}_{P}\{\pi^{+}\pi^{-}\}_{S}$; $\{K^{+}K^{-}\}_{P}\{\pi^{+}\pi^{-}\}_P$ $S$, $P$, and $D$ wave; %%@
$\{K^{+}K^{-}\}_{S}\{\pi^{+}\pi^{-}\}_D$; $\{K^{+}K^{-}\}_{D}\{\pi^{+}\pi^{-}\}_{S}$; %%@
$\{K^{+}K^{-}\}_{P}\{\pi^{+}\pi^{-}\}_D$ $P$ and $D$ wave; $\{K^{+}K^{-}\}_{D}\{\pi^{+}\pi^{-}\}_P$ $P$ and $D$ %%@
wave}
 \item{$\{K^{+}\pi^{-}\}_{S}\{K^{-}\pi^{+}\}_P$; $\{K^{+}\pi^{-}\}_{P}\{K^{-}\pi^{+}\}_{S}$; %%@
$\{K^{+}\pi^{-}\}_{P}\{K^{-}\pi^{+}\}_P$ $S$, $P$, and $D$ wave; $\{K^{+}\pi^{-}\}_{S}\{K^{-}\pi^{+}\}_D$; %%@
$\{K^{+}\pi^{-}\}_{D}\{K^{-}\pi^{+}\}_{S}$; $\{K^{+}\pi^{-}\}_{P}\{K^{-}\pi^{+}\}_D$ $P$ and $D$ wave; %%@
$\{K^{+}\pi^{-}\}_{D}\{K^{-}\pi^{+}\}_P$ $P$ and $D$ wave}
\end{itemize}
\end{itemize} 

\section{Alternative models}
\label{app:bestfits}
The fit fractions and $\chi^2 /\nu$ for the best seven fits to the $D^0 \to K^+K^-\pi^+\pi^-$ (models 1 to 7) data %%@
are shown in Table~\ref{best8Tab}.  Model 1 is chosen as baseline. Models 8 and 9 contain different variations of %%@
the non-resonant amplitude component. 
 
\begin{table}
%\begin{sideways}
%\begin{minipage}[b]{\textheight}
\begin{center}
\scriptsize
\caption{ Fit fractions (in \%) and $\chi^2/\nu$ for  alternative models in a combined fit to all flavor and %%@
$CP$-tagged data sets. The uncertainties are statistical. Where it is necessary to specify the angular momentum %%@
state of pairs of non-resonant particles, this information is given in the subscript. The baseline model adopted in %%@
this paper is number 1.} \label{best8Tab} \vspace*{0.2cm}
%\arraystretch{0.75}
\begin{tabular}{lccccccccc}
\hline\hline
Model   & \textbf{$1$}  & \textbf{$2$}  & \textbf{$3$}  & \textbf{$4$}  & \textbf{$5$}  & \textbf{$6$}  & %%@
\textbf{$7$} &  \textbf{$8$} & \textbf{$9$}\\

\hline
$K_{1}(1270)^{+}( K^{*0}\pi^{+})K^{-}$ 
&   7.3$\pm$0.8        & 7.0$\pm$0.8     &7.0$\pm$0.8       &7.0$\pm$0.8     &7.0$\pm$0.8       &8.7$\pm$0.9        %%@
&6.7$\pm$0.8 & $8.4\pm 1.0$ & $8.7\pm 0.9$ \\

$K_{1}(1270)^{-}( \overline{K^{*0}}\pi^{-})K^{+}$ 
       &0.9$\pm$0.3        &0.7$\pm$0.3&0.8$\pm$0.3&0.8$\pm$0.3&0.9$\pm$0.3&1.7$\pm$0.5&0.8$\pm$0.3 & $1.7\pm 0.5$ %%@
& $0.8\pm 0.4$\\

$K_{1}(1270)^{+}(\rho^{0}K^{+})K^{-}$
   &4.7$\pm$0.7&4.7$\pm$0.7&4.8$\pm$0.7&4.6$\pm$0.7&6.0$\pm$0.8&5.1$\pm$0.7&6.8$\pm$1.1 & $6.0\pm 0.9$ & $4.5\pm %%@
0.8$ \\

$K_{1}(1270)^{-}(\rho^{0}K^{-})K^{+}$ 
&6.0$\pm$0.8&5.8$\pm$0.7&5.9$\pm$0.7&6.0$\pm$0.8&3.7$\pm$0.8&6.3$\pm$0.8&3.5$\pm$0.8 & $7.9\pm 1.3$ & $5.9\pm %%@
0.9$\\

$ K^{*}(1410)^{+} ( K^{*0} \pi^{+})K^{-}$
&4.2$\pm$0.7&4.2$\pm$0.7&3.4$\pm$0.7&4.2$\pm$0.7&4.2$\pm$0.7&---&4.1$\pm$0.7 & $3.2\pm 0.7$ & $3.2\pm 0.7$\\

$ K^{*}(1410)^{-} (  \overline{K^{*0}}  \pi^{-}) K^{+}$
&4.7$\pm$0.6&4.5$\pm$0.6&2.9$\pm$0.6&4.5$\pm$0.6&4.7$\pm$0.6&---&4.5$\pm$0.6 & $5.1\pm 0.8$ & $4.7\pm 0.7$\\

$K^{*0}\left\{K^{-}\pi^{+}\right\}_P$ P wave&---&---&---&---&---&3.9$\pm$0.7&--- & --- & ---\\

$\overline{K^{*0}}\left\{K^{+} \pi^{-}\right\}_P$  P wave&---&---&---&---&---&3.0$\pm$0.5&--- & --- & --- \\

$K^{*0} \overline{K^{*0}}$  S wave  
&6.1$\pm$0.8&6.1$\pm$0.8&6.1$\pm$0.8&6.1$\pm$0.8&6.3$\pm$0.8&4.2$\pm$0.6&6.3$\pm$0.8 & $3.6\pm 0.8$ & $4.4\pm %%@
0.8$\\

$K^{*0}\overline{K^{*0}}$ P wave&---&---&0.9$\pm$0.3&---&---&---&---&---&---\\

$\phi\left\{\pi^{+} \pi^{-}\right\}_{S}$
&10.3$\pm$1.0&9.0$\pm$0.9 &9.1$\pm$0.9&9.6$\pm$1.0&10.1$\pm$0.9&---&8.9$\pm$0.9 & $7.3\pm 0.9$ & $9.1\pm 0.09$\\

$\phi\left\{\pi^{+} \pi^{-}\right\}_P$  S wave 
&---&6.3$\pm$1.1&6.1$\pm$1.1&11.4$\pm$2.5&---&---&5.9$\pm$1.1 & --- & --- \\

$\phi\left\{\pi^{+} \pi^{-}\right\}_P$  P wave &---&---&---&---&---&4.7$\pm$0.6&--- & --- & ---\\

$\phi\left\{\pi^{+} \pi^{-}\right\}_P$  D wave &---&---&---&2.0$\pm$1.1&---&---&--- & ---& ---\\

$\phi\rho^{0}$ S wave 
&38.3$\pm$2.2&21.0$\pm$2.0&21.2$\pm$1.9&15.9$\pm$2.6&38.0$\pm$2.0&26.8$\pm$1.3&20.9$\pm$1.9 & $34.6\pm 2.1$ & %%@
$36.7\pm 0.2$ \\

$\phi\rho^{0}$ D wave
&3.4$\pm$0.7&3.7$\pm$0.7&3.7$\pm$0.7&1.1$\pm$0.6&3.5$\pm$0.7&---&3.8$\pm$0.7 & $2.9\pm 0.7$ & $2.8\pm 0.7$ \\

$\rho^{0}\left\{K^{+}K^{-}\right\}_P$  S wave &---&---&---&---&1.2$\pm$0.6&---&1.2$\pm$0.6 & --- & --- \\

 $\left\{K^-\pi^+\right\}_P \left\{K^+ \pi^-\right\}_S$
&10.9$\pm$1.2&10.9$\pm$1.1&11.0$\pm$1.1&10.8$\pm$1.1&9.3$\pm$1.1&---&9.1$\pm$1.1 & --- & $9.0\pm 1.2$\\

 $\left\{K^+\pi^{-}\right\}_P\left\{ K^{-}\pi^{+}\right\}_S$ &---&---&---&---&---&---&--- & $11.6\pm 1.5$ & $5.0\pm %%@
1.0$\\ 

 $\left\{K^- \pi^+\right\}_P \left\{K^+ \pi^-\right\}_P$  D wave&---&---&---&---&---&13.8$\pm$1.2&---&---&---\\
 $\left\{\pi^+\pi^-\right\}_P \left\{K^+K^-\right\}_P$  D wave&---&---&---&---&---&2.6$\pm$0.7&---&---&--- \\

\hline

 Sum & 96.7$\pm$2.6 & 84.0$\pm$2.0 & 82.9$\pm$2.0 & 84.0$\pm$2.0 & 94.9$\pm$2.3
& 80.9$\pm$1.4& 82.5$\pm$2.0 & $92.4\pm 2.4$ & $94.9\pm 2.4$\\
\hline

 $\chi^2/\nu$ & 1.63 & 1.64 & 1.65 & 1.65& 1.66 & 1.66 & 1.66 & 1.77 & 1.78\\

\hline\hline

\end{tabular}
\end{center}
%\end{minipage}
%\end{sideways}

\end{table}

\section{Accounting for quantum correlations}

In the analysis the CLEO-c flavor-tagged data are
assumed, apart from the proportion of events that were mistagged, to form a pure flavor sample. However, since the %%@
$D^{0}\overline{D^0}$ pair results from the decay of the $J^{PC} = 1^{--}$  $\psi(3770)$ particle, the $D$ mesons %%@
are produced in a correlated state. For example, when both the $D^0$ and $\overline{D^0}$ decay to %%@
$CP$-eigenstates, these states will have opposite $CP$. The antisymmetric
wave function which describes the decay is
\begin{equation}
 \frac{1}{\sqrt{2}}\left[ %%@
\mathcal{A}_{D^{0}}(\mathbf{s})\overline{\mathcal{A}}_j-\mathcal{A}_j\mathcal{A}_{\overline{D^{0}}}
(\mathbf{s})\right] \;.
\label{eq:Aij}
\end{equation}
Here, $\mathcal{A}_j$ $(\overline{\mathcal{A}}_j)$ is the amplitude for the decay of the other $D^{0}$ %%@
$(\overline{D^{0}})$ in the decay.  In this paper the CLEO-c flavor tagging is provided by tagging the non-signal %%@
$D$ in its inclusive decay to kaons under the assumption of a Cabibbo-favored decay. For example
 $D^{0} \rightarrow K^{-}X$ and  $\overline{D}^{0} \rightarrow K^{+}X$. Therefore it is useful to consider the case %%@
where state $j$ of Eq. (\ref{eq:Aij}) is
defined to be a specific tag, labeled by the subscript $t$, so that
 \begin{eqnarray}
A_{j} &=& A(D^{0} \rightarrow  K^{-} X_{t}) = K_{t} \; \mathrm{and} \\ 
\label{eq:Ajdef}
\overline{A}_{j} &=& A(\overline{D}^{0} \rightarrow K^{-} X_{t}) = K_{t} k_{t} e^{i\delta_{t}} \; ,
\label{eq:Abarjdef}
\end{eqnarray}  
where  $K_{t}$,  $k_{t}$, and $\delta_{t}$ are real numbers.
Therefore, from Eq.~(\ref{eq:Aij}) the full decay rate for the decay of interest against the specific tag defined %%@
by Eq.~(\ref{eq:Ajdef}), will be given by
\begin{eqnarray}
\advance \leftskip-3.4cm
\Gamma_{t} &\propto& K^{2}_{t} ( |\mathcal{A}_{D^{0}}|^{2} +k^{2}_{t} |\mathcal{A}_{\overline{D^{0}}}|^{2} %%@
\nonumber \\ 
&-& 2k_{t} [\mathrm{Re}[\mathcal{A}_{D^{0}}\mathcal{A}_{\overline{D^{0}}}] \cos{\delta_{t}} + %%@
\mathrm{Im}[\mathcal{A}_{D^{0}}\mathcal{A}_{\overline{D^{0}}}] \sin{\delta_{t}}] ) \; .
\label{QCformt}
\end{eqnarray}
summing over all possible tags $t$, which will all in general have different values for $k_{t}$, $K_{t}$, and %%@
$\delta_{t}$. Hence, the full decay rate will be given by
\begin{eqnarray}
\Gamma &\propto&  ( \Sigma_{t} K^{2}_{t} ) [ |\mathcal{A}_{D^{0}}|^{2} + \left(\frac{\Sigma_{t}K^{2}_{t} %%@
k^{2}_{t}}{\Sigma_{t}K^{2}_{t}}\right) |\mathcal{A}_{\overline{D^{0}}}|^2 \nonumber \\
&-& 2 (\mathrm{Re}[\mathcal{A}_{D^{0}}\mathcal{A}_{\overline{D^{0}}}]
\left(\frac{\Sigma_{t}K^{2}_{t}k_{t} \cos{\delta_{t}}}{\Sigma_{t}K^{2}_{t}}\right) + %%@
\mathrm{Im}[\mathcal{A}_{D^{0}}\mathcal{A}_{\overline{D^{0}}}] \left(\frac{\Sigma_{t}K^{2}_{t}k_{t}
\sin{\delta_{t}}}{\Sigma_{t}K^{2}_{t}}\right) ) ] \; . 
\label{sumtQC}
\end{eqnarray}
Defining
\begin{eqnarray}
\langle k^{2}\rangle  &\equiv& \left(\frac{\Sigma_{t}K^{2}_{t} k^{2}_{t}}{\Sigma_{t}K^{2}_{t}}\right) \; ,\\
\langle k\cos{\delta}\rangle  &\equiv & %%@
\left(\frac{\Sigma_{t}K^{2}_{t}k_{t}\cos{\delta_{t}}}{\Sigma_{t}K^{2}_{t}}\right) \; , \; \mathrm{and} \;\\
\langle k\sin{\delta}\rangle  &\equiv& %%@
\left(\frac{\Sigma_{t}K^{2}_{t}k_{t}\sin{\delta_{t}}}{\Sigma_{t}K^{2}_{t}}\right) \;, 
\label{qcKdefs}
\end{eqnarray}
Eq. (\ref{sumtQC}) can be rewritten as
\begin{eqnarray}
 \Gamma &\propto& ( \Sigma_{t} K^{2}_{t} ) [|\mathcal{A}_{D^{0}}|^{2} + |\mathcal{A}_{\overline{D^{0}}}|^{2} %%@
\langle k^{2}\rangle   \\ \nonumber 
 &-& 2\left( \mathrm{Re}[\mathcal{A}_{D^{0}}\mathcal{A}_{\overline{D^{0}}}] \langle k\cos{\delta}\rangle  +  %%@
\mathrm{Im}[\mathcal{A}_{D^{0}}\mathcal{A}_{\overline{D^{0}}}] \langle k\sin{\delta}\rangle  \right) ] .
\label{QCform}
\end{eqnarray}
Since the tags in question are dominated by Cabibbo-favored and doubly-Cabibbo-suppressed decays it is noted that %%@
$k_{t} \approx 0.05$. Therefore, the term
$|\mathcal{A}_{\overline{D^{0}}}|^{2} \langle k^{2}\rangle $ can be neglected. It is also assumed that the %%@
parameters relating to the tags - $( \Sigma_{t} K^{2}_{t} )$
$\langle k\cos{\delta}\rangle $, and $\langle k\sin{\delta}\rangle $ - will not vary over the %%@
$K^{+}K^{-}\pi^{+}\pi^{-}$ Dalitz space; this assumption allows the term $( \Sigma_{t} K^{2}_{t} )$ to be absorbed %%@
into the
normalization, leaving only the parameters $\langle k\cos{\delta}\rangle $ and $\langle k\sin{\delta}\rangle $ to %%@
be determined from the data.

A fit to CLEO-c $3770$ flavor-tagged  data is performed with the distribution described by Eq. (D9), with $\langle %%@
k\cos{\delta}\rangle $ and $\langle k\sin{\delta}\rangle $ as additional free parameters. The results of this fit %%@
are $\langle k\cos{\delta}\rangle =0.061\pm 0.042$ and $\langle k\sin{\delta}\rangle =0.029\pm 0.007$. The %%@
difference in the values of $\mathrm{Re}(a_i)$ and $\mathrm{Im}(a_i)$ from the quantum-correlated fit compared to %%@
the nominal fit are used to estimate the systematic uncertainty.

Quantum correlations are also present for data produced with a center-of-mass energy of 4170 MeV. However the %%@
$D^{0}$ and $\overline{D}^{0}$ particles are generally not produced directly at this
 energy but via the decay of higher mass resonances; therefore, the interference effects are expected to not be the %%@
same as for the CLEO-c 3770 data. The measured values of $\langle k\cos{\delta}\rangle $ and $\langle %%@
k\sin{\delta}\rangle $ are $0.051\pm 0.032$ and $0.037\pm 0.011$, respectively.

The systematic shifts due to quantum correlations at the $\psi(3770)$ and $\psi(4170)$ center-of-mass energies are %%@
added in quadrature in order to give the total systematic shift due to these effects.

\label{app:backplots}
\label{app:results}

\section{EPAPS information for the paper ``Amplitude Analysis of $D^{0}\to K^{+}K^{-}\pi^{+}\pi^{-}$"}
\subsection{Additional information about the baseline fit}
Table~\ref{tab:numbering} gives a key to the amplitude and parameter indexing used in this addendum. 
Table~\ref{tab:statcorr} gives the statistical correlation matrix among the fit parameters for the best-fit model %%@
(model 1) described in the main body of the text. Table~\ref{tab:interference} gives the interference fractions for %%@
best-fit model. The interference fraction  between the $m^{\mathrm{th}}$ and $n^{\mathrm{th}}$ amplitudes, %%@
$\mathcal{I}_{m,n}$, is defined as:
\begin{equation}
 \mathcal{I}_{m,n} = \frac{\int 2\mathrm{Re}(\mathcal{A}_m\mathcal{A}_{n}^{*})dp_j}{\int |\mathcal{A}_{D^{0}}|^2 %%@
dp_j} \; .
\end{equation} 
The interference fractions are listed in order of their absolute magnitude and only terms with a magnitude greater %%@
than 0.5\% are listed.

\begin{table}[h]
\begin{center}
\caption{Amplitude and parameter indexing used in this addendum.} \label{tab:numbering}
\begin{tabular}{lc}
\hline\hline
 Amplitude & Index   \\ \hline
$K_{1}(1270)^{+}(K^{*0}\pi^{+}) K^{-}$ & 1  \\
$K_{1}(1270)^{-}(\overline{K^{*0}}\pi^{-})K^{+}$ & 2 \\
$K_{1}(1270)^{+}(\rho^{0} K^{+})K^{-}$ &  3 \\
$K_{1}(1270)^{-}(\rho^{0} K^{-})K^{+}$ & 4 \\
$ K^{*}(1410)^{+} (K^{*0} \pi^{+})K^{-}$ & 5 \\
$ K^{*}(1410)^{-} (\overline{K^{*0}}\pi^{-}) K^{+}$ & 6 \\
$K^{*0} \overline{K^{*0}}$ $S$ wave & 7 \\
$\phi\rho^{0}$ $S$ wave & 8 \\
$\phi\rho^{0}$ $D$ wave & 9 \\
$\phi \left\{\pi^{+}\pi^{-}\right\}_S$ & 10 \\
$\left\{K^{-}\pi^{+}\right\}_P\left\{K^{+}\pi^{-}\right\}_S$ & 11  \\
\hline\hline
\end{tabular}
\end{center}
\end{table}

\begin{table}[p]
\caption{Statistical correlation matrix among parameters for the baseline fit. The amplitude numbering is given in %%@
the text.}\label{tab:statcorr}
\begin{center}
\rotatebox{90}{\mbox{ \scriptsize
\begin{tabular}{lccccccccccccccccccc} \hline\hline
&   $\mathrm{Im}(a_{2})$  
& $\mathrm{Re}(a_{3})$ & $\mathrm{Im}(a_{3})$ & 
$\mathrm{Re}(a_{4})$ & $\mathrm{Im}(a_{4})$ &
$\mathrm{Re}(a_{5})$ & $\mathrm{Im}(a_{5})$ &
$\mathrm{Re}(a_{6})$ & $\mathrm{Im}(a_{6})$ &
$\mathrm{Re}(a_{7})$ & $\mathrm{Im}(a_{7})$ &
$\mathrm{Re}(a_{8})$ & $\mathrm{Im}(a_{8})$ &
$\mathrm{Re}(a_{9})$ & $\mathrm{Im}(a_{9})$ &
$\mathrm{Re}(a_{10})$ & $\mathrm{Im}(a_{10})$ &
$\mathrm{Re}(a_{11})$ & $\mathrm{Im}(a_{11})$ \\ \hline
$\mathrm{Re}(a_{2})$ & $\phantom{-}0.164$ & $-0.225$ & $\phantom{-}0.361$ & $\phantom{-}0.022$ & $\phantom{-}0.089$ %%@
& $\phantom{-}0.155$ & $\phantom{-}0.214$ & $\phantom{-}0.455$ & $\phantom{-}0.048$ & $-0.212$ & $\phantom{-}0.408$ %%@
& $-0.347$ & $-0.120$ & $\phantom{-}0.236$ & $\phantom{-}0.042$ & $\phantom{-}0.297$ & $-0.246$ & %%@
$\phantom{-}0.002$ & $\phantom{-}0.409$ \\
$\mathrm{Im}(a_{2})$ & $\phantom{-}1.000$ &$-0.223$ & $\phantom{-}0.070$ &$-0.218$ & $\phantom{-}0.092$ & %%@
$\phantom{-}0.019$ & $\phantom{-}0.280$ & $\phantom{-}0.196$ & $\phantom{-}0.176$ &$-0.334$ & $\phantom{-}0.080$ %%@
&$-0.036$ &$-0.177$ & $\phantom{-}0.038$ & $\phantom{-}0.175$ & $\phantom{-}0.043$ & $\phantom{-}0.188$ &$-0.175$ & %%@
$\phantom{-}0.188$ \\
$\mathrm{Re}(a_{3})$ &  --- &$\phantom{-}1.000$ &$-0.409$ &$-0.057$ &$-0.139$ &$-0.016$ &$-0.475$ &$-0.339$ %%@
&$-0.367$ & $\phantom{-}0.380$ &$-0.362$ & $\phantom{-}0.359$ & $\phantom{-}0.458$ &$-0.277$ &$-0.272$ &$-0.441$ %%@
&$-0.080$ & $\phantom{-}0.423$ &$-0.460$ \\
$\mathrm{Im}(a_{3})$ & --- &--- &$\phantom{-}1.000$ & $\phantom{-}0.269$ & $\phantom{-}0.076$ & $\phantom{-}0.318$ %%@
& $\phantom{-}0.167$ & $\phantom{-}0.542$ & $\phantom{-}0.008$ & $\phantom{-}0.139$ & $\phantom{-}0.520$ &$-0.335$ %%@
& $\phantom{-}0.078$ & $\phantom{-}0.252$ &$-0.164$ & $\phantom{-}0.308$ &$-0.420$ & $\phantom{-}0.044$ & %%@
$\phantom{-}0.534$ \\
$\mathrm{Re}(a_{4})$ & --- &--- &--- &$\phantom{-}1.000$ &$-0.261$ & $\phantom{-}0.204$ &$-0.260$ & %%@
$\phantom{-}0.113$ &$-0.336$ & $\phantom{-}0.444$ & $\phantom{-}0.079$ &$-0.178$ & $\phantom{-}0.203$ & %%@
$\phantom{-}0.094$ &$-0.252$ & $\phantom{-}0.085$ &$-0.474$ & $\phantom{-}0.272$ & $\phantom{-}0.043$ \\
$\mathrm{Im}(a_{4})$ & --- &--- &--- &--- &$\phantom{-}1.000$ & $\phantom{-}0.220$ & $\phantom{-}0.188$ & %%@
$\phantom{-}0.290$ & $\phantom{-}0.099$ &$-0.061$ & $\phantom{-}0.398$ & $\phantom{-}0.072$ & $\phantom{-}0.180$ %%@
&$-0.120$ &$-0.015$ &$-0.064$ &$-0.011$ & $\phantom{-}0.206$ & $\phantom{-}0.284$ \\
$\mathrm{Re}(a_{5})$ &  --- &--- &--- &--- &--- &$\phantom{-}1.000$ & $\phantom{-}0.024$ & $\phantom{-}0.215$ %%@
&$-0.039$ & $\phantom{-}0.189$ & $\phantom{-}0.304$ &$-0.107$ & $\phantom{-}0.217$ & $\phantom{-}0.025$ &$-0.191$ & %%@
$\phantom{-}0.059$ &$-0.307$ & $\phantom{-}0.217$ & $\phantom{-}0.243$ \\
$\mathrm{Im}(a_{5})$ & --- &--- &--- &--- &--- &--- &$\phantom{-}1.000$ & $\phantom{-}0.368$ & $\phantom{-}0.353$ %%@
&$-0.472$ & $\phantom{-}0.401$ &$-0.367$ &$-0.526$ & $\phantom{-}0.318$ & $\phantom{-}0.351$ & $\phantom{-}0.423$ & %%@
$\phantom{-}0.106$ &$-0.368$ & $\phantom{-}0.419$ \\
$\mathrm{Re}(a_{6})$ & --- &--- &--- &--- &--- &--- &--- &$\phantom{-}1.000$ & $\phantom{-}0.199$ &$-0.101$ & %%@
$\phantom{-}0.653$ &$-0.556$ &$-0.194$ & $\phantom{-}0.443$ & $\phantom{-}0.084$ & $\phantom{-}0.438$ &$-0.383$ %%@
&$-0.001$ & $\phantom{-}0.672$ \\
$\mathrm{Im}(a_{6})$ &--- &--- &--- &--- &--- &--- &--- &--- &$\phantom{-}1.000$ &$-0.478$ & $\phantom{-}0.191$ %%@
&$-0.218$ &$-0.493$ & $\phantom{-}0.216$ & $\phantom{-}0.340$ & $\phantom{-}0.287$ & $\phantom{-}0.252$ &$-0.389$ & %%@
$\phantom{-}0.264$ \\
$\mathrm{Re}(a_{7})$ &  --- &--- &--- &--- &--- &--- &--- &--- &--- &$\phantom{-}1.000$ &$-0.094$ & %%@
$\phantom{-}0.159$ & $\phantom{-}0.619$ &$-0.157$ &$-0.465$ &$-0.277$ &$-0.408$ & $\phantom{-}0.517$ &$-0.173$ \\
$\mathrm{Im}(a_{7})$ & --- &--- &--- &--- &--- &--- &--- &--- &--- &--- &$\phantom{-}1.000$ &$-0.505$ &$-0.182$ & %%@
$\phantom{-}0.382$ & $\phantom{-}0.064$ & $\phantom{-}0.435$ &$-0.328$ &$-0.041$ & $\phantom{-}0.703$ \\
$\mathrm{Re}(a_{8})$ &  --- &--- &--- &--- &--- &--- &--- &--- &--- &--- &--- &$\phantom{-}1.000$ & %%@
$\phantom{-}0.664$ &$-0.848$ &$-0.349$ &$-0.795$ & $\phantom{-}0.401$ & $\phantom{-}0.235$ &$-0.691$ \\
$\mathrm{Im}(a_{8})$ & --- &--- &--- &--- &--- &--- &--- &--- &--- &--- &--- &--- &$\phantom{-}1.000$ &$-0.597$ %%@
&$-0.699$ &$-0.647$ &$-0.150$ & $\phantom{-}0.630$ &$-0.374$ \\
$\mathrm{Re}(a_{9})$ &  --- &--- &--- &--- &--- &--- &--- &--- &--- &--- &--- &--- &--- &$\phantom{-}1.000$ & %%@
$\phantom{-}0.406$ & $\phantom{-}0.554$ &$-0.321$ &$-0.226$ & $\phantom{-}0.540$ \\
$\mathrm{Im}(a_{9})$ & --- &--- &--- &--- &--- &--- &--- &--- &--- &--- &--- &--- &--- &--- &$\phantom{-}1.000$ & %%@
$\phantom{-}0.224$ & $\phantom{-}0.302$ &$-0.381$ & $\phantom{-}0.144$ \\
$\mathrm{Re}(a_{10})$ &  --- &--- &--- &--- &--- &--- &--- &--- &--- &--- &--- &--- &--- &--- &--- %%@
&$\phantom{-}1.000$ &$-0.313$ &$-0.356$ & $\phantom{-}0.637$ \\
$\mathrm{Im}(a_{10})$ & --- &--- &--- &--- &--- &--- &--- &--- &--- &--- &--- &--- &--- &--- &--- &--- %%@
&$\phantom{-}1.000$ &$-0.335$ &$-0.389$ \\
$\mathrm{Re}(a_{11})$ &  --- &--- &--- &--- &--- &--- &--- &--- &--- &--- &--- &--- &--- &--- &--- &--- &--- %%@
&$\phantom{-}1.000$ &$-0.178$ \\ \hline\hline
\end{tabular}
}}
\end{center}
\end{table}

\begin{table}[h]
 \begin{center}
 \caption{Interference fractions for the best-fit model in \%. Only interference fractions greater than 0.5\% are 
 given.}\label{tab:interference}
  \begin{tabular}{lr} \hline\hline
  $\mathcal{I}_{i,j}$ \hspace{1.0cm} & Value (\%) \\ \hline
$\mathcal{I}_{8,9}$ & $-15.7\phantom{00}$  \\
$\mathcal{I}_{4,8}$ & $5.7\phantom{00}$ \\
$\mathcal{I}_{1,7}$ & $3.3\phantom{00}$ \\
$\mathcal{I}_{3,8}$ & $3.0\phantom{00}$ \\
$\mathcal{I}_{3,4}$ & $2.8\phantom{00}$ \\ 
$\mathcal{I}_{10,11}$ & $-2.6\phantom{00}$ \\
$\mathcal{I}_{1,3}$ & $1.5\phantom{00}$ \\
$\mathcal{I}_{3,7}$ & $1.3\phantom{00}$ \\
$\mathcal{I}_{7,8}$ & $1.1\phantom{00}$ \\
$\mathcal{I}_{5,6}$ & $1.0\phantom{00}$ \\
$\mathcal{I}_{1,8}$ & $1.0\phantom{00}$ \\
$\mathcal{I}_{4,9}$ & $-0.9\phantom{00}$ \\
$\mathcal{I}_{4,7}$ & $0.8\phantom{00}$ \\
$\mathcal{I}_{3,9}$ & $-0.6\phantom{00}$ \\ \hline\hline
 \end{tabular}
 \end{center}
\end{table}

\subsection{Fit parameters for alternative models}
Tables~\ref{tab:altmodpar1} and \ref{tab:altmodpar2} contain the real and imaginary parts of $a_i$ fitted for %%@
alternative models 2-5 and 6-9, respectively.   

\begin{table}
\begin{center}
\caption{$a_i$ for alternative models 2-5. Uncertainties are statistical only.} \label{tab:altmodpar1} %%@
\vspace*{0.2cm}
\rotatebox{90}{\mbox{\footnotesize
\begin{tabular}{lcccccccc}
\hline\hline
Model   & \multicolumn{2}{c}{2}  
        & \multicolumn{2}{c}{3}
        & \multicolumn{2}{c}{4}
	    & \multicolumn{2}{c}{5} \\			
		& $\mathrm{Re}(a_i)$ & $\mathrm{Im}(a_i)$
		& $\mathrm{Re}(a_i)$ & $\mathrm{Im}(a_i)$
	    & $\mathrm{Re}(a_i)$ & $\mathrm{Im}(a_i)$
		& $\mathrm{Re}(a_i)$ & $\mathrm{Im}(a_i)$ \\ \hline
$K_{1}(1270)^{+}( K^{*0}\pi^{+})K^{-}$ &
1.0 & 0.0 & 
1.0 & 0.0 & 
1.0 & 0.0 & 
1.0 & 0.0 \\
$K_{1}(1270)^{-}( \overline{K^{*0}}\pi^{-})K^{+}$ &  
$\phantom{-}0.42\pm 0.10$ & $-0.18\pm 0.10$ &
$\phantom{-}0.43\pm 0.09$ & $-0.18\pm 0.10$ &
$\phantom{-}0.42\pm 0.09$ & $-0.20\pm 0.10$ &
$\phantom{-}0.48\pm 0.10$ & $-0.18\pm 0.10$ \\
$K_{1}(1270)^{+}(\rho^{0}K^{+})K^{-}$ & 
$\phantom{-}1.39\pm 1.33$ & $\phantom{-}8.33\pm 0.95$ &
$\phantom{-}0.86\pm 1.37$ & $\phantom{-}8.50\pm 0.95$ &
$\phantom{-}1.49\pm 1.20$ & $\phantom{-}8.24\pm 0.97$ &
$\phantom{-}1.98\pm 1.39$ & $\phantom{-}9.44\pm 0.91$ \\
$K_{1}(1270)^{-}(\rho^{0}K^{-})K^{+}$ &
$\phantom{-}6.28\pm 1.29$ & $\phantom{-}7.58\pm 1.10$ &
$\phantom{-}6.32\pm 1.30$ & $\phantom{-}7.64\pm 1.11$ &
$\phantom{-}6.48\pm 1.37$ & $\phantom{-}7.68\pm 1.24$ &
$\phantom{-}4.78\pm 1.27$ & $\phantom{-}6.19\pm 1.13$ \\
$ K^{*}(1410)^{+} ( K^{*0} \pi^{+})K^{-}$ &
$\phantom{-}9.05\pm 0.93$ & $\phantom{-}0.07\pm 1.11$ &
$\phantom{-}7.81\pm 1.00$ & $\phantom{-}2.61\pm 1.11$ &
$\phantom{-}9.01\pm 0.90$ & $-0.08\pm 1.07$ &
$\phantom{-}8.85\pm 0.91$ & $\phantom{-}0.50\pm 1.07$ \\
$ K^{*}(1410)^{-} (  \overline{K^{*0}}  \pi^{-}) K^{+}$ &
$\phantom{-}8.77\pm 1.06$ & $-4.24\pm 1.23$ &
$\phantom{-}7.43\pm 0.97$ & $-2.66\pm 1.15$ &
$\phantom{-}8.74\pm 1.06$ & $-4.28\pm 1.17$ &
$\phantom{-}9.09\pm 0.99$ & $-3.77\pm 1.22$ \\
$K^{*0}\left\{K^{-}\pi^{+}\right\}_P$ P wave &
 --- & --- & 
 --- & --- & 
 --- & --- & 
 --- & --- \\
$\overline{K^{*0}}\left\{K^{+} \pi^{-}\right\}_P$  P wave &
 --- & --- & 
 --- & --- & 
 --- & --- & 
 --- & --- \\
$K^{*0} \overline{K^{*0}}$  S wave  & 
$\phantom{-}0.23\pm 0.06$ & $\phantom{-}0.44\pm 0.06$ &
$\phantom{-}0.20\pm 0.06$ & $\phantom{-}0.46\pm 0.06$ &
$\phantom{-}0.24\pm 0.05$ & $\phantom{-}0.44\pm 0.06$ &
$\phantom{-}0.20\pm 0.06$ & $\phantom{-}0.46\pm 0.06$ \\
$K^{*0}\overline{K^{*0}}$ P wave  &
 --- & --- &
 $-0.03\pm 0.06$ & $\phantom{-}0.27\pm 0.04$ &
--- & --- &
--- & --- \\
$\phi\left\{\pi^{+} \pi^{-}\right\}_{S}$ &
$\phantom{-}3.52\pm 0.97$ & $-7.66\pm 0.71$ &
$\phantom{-}3.86\pm 1.02$ & $-7.58\pm 0.74$ &
$\phantom{-}3.74\pm 0.95$ & $-7.85\pm 0.64$ &
$\phantom{-}4.35\pm 1.06$ & $-7.72\pm 0.81$ \\
$\phi\left\{\pi^{+} \pi^{-}\right\}_P$  S wave &
$-3.04\pm 0.42$ & $-1.45\pm 0.40$ &
$-2.91\pm 0.45$ & $-1.60\pm 0.39$ &
$-4.03\pm 0.46$ & $-2.05\pm 0.64$ &
--- & --- \\
$\phi\left\{\pi^{+} \pi^{-}\right\}_P$  P wave & 
 --- & --- &
 --- & --- &
 --- & --- &
 --- & --- \\
 $\phi\left\{\pi^{+} \pi^{-}\right\}_P$  D wave &
 --- & --- &
 --- & --- &
 $\phantom{-}8.97\pm 2.15$ & $\phantom{-}3.44\pm 2.42$ &
 --- & --- \\
$\phi\rho^{0}$ S wave & 
$-0.80\pm 0.13$ & $\phantom{-}0.76\pm 0.13$ &
$-0.84\pm 0.13$ & $\phantom{-}0.71\pm 0.15$ &
$-0.61\pm 0.09$ & $\phantom{-}0.74\pm 0.11$ &
$-1.33\pm 0.12$ & $\phantom{-}0.68\pm 0.21$ \\ 
$\phi\rho^{0}$ D wave & 
$\phantom{-}1.88\pm 0.31$ & $-1.06\pm 0.37$ & 
$\phantom{-}1.97\pm 0.30$ & $-0.92\pm 0.38$ &
$\phantom{-}0.67\pm 0.31$ & $-0.98\pm 0.33$ &
$\phantom{-}1.70\pm 0.30$ & $-1.27\pm 0.42$ \\ 
$\rho^{0}\left\{K^{+}K^{-}\right\}_P$  S wave & 
--- & --- &
--- & --- &
--- & --- &
$-0.39\pm 0.91$ & $-4.32\pm 1.12$ \\ 
$\left\{K^-\pi^+\right\}_P \left\{K^+ \pi^-\right\}_S$ &
$\phantom{-}81.8\pm 11.8$ & $\phantom{-}91.2\pm 11.6$ &
$\phantom{-}80.0\pm 12.2$ & $\phantom{-}93.5\pm 11.8$ &
$\phantom{-}83.0\pm 10.5$ & $\phantom{-}89.8\pm 11.3$ &
$\phantom{-}70.4\pm 11.7$ & $\phantom{-}86.3\pm 10.6$ \\
$\left\{K^+\pi^{-}\right\}_P\left\{ K^{-}\pi^{+}\right\}_S$ &  
--- & --- &
--- & --- &
--- & --- &
--- & --- \\ 
$\left\{K^- \pi^+\right\}_P \left\{K^+ \pi^-\right\}_P$  &
--- & --- &
--- & --- &
--- & --- &
--- & --- \\ 
$\left\{\pi^+\pi^-\right\}_P \left\{K^+K^-\right\}_P$  &
--- & --- &
--- & --- &
--- & --- &
--- & --- \\ 
\hline\hline
\end{tabular}
}}
\end{center}
\end{table}

\begin{table}
\begin{center}
\caption{$a_i$ for alternative models 6-9. Uncertainties are statistical only.} \label{tab:altmodpar2} %%@
\vspace*{0.2cm}
\rotatebox{90}{\mbox{\footnotesize
\begin{tabular}{lcccccccc}
\hline\hline
Model   & \multicolumn{2}{c}{6}  
        & \multicolumn{2}{c}{7}
        & \multicolumn{2}{c}{8}
	    & \multicolumn{2}{c}{9} \\			
		& $\mathrm{Re}(a_i)$ & $\mathrm{Im}(a_i)$
		& $\mathrm{Re}(a_i)$ & $\mathrm{Im}(a_i)$
	    & $\mathrm{Re}(a_i)$ & $\mathrm{Im}(a_i)$
		& $\mathrm{Re}(a_i)$ & $\mathrm{Im}(a_i)$ \\ \hline
$K_{1}(1270)^{+}( K^{*0}\pi^{+})K^{-}$ &
1.0 & 0.0 & 
1.0 & 0.0 & 
1.0 & 0.0 & 
1.0 & 0.0 \\
$K_{1}(1270)^{-}( \overline{K^{*0}}\pi^{-})K^{+}$ &  
$\phantom{-}0.51\pm 0.09$ & $-0.36\pm 0.09$ &
$\phantom{-}0.44\pm 0.10$ & $-0.21\pm 0.10$ &
$\phantom{-}0.36\pm 0.09$ & $-0.27\pm 0.11$ &
$\phantom{-}0.17\pm 0.08$ & $-0.25\pm 0.07$ \\
$K_{1}(1270)^{+}(\rho^{0}K^{+})K^{-}$ & 
$\phantom{-}3.81\pm 0.73$ & $\phantom{-}6.97\pm 0.74$ &
$\phantom{-}3.11\pm 1.41$ & $\phantom{-}9.83\pm 1.08$ &
$\phantom{-}2.26\pm 0.74$ & $\phantom{-}5.75\pm 0.77$ &
$\phantom{-}2.92\pm 0.60$ & $\phantom{-}4.35\pm 0.75$ \\
$K_{1}(1270)^{-}(\rho^{0}K^{-})K^{+}$ &
$\phantom{-}7.67\pm 1.01$ & $\phantom{-}5.24\pm 0.84$ &
$\phantom{-}5.04\pm 1.15$ & $\phantom{-}5.92\pm 1.21$ &
$\phantom{-}7.44\pm 0.76$ & $-0.14\pm 0.99$ &
$\phantom{-}6.27\pm 0.60$ & $\phantom{-}0.79\pm 0.90$ \\
$ K^{*}(1410)^{+} ( K^{*0} \pi^{+})K^{-}$ &
--- & --- &
$\phantom{-}9.15\pm 0.91$ & $\phantom{-}0.10\pm 1.06$ &
$\phantom{-}3.64\pm 0.69$ & $-3.63\pm 0.85$ &
$\phantom{-}3.77\pm 0.58$ & $-3.31\pm 0.72$ \\
$ K^{*}(1410)^{-} (  \overline{K^{*0}}  \pi^{-}) K^{+}$ &
--- & --- &
$\phantom{-}8.99\pm 1.04$ & $-4.23\pm 1.17$ &
$\phantom{-}5.03\pm 0.80$ & $-4.32\pm 0.72$ &
$\phantom{-}3.56\pm 0.80$ & $-5.19\pm 0.62$ \\
$K^{*0}\left\{K^{-}\pi^{+}\right\}_P$ P wave &
 $-9.28\pm 1.07$ & $-3.44\pm 0.84$ & 
 --- & --- & 
 --- & --- & 
 --- & --- \\
$\overline{K^{*0}}\left\{K^{+} \pi^{-}\right\}_P$  P wave &
 $-8.48\pm 0.95$ & $\phantom{-}2.90\pm 0.99$ & 
 --- & --- & 
 --- & --- & 
 --- & --- \\
$K^{*0} \overline{K^{*0}}$  S wave  & 
$\phantom{-}0.12\pm 0.04$ & $\phantom{-}0.35\pm 0.04$ &
$\phantom{-}0.24\pm 0.06$ & $\phantom{-}0.46\pm 0.06$ &
$\phantom{-}0.21\pm 0.04$ & $\phantom{-}0.13\pm 0.04$ &
$\phantom{-}0.24\pm 0.03$ & $\phantom{-}0.13\pm 0.04$ \\
$K^{*0}\overline{K^{*0}}$ P wave  &
 --- & --- &
 --- & --- &
--- & --- &
--- & --- \\
$\phi\left\{\pi^{+} \pi^{-}\right\}_{S}$ &
--- & --- &
$\phantom{-}4.00\pm 0.98$ & $-7.57\pm 0.78$ &
$\phantom{-}1.30\pm 1.02$ & $-4.73\pm 0.44$ &
$-0.81\pm 0.66$ & $-5.29\pm 0.42$ \\
$\phi\left\{\pi^{+} \pi^{-}\right\}_P$  S wave &
--- & --- &
$-2.94\pm 0.45$ & $-1.54\pm 0.37$ &
--- & --- &
--- & --- \\
$\phi\left\{\pi^{+} \pi^{-}\right\}_P$  P wave & 
 $-2.87\pm 0.49$ & $\phantom{-}6.03\pm 0.58$  &
 --- & --- &
 --- & --- &
 --- & --- \\
 $\phi\left\{\pi^{+} \pi^{-}\right\}_P$  D wave &
 --- & --- &
 --- & --- &
 --- & --- &
 --- & --- \\
$\phi\rho^{0}$ S wave & 
$-1.07\pm 0.06$ & $\phantom{-}0.32\pm 0.08$ &
$-0.86\pm 0.13$ & $\phantom{-}0.72\pm 0.13$ &
$-0.78\pm 0.09$ & $\phantom{-}0.48\pm 0.13$ &
$-0.53\pm 0.10$ & $\phantom{-}0.76\pm 0.09$ \\ 
$\phi\rho^{0}$ D wave & 
--- & --- & 
$\phantom{-}1.99\pm 0.28$ & $-1.04\pm 0.35$ &
$\phantom{-}0.96\pm 0.22$ & $-0.78\pm 0.23$ &
$\phantom{-}0.70\pm 0.24$ & $-0.98\pm 0.20$ \\ 
$\rho^{0}\left\{K^{+}K^{-}\right\}_P$  S wave & 
--- & --- &
$-0.37\pm 1.05$ & $-4.53\pm 1.16$ &
--- & --- &
--- & --- \\ 
$\left\{K^-\pi^+\right\}_P \left\{K^+ \pi^-\right\}_S$ &
--- & --- &
$\phantom{-}78.1\pm 11.2$ & $\phantom{-}83.5\pm 11.2$ &
--- & --- &
$\phantom{-}68.1\pm 6.5$ & $\phantom{-}16.7\pm 7.7$ \\
$\left\{K^+\pi^{-}\right\}_P\left\{ K^{-}\pi^{+}\right\}_S$ &  
--- & --- &
--- & --- &
$\phantom{-}69.2\pm 6.7$ & $\phantom{-}40.8\pm 8.3$ &
$\phantom{-}45.6\pm 5.7$ & $\phantom{-}25.2\pm 6.0$ \\ 
$\left\{K^- \pi^+\right\}_P \left\{K^+ \pi^-\right\}_P$ D wave &
$-401.0\pm 32.7$ & $-344.0\pm 32.8$ &
--- & --- &
--- & --- &
--- & --- \\ 
$\left\{\pi^+\pi^-\right\}_P \left\{K^+K^-\right\}_P$  D wave &
$-105.0\pm 26.9$ & $-151.0\pm 23.3$ &
--- & --- &
--- & --- &
--- & --- \\ 
\hline\hline
\end{tabular}
}}
\end{center}
\end{table}

\end{document}